\documentclass[fleqn,usenatbib]{mnras}

\usepackage{newtxtext,newtxmath}
\usepackage[T1]{fontenc}

\DeclareRobustCommand{\VAN}[3]{#2}
\let\VANthebibliography\thebibliography
\def\thebibliography{\DeclareRobustCommand{\VAN}[3]{##3}\VANthebibliography}

\usepackage{graphicx}	% Including figure files
\usepackage{amsmath}	% Advanced maths commands
\usepackage[dvipsnames]{xcolor}
\usepackage[normalem]{ulem}
\usepackage{booktabs}
\usepackage{xspace}
\newcommand{\LCDM}{$\Lambda$CDM\xspace} %the xspace command adds a space where needed (everywhere except before a full-stop or comma or closing bracket)
\newcommand{\pyLDT}{\texttt{pyLDT}}
\newcommand{\Mpch}{$\mathrm{Mpc}/h$}
\newcommand{\hMpc}{$h\mathrm{Mpc}^{-1}$}
\newcommand{\fR}[1]{|f_{R#1}|}

\newcommand{\mP}{{\cal P}}
\newcommand{\Om}{\Omega_{\mathrm{m}}}
\newcommand{\Ob}{\Omega_{\mathrm{b}}}
\newcommand{\Ol}{\Omega_{\Lambda}}
\newcommand{\Orc}{\Omega_{\mathrm{rc}}}
\newcommand{\Oeff}{\Omega_{\mathrm{eff}}}
\newcommand{\fr}{$f(R)$ gravity\xspace}
\newcommand{\newtau}{\delta_{\rm L}}

\definecolor{darkgreen}{cmyk}{0.85,0.2,1.00,0.05}

\newcommand\gsout{\bgroup\markoverwith{\textcolor{darkgreen}{\rule[0.5ex]{2pt}{0.6pt}}}\ULon}

\title[Matter density PDF for modified gravity and dark energy]{The matter density PDF for modified gravity and dark energy with Large Deviations Theory}

\author[M. Cataneo et al.]{
Matteo Cataneo$^{1}$\thanks{E-mail: matteo@roe.ac.uk},
Cora Uhlemann$^{2}$,
Christian Arnold$^{3}$,
Alex Gough$^{2}$,
Baojiu Li$^{3}$,
Catherine Heymans$^{1,4}$
\\
$^{1}$Institute for Astronomy, University of Edinburgh, Royal Observatory, Blackford Hill, Edinburgh, EH9 3HJ, U.K.\\
$^{2}$ {School of Mathematics, Statistics and Physics, Newcastle University, Herschel Building, NE1 7RU Newcastle-upon-Tyne, U.K.}\\
$^{3}$Institute for Computational Cosmology, Department of Physics, Durham University, South Road, Durham DH1 3LE, U.K.\\
$^{4}$Ruhr-Universität Bochum, Astronomisches Institut, German Centre for Cosmological Lensing (GCCL), Universitätsstr. 150, 44801, Bochum, Germany\\
}

\date{Accepted XXX. Received YYY; in original form ZZZ}

\pubyear{2021}

\begin{document}
\label{firstpage}
\pagerange{\pageref{firstpage}--\pageref{lastpage}}
\maketitle

\begin{abstract}
% We present an analytical description of the probability distribution function (PDF) of the smoothed three-dimensional matter density field for non-standard cosmological models. In a comparison to numerical simulations for $f(R)$, DGP and evolving dark energy theories, we find percent level accuracy for the model with smoothing scales of $R \gtrsim 10$ \Mpch. A Fisher forecast of an idealised experiment with a {\it Euclid}-like survey demonstrates the power of combining measurements of the 3D matter PDF with the 3D matter power spectrum. This combination is shown to half the uncertainty on parameters for an evolving dark energy model, relative to a power spectrum analysis on its own. The PDF is also found to increase the detection significance for modified gravity models to the $5\sigma/13\sigma$ level for DGP and F6 $f(R)$ respectively. Our approach, based on the principles of Large Deviation Theory, is applicable to general models. We show that changes to the standard cosmology can be included through Einstein-de Sitter spherical collapse dynamics combined with linear theory calculations and a measurement of the non-linear variance of the smoothed density field, which can be calibrated through a simple numerical simulation. This analysis is therefore very promising for future beyond-two-point statistical studies, as it has the potential to alleviate the reliance of these analyses on expensive high resolution simulations and emulators. 
%
We present an analytical description of the probability distribution function (PDF) of the smoothed three-dimensional matter density field for modified gravity and dark energy. Our approach, based on the principles of Large Deviations Theory, is applicable to general extensions of the standard \LCDM{} cosmology. We show that late-time changes to the law of gravity and background expansion can be included through Einstein-de Sitter spherical collapse dynamics combined with linear theory calculations and a calibration measurement of the non-linear variance of the smoothed density field from a simple numerical simulation. In a comparison to $N$-body simulations for $f(R)$, DGP and evolving dark energy theories, we find percent level accuracy around the peak of the distribution for predictions in the mildly non-linear regime. A Fisher forecast of an idealised experiment with a {\it Euclid}-like survey volume demonstrates the power of combining measurements of the 3D matter PDF with the 3D matter power spectrum. This combination is shown to halve the uncertainty on parameters for an evolving dark energy model, relative to a power spectrum analysis on its own. The PDF is also found to substantially increase the detection significance for small departures from General Relativity, with improvements of up to six times compared to the power spectrum alone. This analysis is therefore very promising for future studies including non-Gaussian statistics, as it has the potential to alleviate the reliance of these analyses on expensive high resolution simulations and emulators.

\end{abstract}


%%%%%%%%%%%%%%%%%%%%%%%%%%%%%%%%%%%%%%%%%%%%%%%%%%

%%%%%%%%%%%%%%%%% BODY OF PAPER %%%%%%%%%%%%%%%%%%

\section{Introduction}

Over the past two decades an extraordinary and diverse array of experimental evidence has earned the Lambda-Cold Dark Matter (\LCDM{}) paradigm the status of standard cosmological model \citep[see, e.g.,][for recent analyses]{Planck:2018,Aiola:2020,Hamana:2020,Dutcher:2021,Heymans:2021,DES-y3kp:2021,Alam:2021}. However, in recent years mild to severe tensions between early- and late-time probes of the growth of structure and background expansion have put a strain on the ability of \LCDM{} to explain our universe \citep[see, e.g.,][for reviews]{Douspis:2019,DiValentino:2021b,DiValentino:2021,Perivolaropoulos:2021}. Furthermore, Einstein’s general relativity (GR)—the theory of gravity at the foundation of \LCDM{}—has been thoroughly tested only on small astrophysical scales and in the strong field regime \citep{Will:2014,Abbott:2017,Abbott:2019}, leaving ample room for modifications to the field equations on cosmological scales \citep{DES-y1MG:2019,Ishak:2019,Ferreira:2019,Troester:2021,Raveri:2021,Pogosian:2021}. Together with the yet unexplained nature of the observed accelerated cosmic expansion \citep{Riess:1998,Perlmutter:1999}, these considerations motivate the exploration of alternatives to the cosmological constant ($\Lambda$) and standard gravity. In this paper, we focus on extensions of \LCDM{} that include modified gravity and (late-time) dark energy, which we will concisely refer to as `extended models'.

Two-point statistics are central to many of the leading cosmological analyses of the large-scale structure searching for deviations from \LCDM{} \citep{Simpson:2013,Song:2015,Amon_2018_Eg,DES-y1MG:2019,Troester:2021,Lee:2021,Muir:2021,Chudaykin:2021,Vazsonyi:2021}, and a great deal of effort has gone into accurately modelling the non-linear matter power spectrum in modified gravity and dark energy cosmologies—a theoretical ingredient essential to extract the cosmological information locked in small scales \citep[e.g.,][]{Koyama:2009,Takahashi:2012,Brax:2012,Heitmann:2014,Zhao:2014,Mead:2016,Casarini:2016,Cusin:2018,Cataneo:2019,Winther:2019,Ramachandra:2021,EE2:2021}. However, non-linear gravitational clustering converts the nearly Gaussian initial density field \citep{Planck:2018} to a late-time density field with significant non-Gaussian features that these standard analyses are unable to access \citep{Bernardeau:2002}. Non-Gaussian statistics, such as the bispectrum \citep{Brax:2012,Munshi:2017,Yamauchi:2017,Crisostomi:2020,Bose:2020b}, higher-order weak lensing spectra \citep{Munshi:2020}, the halo mass function \citep{Lam_2012,Cataneo:2016,Hagstotz:2019,McClintock:2019,Bocquet:2020}, the void size function \citep{Perico:2019,Verza:2019,Contarini:2021} and Minkowski functionals \citep{Kratochvil:2012,Fang:2017}, respond strongly to modified gravity and dark energy through the induced changes in the higher moments of the cosmic density field, and their remarkable complementarity to traditional two-point functions leads to tighter joint constraints on the extra non-standard parameters \citep{Shirasaki:2017,Peel:2018,Sahlen:2019,Liu:2021}. 

The probability distribution function (hereafter PDF) of the three-dimensional matter density field smoothed on a given scale is one of the simplest non-Gaussian statistics, and accurate predictions allow us to extract additional cosmological information \citep{Uhlemann:2020}. Modified gravity and evolving dark energy leave distinctive imprints on the skewness, kurtosis and higher cumulants of the PDF, which have been observed in $N$-body simulations \citep{Li_2012halosvoidsfR,Hellwing:2013,Hellwing:2017,Shin:2017}. Thus far, theoretical predictions of the PDF for modified gravity have required either sophisticated and time-consuming approaches for the spherical collapse \citep{Brax:2012} or computationally costly simulations \citep{Li_2012halosvoidsfR,Hellwing:2013,Hellwing:2017}. Similarly, for the PDF response to dark energy beyond a cosmological constant, so far only ad-hoc fitting functions obtained from simulations are available \citep{Shin:2017,Mandal_2020,Wen_2020}. In this work, we apply the principles of Large Deviation Theory (LDT) to the cosmic density field to derive a general and analytical prescription for the 3D matter PDF in modified gravity and dark energy cosmologies. We build on the formalism developed in \cite{LDPinLSS} and \cite{Uhlemann16}, and show that the Einstein-de Sitter spherical collapse dynamics together with linear theory calculations can reproduce the 3D matter PDF measured from state-of-the-art simulations to a few percent accuracy in the mildly non-linear regime. Remarkably, the matter PDF can be accurately predicted requiring only linear information from extended cosmologies, which sources the characteristic differences in the non-linear variance and higher cumulants. Through Fisher forecasts we quantify, for the first time, the ability of the PDF to detect distinct departures from GR and to constrain the dark energy equation of state, especially when combined with the matter power spectrum. Our method implemented in the public code \pyLDT{}\footnote{\url{https://github.com/mcataneo/pyLDT-cosmo}} delivers fast predictions for the cosmology-dependence of the PDF, and it paves the way for the modelling of observable statistics of the large-scale structure in general theories of gravity and dark energy \citep[see][for a review]{Frusciante:2020}. Our framework can be applied to predict the weak lensing convergence PDF \citep{Barthelemy:2020,Boyle_2020}, galaxy counts-in-cells \citep{Uhlemann:2018b,Repp_2020,Friedrich_2021}, and density-split statistics \citep{Friedrich_2018,Gruen:2018}. 

The paper is structured as follows: Section \ref{sec:LDT_PDF} reviews the LDT framework and provides a simple extension to include the effects of modified gravitational couplings and background expansion on the PDF. Section \ref{sec:sims} describes the $N$-body simulations and PDF measurements used to validate the theoretical predictions. In Section \ref{sec:results} we demonstrate the accuracy of our methodology, as well as the complementarity of the matter PDF and the power spectrum for modified gravity and dark energy parameters using Fisher matrix analyses. We summarise our results and give an outlook on future research in Section \ref{sec:conclusions}.

\section{The matter density PDF in Large Deviations Theory}
\label{sec:LDT_PDF}

\subsection{Large deviations theory framework}
\label{sec:LDT}

Large deviations theory \citep[see][for a basic introduction]{Touchette_2012} provides a means to predict the probability density function (PDF) of non-linear matter densities in spheres \citep{Bernardeau94,Valageas02,Bernardeau14,LDPinLSS,Uhlemann16}. The formalism can be applied on mildly non-linear scales quantified by the value of the non-linear variance $\sigma_{\rm NL}^2$ at the redshift $z$ and radius $R$ of interest as long as $\sigma_{\rm NL}^2(R,z)<1$. For Gaussian initial conditions\footnote{For extensions that include primordial non-Gaussianity see \cite{Uhlemann18pNG,Friedrich20pNG}.}, the PDF, $\mP(\newtau)$, of the linear matter density contrast, $\newtau$, in a sphere of radius $r$ is a Gaussian distribution where the width is fully specified by the linear variance, $\sigma_{\rm L}^2$ at that scale $r$ and redshift $z$
\begin{align}
\label{eq:PDFlin}
\mathcal P_{r,z}^{\rm lin}(\newtau) &= \sqrt{\frac{1}{2\pi\sigma_{\rm L}^2(r,z)}} \exp\left[-\frac{ \newtau^2}{2\sigma^2_{\rm L}(r,z)}\right]\,.
\end{align}
The linear variance at scale $r$ is obtained from an integral over the linear power spectrum, $P_{\rm L}$, with a spherical top-hat filter in position space
% \begin{equation}
% \sigma^2_{\rm L}(r,z) 
% = \int \frac{dk}{2\pi^2} P_{\rm L}(k,z)  k^2 W^2_{\rm 3D}(k r)\,,
% W_{\rm 3D}(k)=3\sqrt{\frac{\pi}{2}}\frac{J_{3/2}(k)}{k^{3/2}}\,,
% \label{eq:defSigma2lin}
% \end{equation}
\begin{equation}
\sigma^2_{\rm L}(r,z) 
= \int \frac{dk}{2\pi^2} P_{\rm L}(k,z)  k^2 W^2_{\rm 3D}(k r)\,,
\label{eq:defSigma2lin}
\end{equation}
where  $W_{\rm 3D}(k)$ is the Fourier transform of the 3D spherical top-hat filter. 
%and $J_{3/2}(k)$ is the Bessel function of the first kind of order $3/2$. 

To describe the impact of non-linear gravitational dynamics on the shape of the initially Gaussian matter PDF, it is informative to look at the exponential decay of the PDF with increasing density contrast. To formalise this argument, one considers the exponential decay of the PDF in equation~\eqref{eq:PDFlin} encoded in the decay-rate function
\begin{align}
\label{eq:PsiL}
\Psi_{r,z}^{\rm lin}(\newtau)&= \frac{ \newtau^2}{2\sigma^2_{\rm L}(r,z)}\,.
\end{align}
In general, the non-linear matter PDF can be written as a path integral over all possible ways to realise a non-linear normalised density $\rho=1+\delta$ from a given linear density contrast. But since large deviations are exponentially unlikely, there is only one path, namely the least unlikely one, which dominates this complex integral. The dominant contribution is a saddle point of the corresponding functional integral, which is given by the spherical collapse dynamics thanks to the spherical symmetry of the cells and statistical isotropy ensuring average density profiles to be spherical \citep{Bernardeau94,Valageas02,Ivanov_2019}. 
This idea leads to the contraction principle of large deviation statistics \citep{LDPinLSS}, which states that the decay-rate function of the final sphere density $\rho$ (at scale $R$ and redshift $z$) can be obtained from the initial one by using the spherical collapse mapping $\rho=\rho_{\rm SC}(\newtau)$ to obtain the associated most likely linear density contrast $\newtau(\rho)$ and mass conservation for the initial radius $r=R\rho^{1/3}$, such that 
\begin{align}
\label{eq:PsiNL}
 \Psi_{R,z}(\rho)&
= \frac{\sigma^2_{\rm L}(R,z)}{\sigma_{\rm NL}^2(R,z)} \frac{\newtau^2(\rho)}{2\sigma^2_{\rm L}(R\rho^{1/3},z)}\,.
\end{align}
The prefactor arises from performing calculations with a decay-rate function rescaled with $\sigma_{\rm L}^2(R,z)$, which renders it a proper rate function described by large deviation statistics \citep{LDPinLSS,Uhlemann16} and ensures a well-defined $\sigma\rightarrow 0$ limit, and then restoring the desired final non-linear variance $\sigma_{\rm NL}^2$.\footnote{Physically speaking, this procedure amounts to asserting that the reduced cumulants (discussed later), encoded in the large deviation statistics rate function and predicted from spherical collapse in the limit $\sigma^2\rightarrow 0$, can reliably be extrapolated to small, nonzero variances \citep[as demonstrated with simulated data in][]{Uhlemann16}. The $\sigma_{\rm NL}^2$ factor in the denominator then plays the role of converting the reduced cumulants back to the cumulants using the correct nonlinear variance, which controls the width of the PDF.}
% where $\sigma_{\rm NL}^2$ is the non-linear variance at scale $R$ and redshift $z$. 

From the decay-rate function in Equation~\eqref{eq:PsiNL} one can reconstruct the full PDF. This can be achieved by 
computing the cumulant generating function via a Legendre transform, which in turn allows to compute the final PDF  via an inverse Laplace transform. This integral can be computed numerically \citep{Bernardeau14}, but an
excellent analytical approximation can be obtained from a saddle-point approximation for the log-density $\mu=\ln\rho$ \citep{Uhlemann16}. To achieve this, the decay-rate function in Equation~\eqref{eq:PsiNL} is rewritten in terms of the logarithmic density
$\mu$ and its non-linear variance $\sigma^2_\mu$
\begin{subequations}
\label{eq:PDFpred}
\begin{align}
\label{eq:PsiNLlog}
 \Psi_{R,z}(\mu)&
= \frac{\sigma^2_{\rm L}(R,z)}{\sigma_{\rm \mu}^2(R,z)} \frac{\newtau^2(\rho(\mu))}{2\sigma^2_{\rm L}(R\rho^{1/3},z)}\,.
\end{align}
Then the matter density PDF, $\mP_{R,z}(\rho)$, is obtained from
\begin{align}
\label{eq:PDFfromPsilog}
\tilde\mP_{R,z}(\tilde\rho) = \sqrt{\frac{\Psi''_{R,z}(\tilde\rho)+\Psi'_{R,z}(\tilde\rho)/\tilde\rho}{2\pi}} \exp\left[-\Psi_{R,z}(\tilde\rho)\right]\,.
\end{align}
The prefactor arises from a combination of the second derivative of the decay-rate function with respect to the logarithmic density and the Jacobian of the nonlinear transformation.

Because of the use of the log-transform, one has to ensure the correct mean density $\langle\rho\rangle=\int d\rho\, \rho\, \mP(\rho)=1$ by specifying the mean of the log-density $\langle\ln\rho\rangle$. This can be implemented by rescaling the `raw' PDF, $\tilde\mP_{R,z}(\tilde{\rho})$, from Equation~\eqref{eq:PDFfromPsilog} as
\begin{align}
\label{eq:PDFnorm}
\mP_{R,z}(\rho) &= \tilde\mP_{R,z}\left(\rho \cdot \frac{\langle\tilde\rho\rangle}{\langle 1\rangle}\right)  \cdot \frac{\langle\tilde\rho\rangle}{\langle 1\rangle^2}\,,
\end{align}
\end{subequations}
where $\langle f(\tilde\rho) \rangle=\int d\tilde\rho\, f(\tilde\rho) \tilde\mP(\tilde\rho)$ for any function $f(\tilde\rho)$, such as $f=1$ and $f=\tilde\rho$ here.

Remarkably, for a standard $\Lambda$CDM universe, there are only three ingredients that enter this theoretical model for the matter PDF,
\begin{enumerate}
    \item the time- and scale-dependence of the linear variance, $\sigma_{\rm L}^2(r,z)$,
    \item the mapping from initial to final densities in spheres, $\rho_{\rm SC}(\newtau)$,
    \item the non-linear variance of the log-density at the sphere radius and redshift of interest, $\sigma_{\rm NL}^2(R,z) \rightarrow \sigma_{\mu}^2(R,z)$ \,.
\end{enumerate}
The linear variance and its cosmology dependence can be readily obtained from the linear power spectrum computed from Einstein-Boltzmann codes like CAMB \citep{CAMB} or CLASS \citep{CLASS}.

The spherical collapse mapping entering the matter PDF was shown to be very mildly cosmology-dependent and well-approximated by the redshift-independent Einstein-de Sitter (EdS) result in \cite{Uhlemann:2020}. This can also be seen in Figure~\ref{fig:tau}, where the fractional difference between the \LCDM{} and the EdS solution remains within 0.25\% for all non-linear densities considered. We develop a simple yet accurate EdS-based approximation for spherical collapse within modified gravity in the following section.

The non-linear variance of the log-density at the sphere radius and redshift of interest, $\sigma_{\mu}^2(R,z)$, can be considered a free parameter and measured directly from simulations. Once measured from a single (or small set of) simulations at the fiducial cosmology, its changes with cosmology can be predicted using a phenomenological approximation inspired from the lognormal model (see Equation \ref{eq:siglogcos}). Alternatively, the non-linear variance of the log-density could also be chosen to reproduce a predicted non-linear variance of the density, $\sigma_{\rho}^2(R,z)$, obtained from matter power spectrum fitting functions such as \textsc{halofit}  \citep{halofit}, {\sc hmcode} \citep{Mead:2021} or \textsc{respresso} \citep{Nishimichi17}.

\subsection{Large-deviations statistics in modified gravity and dark energy}
\label{sec:LDTwMG}

For scalar-tensor theories within the Horndeski class \citep{Horndeski:1974} the late-time growth of linear matter perturbations on sub-horizon scales in a spatially flat universe is governed by \citep{Gleyzes:2013,Bellini:2014}
\begin{align}\label{eq:growth}
    D^{\prime\prime} + \frac{3}{2a}\left[ 1-w_{\rm eff}(a)\Oeff(a) \right] D^{\prime} - \frac{3\Om(a)}{2a^2} [ 1+ \epsilon(k,a) ] D = 0 \, ,
\end{align}
where $D$ is the linear growth function such that the final linear density fluctuation $\hat\delta_{\rm L}(k,z) = D(k,z)\delta_{\rm ini}$, primes denote derivatives with respect to the scale factor, $a$, $\Oeff(a)$ and $w_{\rm eff}(a)$ are the energy density and equation of state, respectively, of the effective dark energy fluid driving the background acceleration, and $\epsilon(k,a)$ represents the scale- and time-dependent fractional deviation from the gravitational constant. We can recover the well-known result for \LCDM{} by setting $w_{\rm eff} = -1$ and $\epsilon = 0$. In what follows we shall consider only late-time extensions (`ext') to the standard cosmology, that is, $D_{\rm ext} \rightarrow D_{\Lambda}$ at sufficiently early times such that the initial conditions and the primary CMB anisotropies remain unchanged. Hence, the linear matter power spectrum in any of these extensions can be obtained by rescaling the initial \LCDM{} power spectrum as   
\begin{align}\label{eq:linear_pk}
    P_{\rm L}^{\rm ext}(k,z) = \left[ \frac{D_{\rm ext}(k, z)}{D_\Lambda(z_{\rm i})} \right]^2 P_{\rm L}^{\Lambda}(k,z_{\rm i}) \, ,
\end{align}
where $z_{\rm i}$ is taken deep in the matter-dominated era. This equation can then be used together with Equation~\eqref{eq:defSigma2lin} to provide the linear variance as a function of smoothing scale and redshift.

The next ingredient required in Equation~\eqref{eq:PsiNL} is the function mapping the final density, $\rho$, to the linearly forward-propagated initial density, $\newtau$. A spherical top-hat density fluctuation, $\delta$, evolves as \citep[see, e.g.,][]{Schmidt:2009b}
\begin{align}\label{eq:SCeqn}
    \ddot\delta + 2 H \dot\delta - \frac{4}{3} \frac{\dot{\delta}^{2}}{(1+\delta)} = \frac{3}{2} H^2 \Om (1 + {\cal F}) (1+\delta)\delta \, ,
\end{align}
where dots denote derivatives with respect to cosmic time, $H$ is the Hubble parameter, and for simplicity we have omitted the time dependence from all quantities. Here, ${\cal F}$ is a function describing departures from GR which also incorporates a generic screening mechanism to restore standard gravity in high-density environments \citep[see, e.g.,][]{Koyama:2018,Lombriser:2018}. Note that in the limit of small linear fluctuations ${\cal F} \rightarrow \epsilon$, and Equation~\eqref{eq:SCeqn} reduces to Equation~\eqref{eq:growth}. In the rest of this work we will neglect any non-linear screening mechanism and, in fact, we will argue that in the mildly non-linear regime ($R \gtrsim$ 10 \Mpch) any modified gravity and dark energy effect on the spherical collapse/expansion can be accurately captured by the following approximation
\begin{align}\label{eq:SCapprox}
    \newtau^{\rm ext}(\rho,z) \approx \frac{\sigma_{\rm L}^{\Lambda}(R\rho^{1/3},z)}{\sigma_{\rm L}^{\rm ext}(R\rho^{1/3},z)} \newtau^{\rm EdS}(\rho) \, ,
\end{align}
where $\newtau^{\rm EdS}$ corresponds to the mapping between the final and the initial density fluctuations in an Einstein-de Sitter universe, i.e. $\Om(a) = 1$ and ${\cal F} = 0$ in Equation~\eqref{eq:SCeqn}. For reasons that will be discussed in Sec.~\ref{sec:MGtheo}, our definition of $\newtau^{\rm ext}$ in Eq.~\eqref{eq:SCapprox} does not match the linear density contrast solution to Eq.~\eqref{eq:growth}, in that we use the \LCDM{} linear growth, $D_{\Lambda}$, to extrapolate the initial density fluctuation, $\delta_{\rm ini,ext}$, to the final redshift rather than the modified growth, $D_{\rm ext}$. For scale-independent late-time extensions (i.e. $\sigma_{\rm L}^{\Lambda}/\sigma_{\rm L}^{\rm ext}=D_\Lambda/D_{\rm ext}$ and $\sigma_{\rm L}^{\rm ext, ini} \approx \sigma_{\rm L}^{\rm \Lambda, ini}$), one can alternatively use the modified growth for the extrapolation, i.e. $\tilde\delta_{\rm L}^{\rm ext} \equiv D_{\rm ext}\delta_{\rm ini,ext}$, and arrive at the following approximation ${\tilde\delta_{\rm L}}^{\rm ext}(\rho,z) \approx \newtau^{\rm EdS}(\rho)$\footnote{At first glance this result seems at odds with the notion that dark energy and modified gravity affect the growth of structure. However, here we are fixing the final non-linear density, $\rho$, such that enhancements (suppressions) of the linear growth require lower (higher) initial density contrasts, $\delta_{\rm ini}$, to match that particular $\rho$. In other words, adjustments to the initial conditions compensate for the linear growth modifications to a very good approximation.}. It is easy to show that these two approximations are equivalent and provide the same rate function--we opt for Eq.~\eqref{eq:SCapprox} simply because it explicitly accounts for scale-dependent modifications as well.

\subsubsection{Modified gravity}\label{sec:MGtheo}

\begin{figure}
    \centering
    \includegraphics[width=\columnwidth]{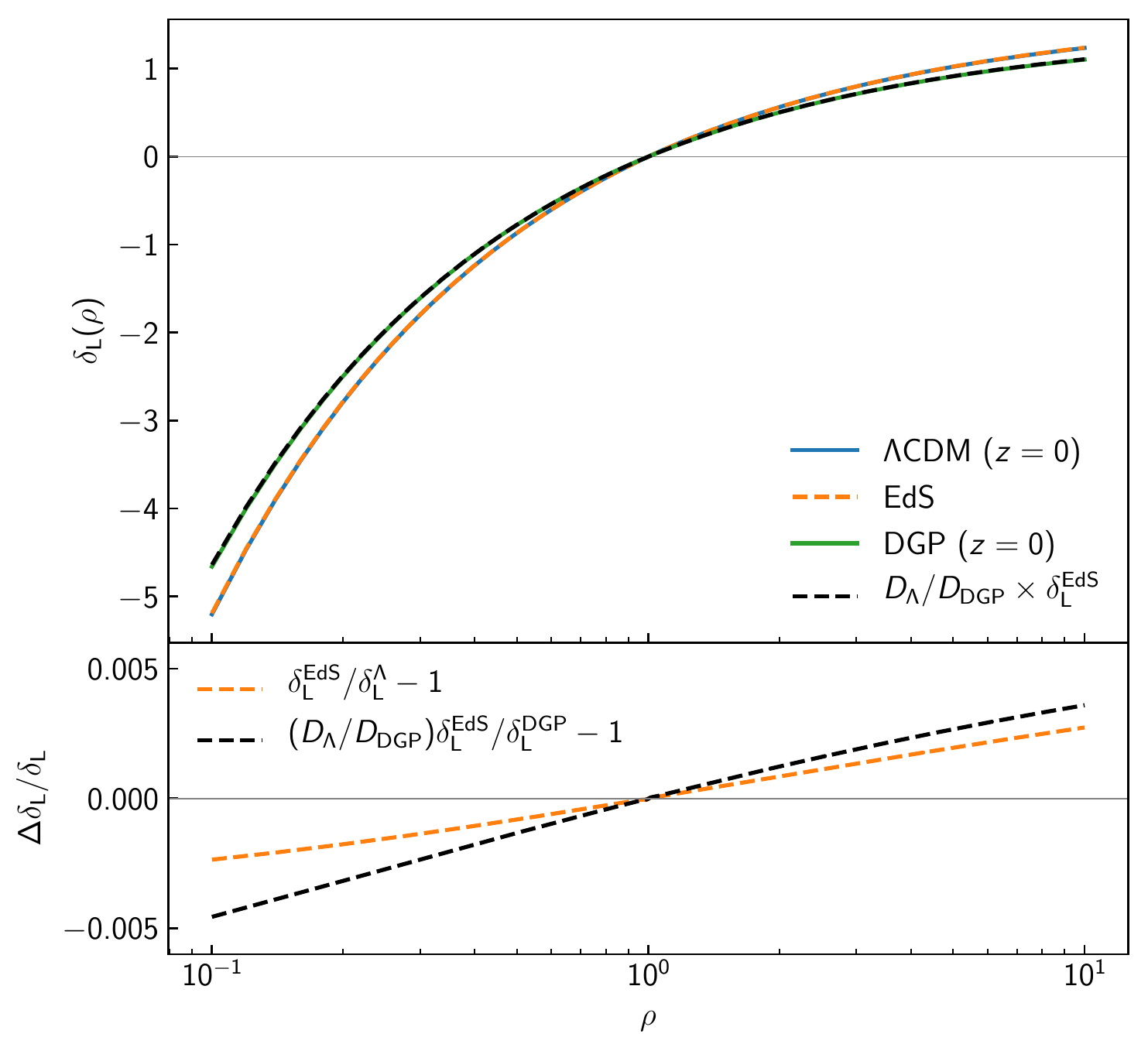}
    \caption{Mapping between the normalised final density, $\rho$, and the initial linearly-scaled density fluctuation, $\newtau$, for a spherical top-hat perturbation. \textit{Upper panel:} the curves show the density evolution in different background/gravity models. For \LCDM\ (blue) and Einstein-de Sitter (dashed orange) cosmologies gravity is GR, while for DGP (green) the gravitational constant is modified as in Equation~\eqref{eq:eps_dgp} with $r_{\rm c}H_0 = 0.5$ (or $\Orc = 0.25$). Here, both \LCDM\ and DGP are evaluated at $z=0$. The dashed black line represents the Einstein-de Sitter mapping rescaled by the ratio of the \LCDM{}-to-DGP linear growth ratio at $z=0$. \textit{Lower panel:} fractional difference of the Einstein-de Sitter mapping from the \LCDM\ prediction (dashed orange) and that of the rescaled Einstein-de Sitter from the DGP evolution (dashed black). The rescaled $\newtau^{\rm EdS}$ can reproduce the modified gravity phenomenology to better than 0.5\%, and it can be seen that most of the difference comes from the discrepancy between the Einstein-de Sitter and the \LCDM\ predictions.}
    \label{fig:tau}
\end{figure}
\begin{figure*}
    \begin{center}
    \includegraphics[width=\columnwidth]{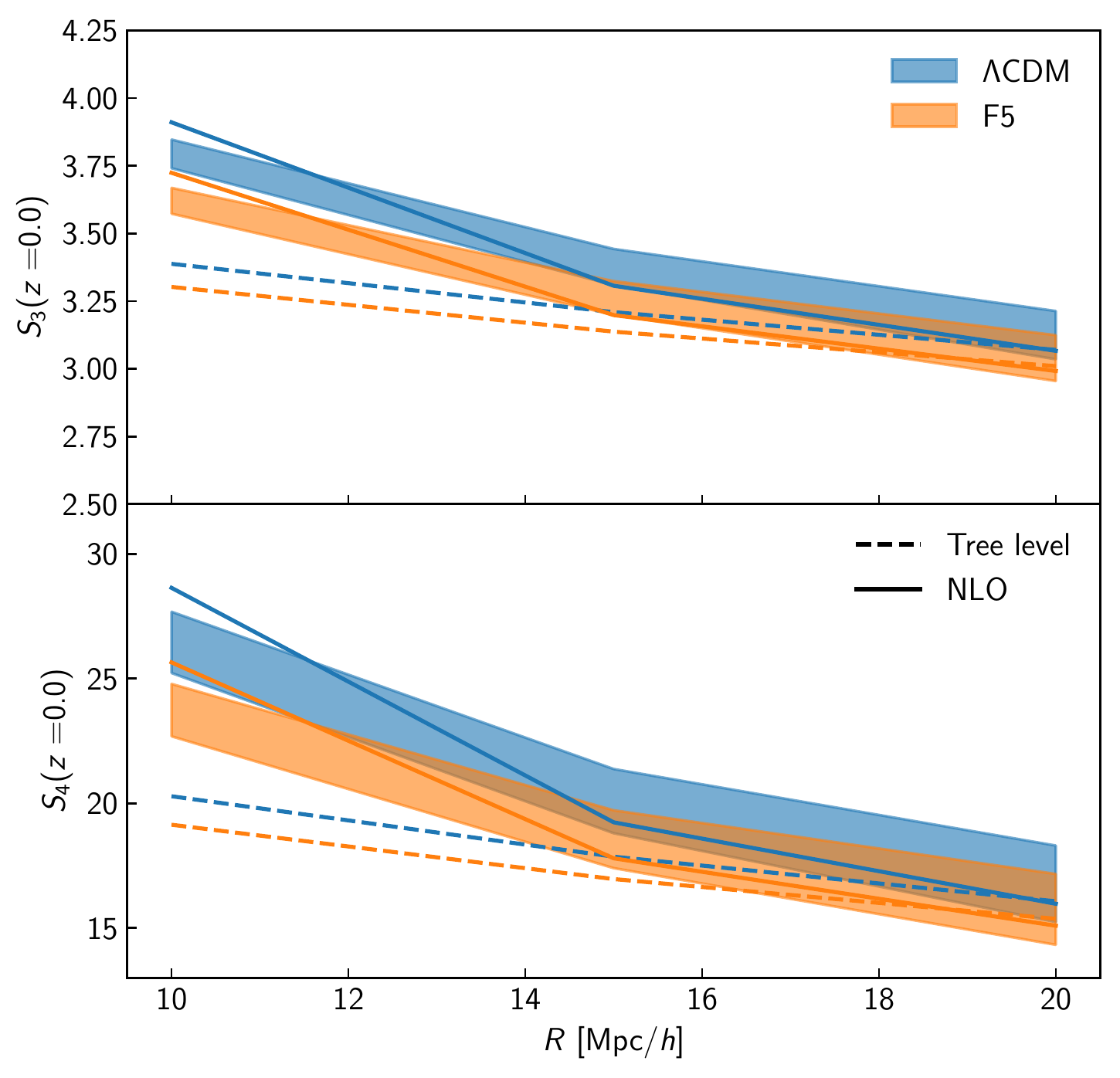}
    \quad
    \includegraphics[width=\columnwidth]{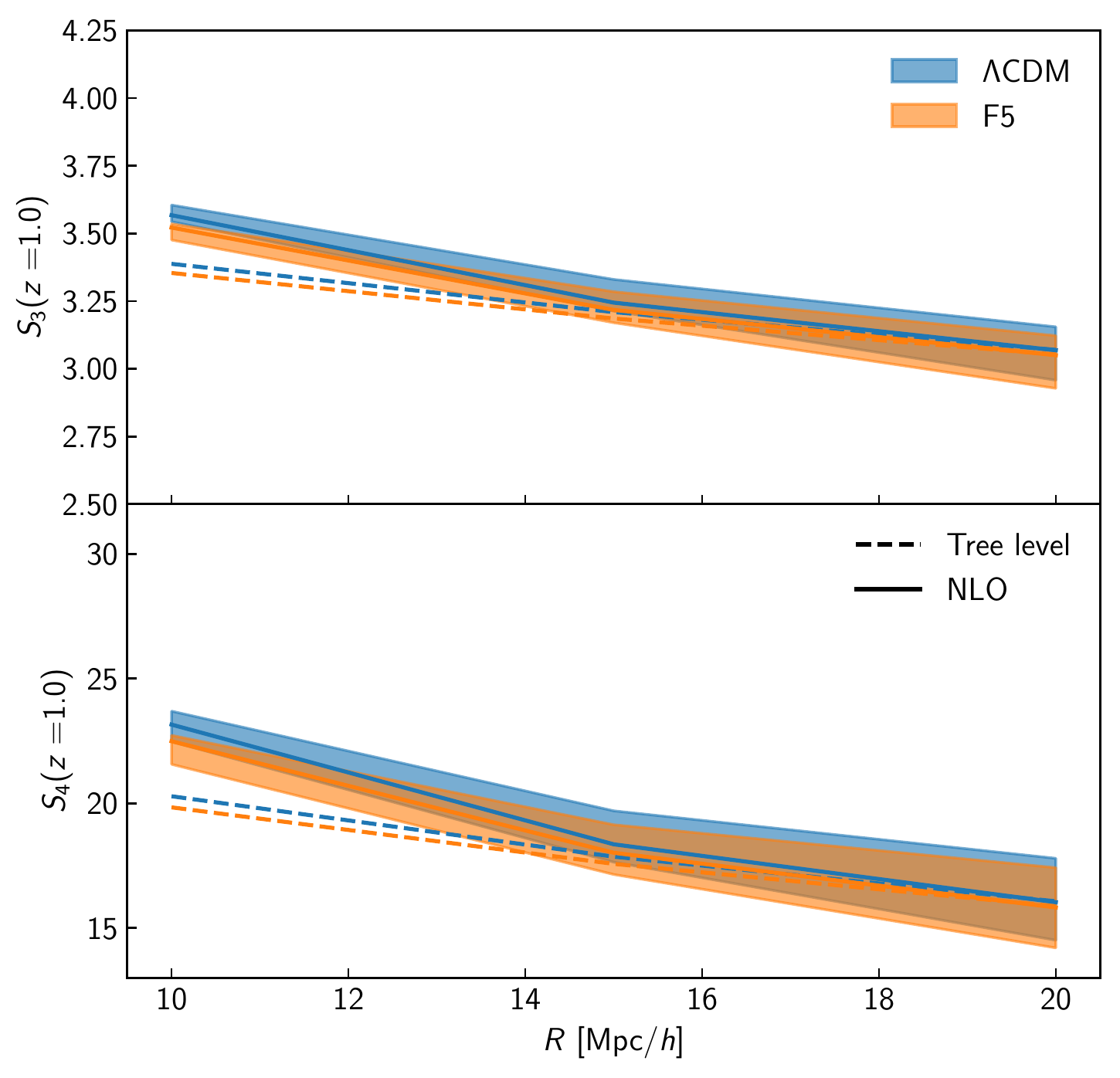}
    \end{center}
    \caption{Reduced cumulants associated with skewness (top) and kurtosis (bottom) of the smoothed matter density PDF at $z=0$ (left) and $z=1$ (right) for \LCDM\ (blue) and $f(R)$ gravity with $|f_{R0}| = 10^{-5}$ and $n=1$ (orange). For each smoothing radius, $R = 10, 15, 20$ \Mpch, the colored bands represent the mean and error on the mean across 8 $N$-body realisations, and the lines correspond to the theoretical predictions. Solid lines are computed with the tree-level approximation and dashed lines use the next-to-leading order (NLO) correction Equation~\eqref{eq:cumulants}, with all vertices derived from the Einstein-de Sitter spherical collapse dynamics. The cosmology-dependence and modified gravity effects are mostly sourced by the linear variance, while the non-linear variance contributes only to a minor extent.}
    \label{fig:fr_cumulants}
\end{figure*}
Since Horndeski gravity encompasses a large number of extensions to GR, here we focus on two well-studied models within this class displaying very different phenomenology: DGP braneworld gravity~\footnote{Technically speaking, DGP is a higher-dimensional theory of gravity that falls outside the Horndeski theory landscape. However, the scales relevant for structure formation are well within the regime in which DGP can be treated as a 4-dimensional scalar-tensor theory \citep{Nicolis:2004,Park:2010}.} \citep{Dvali:2000} and $f(R)$ gravity\footnote{Here $R$ denotes the Ricci scalar and it must not be confused with the smoothing radius defined above. In what follows we shall keep the same notation for both quantities as their meaning should be clear from context.} \citep[see, e.g.,][for a review]{DeFelice:2010}. In particular, we will consider the \emph{normal} branch of DGP with an additional smooth dark energy component such that the background expansion is identical to \LCDM{} \citep{Schmidt:2009}, that is,
\begin{align}\label{eq:hubble}
    \left( \frac{H}{H_0} \right)^2 = \Om a^{-3} + \Ol \, ,
\end{align}
with $\Ol = 1-\Om$, and the subscript `0' denotes present-day values here and throughout. For $f(R)$ gravity we will use the functional form of \citet{Hu:2007}, which has an expansion history also well described by Equation~\eqref{eq:hubble} for viable parameter values. 

The linear growth of structure in DGP is modified by time-varying changes to the gravitational constant given by
\begin{align}\label{eq:eps_dgp}
    \epsilon_{\rm DGP}(a) = \frac{1}{3\beta(a)} \, ,
\end{align}
where
\begin{align}\label{eq:beta}
    \beta(a) \equiv 1 + 2 r_{\rm c}H \left( 1 + \frac{aH^\prime}{3H} \right) \, ,
\end{align}
with $r_{\rm c}$ being the \emph{crossover scale} parameter. Deviations from GR in this model can be parametrised in terms of the effective energy density contribution \citep[see, e.g.,][]{Lombriser:2009}
\begin{align}\label{eq:Om_rc}
    \Orc \equiv \frac{1}{4(r_{\rm c}H_0)^2} \, ,
\end{align}
such that for $\Orc \rightarrow 0$ we recover the standard growth. 

In the non-linear regime the evolution of spherical top-hat over-densities in DGP is correctly described by Equation~\eqref{eq:SCeqn}. For under-densities, instead, the same function $\cal F$ incorporating the Vainshtein screening \citep[see, e.g.,][]{Schmidt:2010} produces either unphysical solutions or a strength of the fifth force exceeding the expected linear limit for voids \citep{Falck:2015}. Here, we neglect the Vainshtein screening by linearising the modification to gravity and show in Section~\ref{sec:results} that this approach accounts for most of the difference between EdS and DGP spherical evolution. In practice, to ensure the distribution of the matter density peaks/troughs, $\nu = \delta/\sigma$, defined at the initial time is preserved at later epochs even for scale-dependent modifications \citep{Kopp:2013,Lombriser:2013}, and to effectively separate the impact of new physics from changes to the standard cosmological parameters \citep{Brax:2012}, the mapping $\newtau^{\rm DGP}(\rho,z)$ is obtained by setting ${\cal F} = \epsilon_{\rm DGP}$ and by extrapolating the initial density fluctuation, $\delta_i(\rho)$, to an arbitrary redshift $z<z_i$ as\footnote{Note that the linearly extrapolated top-hat density fluctuation, $\newtau$, so defined (see also Eq.~\ref{eq:SCapprox}) is just an effective quantity, and in \LCDM extensions will in general differ from the linear theory $\hat\delta_{\rm L}$ defined below Eq.~\eqref{eq:growth}.}
\begin{align}
    \frac{{\delta_i(\rho,z)}}{\sigma_{\rm L}^\Lambda(R\rho^{1/3},z_i)} = \frac{D_\Lambda(z)\delta_i(\rho,z)}{D_\Lambda(z)\sigma_{\rm L}^\Lambda(R\rho^{1/3},z_i)} = \frac{\newtau^{\rm DGP}(\rho,z)}{\sigma_{\rm L}^\Lambda(R\rho^{1/3},z)} \, .
\end{align}
Figure~\ref{fig:tau} shows that we can further approximate $\newtau^{\rm DGP}$ to better than 0.5\% accuracy with the EdS-based approximation from Equation~\eqref{eq:SCapprox}, thus removing entirely the necessity for solving the spherical evolution dynamics beyond the Einstein-de Sitter cosmology. 

In $f(R)$ gravity the new scalar degree of freedom acquires a mass, $m_{f_R}$, defining the effective range of the fifth force interaction (i.e. the Compton wavelength $\lambda_{\rm C}$), which for linearised fluctuations reads
\begin{align}\label{eq:compton}
    \lambda_{\rm C}(a) \equiv m_{f_R}^{-1} = \sqrt{3 c^2 (n+1) |f_{R0}| \frac{\bar R^{n+1}(a=1)}{\bar R^{n+2}(a)}} \, ,
\end{align}
where overbars represent background quantities, $n$ and $f_{R0}$ are free parameters of the theory, $c$ is the speed of light and the Ricci scalar is given by 
\begin{align}\label{eq:ricci}
    \bar R(a) = 12H^2 + 6a HH^\prime \, .
\end{align}
The dynamics of the linear growth modifications is controlled by 
\begin{align}\label{eq:eps_fr}
    \epsilon_{f(R)}(k,a) = \frac{(k \lambda_{\rm C}/a)^2}{3[ 1 +  (k \lambda_{\rm C}/a)^2]} \, ,
\end{align}
with $\epsilon_{f(R)} \approx 0$ for $k\lambda_{\rm C}/a \ll 1$, and reaches a maximum of $\epsilon_{f(R)} \approx 1/3$ for $k\lambda_{\rm C}/a  \gg 1$. GR is restored on all scales for $|f_{R0}| = 0$. 

The non-linear evolution of top-hat density fluctuations in $f(R)$ gravity is complicated by the violation of mass conservation and shell-crossing \citep{Brax:2012,LiEfstathiou_2012, Borisov:2012,Kopp:2013,Lombriser:2013}. Therefore, Equation~\eqref{eq:SCeqn} cannot be used to find the exact $\newtau^{f(R)}(\rho)$ mapping even when neglecting the chameleon screening. However, we can gauge the accuracy of the approximation in Equation~\eqref{eq:SCapprox} by looking at how well the reduced cumulants, $S_n=\langle\delta^n\rangle_c/\sigma^{2(n-1)}$, of the modified gravity PDF can be predicted in the assumption of Einstein-de Sitter evolution. The next-to-leading order (NLO) predictions for the first two non-trivial reduced cumulants can be derived as discussed in \citet{Uhlemann16}, 
%
% \begin{subequations}\label{eq:cumulants}
% \begin{align}
%     S_3^{\rm NLO} = S_3^{\rm tree} + \sigma^2_{\rho} \left[ \frac{3}{2}S_4^{\rm tree} - 4 S_3^{\rm tree} - 2(S_3^{\rm tree})^2 + 7 \right] \, ,
% \end{align}
% %
% \begin{align}
%     S_4^{\rm NLO} = S_4^{\rm tree} + \sigma^2_{\rho} \left[ 2 S_5^{\rm tree} - \frac{17}{2}S_4^{\rm tree} + 66 S_3^{\rm tree} - 12(S_3^{\rm tree})^2 \right. \\ \left. - 3 S_4^{\rm tree} S_3^{\rm tree} - 45 \right] \nonumber \, ,
% \end{align}
% \end{subequations}
\begin{subequations}\label{eq:cumulants}
\begin{align}
    S_3^{\rm NLO} = S_3^{\rm tree} + \sigma^2_{\rho} &\left[ \frac{3}{2}S_4^{\rm tree} - 4 S_3^{\rm tree} - 2(S_3^{\rm tree})^2 + 7 \right] \, , \\
    S_4^{\rm NLO} = S_4^{\rm tree} + \sigma^2_{\rho} &\left[ 2 S_5^{\rm tree} - \frac{17}{2}S_4^{\rm tree} + 66 S_3^{\rm tree} - 12(S_3^{\rm tree})^2 \right. \\ 
    &\left. - 3 S_4^{\rm tree} S_3^{\rm tree} - 45 \right] \nonumber \, ,
\end{align}
\end{subequations}
where all quantities vary with smoothing scale and redshift, $\sigma^2_{\rho}$ is the non-linear variance of the density field, and the standard tree-level (or leading order) expressions can be found in, e.g., \citet{Bernardeau:2002}. In Figure~\ref{fig:fr_cumulants} we compare these predictions against the reduced cumulants measured from the $f(R)$ and \LCDM{} simulations described in Section~\ref{sec:sims} \citep[see also][for similar measurements]{Hellwing:2013}, with the non-linear variance entering Equation~\eqref{eq:cumulants} also computed from the same simulations (values can be found in Table~\ref{tab:summary_stats}). The striking similarity between the performance in $f(R)$ gravity and that in \LCDM{} suggests that the Einstein-de Sitter prescription works equally well for the two cosmologies on mildly non-linear scales. The tree level predictions for $S_N$ contain a constant `raw value' along with smoothing corrections from  logarithmic derivatives of the linear variance $d\log\sigma_{\rm L}(R,z)/d\log R$. Departures from GR are largely captured by changes to the linear variance entering the tree-level terms. 
Changes to the raw value of $S_3$ are negligible compared to this, as was explicitly shown in \cite{Bernardeau_2011} for the Linder $\gamma$-model \citep{Linder:2005}.

In summary, the matter PDF in modified gravity can be predicted using the LDT formalism discussed in Section~\ref{sec:LDT} with the following replacements to the decay-rate function in Equation~\eqref{eq:PsiNL}
\begin{subequations}\label{eq:rate_new}
\begin{align}
    &\sigma_{\rm L} \longrightarrow \sigma_{\rm L}^{\rm ext} \, , \\
    &\sigma_{\rm NL} \longrightarrow \sigma_{\ln\rho}^{\rm ext} \, , \\
    &\newtau(\rho,z) \longrightarrow \newtau^{\rm EdS}(\rho) \, ,
\end{align}
\end{subequations}
From a practical perspective, by approximating $\newtau(\rho,z)$ with the Einstein-de Sitter mapping we can substantially accelerate the calculations of the PDF in exchange for only a minor loss in accuracy--a welcomed feature for applications requiring a large number of evaluations. Note that our approach differs from the method developed in \citet{Brax:2012}, in that they solely focus on modifications to the spherical dynamics by evolving a ``typical'' density profile whose shape is approximated by the linear power spectrum, while neglecting the effect of the fifth force on the linear variance. 

\subsubsection{Evolving dark energy}

As evident from Equation~\eqref{eq:growth}, although gravity in smooth dark energy cosmologies is still described by GR ($\epsilon = 0$), the growth of structure can deviate from \LCDM{} through changes in the expansion history ($w_{\rm eff} \neq -1$). Here, we will consider equations of state parametrised by \citep{Chevallier:2001,Linder:2003}
\begin{align}\label{eq:DEeos}
    w_{\rm eff}(a) = w_0 + w_a(1-a) \, ,
\end{align}
where $\{w_0, w_a\}$ are phenomenological parameters. In particular, we will refer to models with vanishing $w_a$ as $w_0$CDM cosmologies, while referring to models with an evolving equation of state as $w_0w_a$CDM cosmologies.

The non-linear growth of spherical top-hat fluctuations is also affected by the evolving dark energy density via the Hubble parameter in Equation~\eqref{eq:SCeqn}. However, we follow the approach proposed by \citet{Codis:2016}  (i.e. keeping the spherical evolution fixed as in Einstein-de Sitter) and compute the matter PDF by means of Equation~\eqref{eq:rate_new}. We quantify \emph{a posteriori} the goodness of this choice by comparing our predictions against state-of-the-art cosmological simulations in Section~\ref{sec:results}. 
 
\subsubsection{\pyLDT}

We have implemented the large-deviation theory predictions described in Section~\ref{sec:LDT} together with Equations~\eqref{eq:rate_new} in \pyLDT{}, a modularised and user-friendly Python code that takes advantage of the PyJulia interface for computationally intensive tasks. The linear growth for $f(R)$ gravity and DGP is obtained by solving Equation~\eqref{eq:growth}, while the linear power spectrum for the standard cosmology, as well as for the evolving dark energy models, is computed with CAMB\footnote{Note that the common approximation for the linear growth $D(z)\propto H(a)\int_0^a {\rm d} a' (a' H(a'))^{-3}$ [quoted in equation (6) of \cite{Codis:2016} and (A1) of \cite{Uhlemann:2020}] is not accurate enough to estimate the response of the PDF to changing $w$ beyond a cosmological constant.} \citep{CAMB}. Extensions to other modified gravity theories only require either to add a specific function describing changes to the gravitational constant, $\epsilon(k,a)$, or to couple the code to dedicated Einstein-Boltzmann solvers such as \texttt{hi\_class} \citep{Zumalacarregui:2017,Bellini:2020} and EFTCAMB \citep{Hu:2014}.

By default, \pyLDT{} uses an empirical parametrisation of the log-density field non-linear variance in terms of the corresponding linear variance given by \citep{Uhlemann:2020}
\begin{equation}
\label{eq:siglogcos}
    \sigma_{\rm NL}^2\rightarrow \sigma_{\ln \rho}^2(R,z) \simeq \frac{\ln\left[1 + \sigma^2_{\rm L}(R,z)\right]}{\ln\left[1 + \sigma^2_{\rm L, fid}(R,z)\right]} \sigma_{\ln \rho, \rm fid}^2(R,z) \,.
\end{equation}
This relation allows us to predict the non-linear variance for arbitrary cosmologies given the measured non-linear variance at one fiducial \LCDM{} cosmology, $\sigma^2_{\rm L}$, with a typical accuracy of 0.2--1\% for the extensions studied in this work. In terms of the matter PDF, for densities $|\ln\rho - \langle \ln\rho \rangle| < 2\sigma_{\ln\rho}$ the log-normal approximation above returns predictions that are within 2\% of those based on the non-linear variance measured from the simulations. Unless stated otherwise, direct comparisons to simulations performed in Section~\ref{sec:results} are the output of \pyLDT{} with Equation~\eqref{eq:siglogcos} replaced by the actual non-linear variance extracted from the simulations. For the Fisher forecasts presented in Section~\ref{sec:Fisher}, instead, we rely on the parametrisation in Equation~\eqref{eq:siglogcos} to compute the response to changing cosmological parameters and MG scenarios.
  
\section{Simulations}\label{sec:sims}

\begin{table}
\begin{tabular}{@{}lllllll@{}}
\toprule
    & $\Om$ & $\Ob$ & $h$ & $n_s$ & $A_s \times 10^{9}$ & $\sigma_8^\Lambda$ \\ \midrule
DGP & 0.3072      & 0.0481      & 0.68     & 0.9645      &    2.085   &  0.821  \\
$f(R)$ & 0.31315      & 0.0492      & 0.6737     &  0.9652     &   2.097 &  0.822      \\
DE    & 0.26      & 0.044      & 0.72     & 0.96      & 2.082   &  0.79    \\ \bottomrule
\end{tabular}
\caption{Baseline \LCDM{} cosmological parameters for the three simulation suites used in this work. The first column refers to the extension investigated within that particular suite. $\Om$ and $\Ob$ are, respectively, the present-day background total matter and baryon density in units of the critical density, $h = H_0/100$ is the dimensionless Hubble constant, $A_s$ and $n_s$ are the amplitude and slope of the primordial power spectrum, and $\sigma_8^\Lambda$ is the amplitude of mass fluctuations for the baseline \LCDM{} cosmology.}
\label{tab:lcdm_cosmos}
\end{table}

\subsection{\texorpdfstring{$f(R)$}{f(R)} gravity simulations}

The simulations in $f(R)$ gravity used for the analysis in this work were carried out with the \textsc{Arepo} cosmological simulation code \citep{springel2010, weinberger2020} employing the MG extension introduced in \cite{arnold2019}. The simulation suite consists of 8 independent realisations, each run for a baseline \LCDM cosmology (see Table~\ref{tab:lcdm_cosmos} for the selected parameter values), and for  $f(R)$ Hu-Sawicki models with $n=1$ and $\fR0 = 10^{-5}$ (F5), $10^{-6}$ (F6). The suite is completed by two \emph{pseudo} cosmology runs per $f(R)$ model, one for the final output redshift $z_{\rm f} = 0$ and the other for $z_{\rm f} = 1$. In short, a pseudo cosmology is a \LCDM{} cosmology with initial conditions adapted so that its linear matter power spectrum at a later epoch, $z_{\rm f}$, matches that of the \emph{real} beyond-\LCDM{} cosmology of interest \citep{Mead:2017,Cataneo:2019}, 
\begin{align}\label{eq:pseudo_cosmo}
    P_{\rm L}^{\rm pseudo}(k,z_{\rm f}) = P_{\rm L}^{\rm real}(k,z_{\rm f}) \, . 
\end{align} 
Each simulation uses $N_{\rm p} = 1024^3$ dark matter particles in a $L_{\rm box} = 500$ \Mpch{} side-length box.

The initial conditions (ICs) of the independent realisations were selected such that the large-scale sample, or cosmic, variance in the 3D matter power spectrum is minimal when averaged over the simulations. In order to implement this we created 100 independent initial conditions using \textsc{2lptic} \citep{lpt} and measured their 3D matter power spectrum. We then considered all possible pairs of these ICs and selected the four `best' pairs according to the following criteria (this follows the procedure outlined in \citealt{slics} to find ICs with approximately opposite modes on large scale):
\begin{itemize}
    \item each individual power spectrum of a selected pair, as well as their average power spectrum, should deviate as little as possible from the desired linear theory power spectrum for $k < k_{\rm Ny}/2 = \pi N_{\rm p}/ 2 L_{\rm box}$;
    \item and the relative difference of each individual power spectrum to the theory spectrum should fluctuate around zero on large scales rather than being positive or negative over large $k$-ranges to avoid a leakage of power from large to small scales.
\end{itemize}

To simulate structure formation in \fr the simulation code has to solve both the standard Newtonian forces and the fifth force. \textsc{Arepo} computes the standard gravity forces using a Tree Particle-Mesh algorithm in our simulations. The \fr forces are computed employing an iterative solver on an adaptively refining mesh which ensures increased resolution in high density regions \citep[see][for details]{arnold2019}. 

Due to the very non-linear behaviour of the scalar field in \fr{}, tracking its evolution is computationally very expensive. To keep the computational cost of the simulations as small as possible, \textsc{Arepo} therefore employs an adaptive timestepping scheme which only updates the MG forces when necessary \citep{arnold2019}. The standard gravity accelerations are largest (and change most frequently) within large halos, so that they have to be updated with a very small timestep. However, these very same regions in \fr{} are largely screened for $\fR0 \lesssim 10^{-5}$. Therefore, the maximum MG acceleration will typically be much smaller than the maximum standard gravity acceleration, allowing a larger MG timestep without compromising the accuracy of the simulations.

\subsection{DGP simulations}

The DGP simulations used in this work were first presented in \cite{Cataneo:2019}, and they were carried out using the \textsc{Ecosmog} code \citep{Li:2013nua,Li:2011vk}, which is based on the publicly-available Newtonian cosmological $N$-body and hydrodynamical simulation code \textsc{Ramses} \citep{Teyssier:2001cp}. This code solves the non-linear equation of motion of the scalar field in the DGP model using adaptively refined meshes, where a cell in the mesh splits into 8 son cells when the effective particle number of simuation particles in it exceeds 8. We have run one realisation with box size $L_{\rm box}=512$ Mpc$/h$ and particle number $N_{\rm p}=1024^3$ for each of the following: a baseline \LCDM{} cosmology with cosmological parameters listed in Table~\ref{tab:lcdm_cosmos}, two DGP models with $\Orc = 0.25$ (DGPm) and $\Orc = 0.0625$ (DGPw), and the corresponding pseudo cosmologies with final output redshifts $z_{\rm f} = 0$ and $z_{\rm f} = 1$. These runs adopt a domain grid, i.e., a regular base grid with uniform resolution that covers the entire simulation domain, with $1024^3$ cells. Although it has been shown that, for many of the usual statistics of matter and dark matter halo fields, very fine simulation meshes are not necessary for the DGP model \citep[][]{Barreira:2015xvp}, in these runs we have not set an upper limit of the highest refinement level, given that they were designed to be used to study novel statistics. At late times, the most refined regions in the simulation domain have a cell size that is $1/2^6$ times the domain grid cell size; this corresponds to an effective force resolution (twice the cell size) of $\simeq15.3$ kpc$/h$ in those regions. 

The ICs of these simulations are again generated using 2\textsc{lptic}, with an initial redshift $z_{\rm ini}=49$. This is lower than the initial redshift used for the $f(R)$ runs described above ($z_{\rm ini}=127$), but the second-order Lagrangian perturbation theory is still a good approximation at $z=49$. Since the effect of modified gravity is negligible at $z>49$, it is neglected in the ICs.

\subsection{Evolving dark energy simulations}

For the evolving dark energy cosmologies we used the publicly available matter density PDFs\footnote{\url{https://astro.kias.re.kr/jhshin/}} measured from a suite of single-realisation $N$-body simulations with $N_{\rm p} = 2048^3$ and $L_{\rm box} = 1024$ \Mpch{} described in \citet{Shin:2017}. The baseline flat \LCDM{} cosmology has the parameters listed in Table~\ref{tab:lcdm_cosmos}, and for the $w_0w_a$CDM cosmologies we have the four pairs $\{w_0,w_a\} = \{-1.5,0\}$, $\{-0.5,0\}$, $\{-1,-1\}$, and $\{-1,+1\}$. The power spectrum normalisation at $z=0$ is fixed to its baseline value for all dark energy extensions except for $\{w_0,w_a\} = \{-1,+1\}$, which we found to have a somewhat smaller  $\sigma_8$\footnote{Because the linear theory normalisation cancels out in Equation~\eqref{eq:PsiNL}, knowledge of $\sigma_8$ is irrelevant for the LDT predictions when measurements of the variance of the simulated density field are available. In fact, the non-linear variance carries information on $\sigma_8$ so that, ultimately, its impact on the theory PDF is properly accounted for.}.

\subsection{PDF measurements from the simulations}

For our $f(R)$ gravity, DGP and corresponding pseudo and \LCDM{} simulations we measured the PDFs of the smoothed matter density field as follows. Firstly, for each snapshot we reconstruct the continuous density field using the Delaunay Tassellation Field Estimator method \citep{Schaap:2000} and sample it over a $1024^3$ mesh, all of which is automatically performed by the public code {\sc dtfe}\footnote{\url{https://github.com/MariusCautun/DTFE}} \citep{Cautun:2011}. Next, we convolve the sampled density field with spherical top-hat filters of radii $R = 10, 15$ and 20 \Mpch{} (an operation we do in Fourier space). Lastly, we construct the PDF by collecting the normalised density values, $\rho_{R} = 1 + \delta_{R}$, in 99 logarithmically spaced bins in the range $[0.01,100]$. In Appendix~\ref{sec:CiCvDTFE} we show that this method produces PDFs in excellent agreement with those obtained by applying the Cloud-in-Cell (CiC) mass assignment scheme. We report variances and means extracted from the simulations for both the density and the log-density fields in Appendix~\ref{app:stats}.

For the DE simulation suite, instead, the smoothed density field was obtained by summing over the mass of all the particles contained in spheres centered at the $2048^3$ nodes of a regular grid and dividing by the volume of the spheres. In this work, we consider the PDFs measured in spheres of radius $R = 10$ and 25 \Mpch{} for the $z=0,0.5$ and 1 snapshots.

\section{Results}\label{sec:results}

In the following, we first present our results for the modified gravity and dark energy cosmologies discussed in Section~\ref{sec:LDTwMG}, and then examine the detection potential of departures from \LCDM{}
for idealised statistical analyses combining the full shape of the PDF and the matter power spectrum. 

\begin{figure*}
    \centering
    \includegraphics[width=\columnwidth]{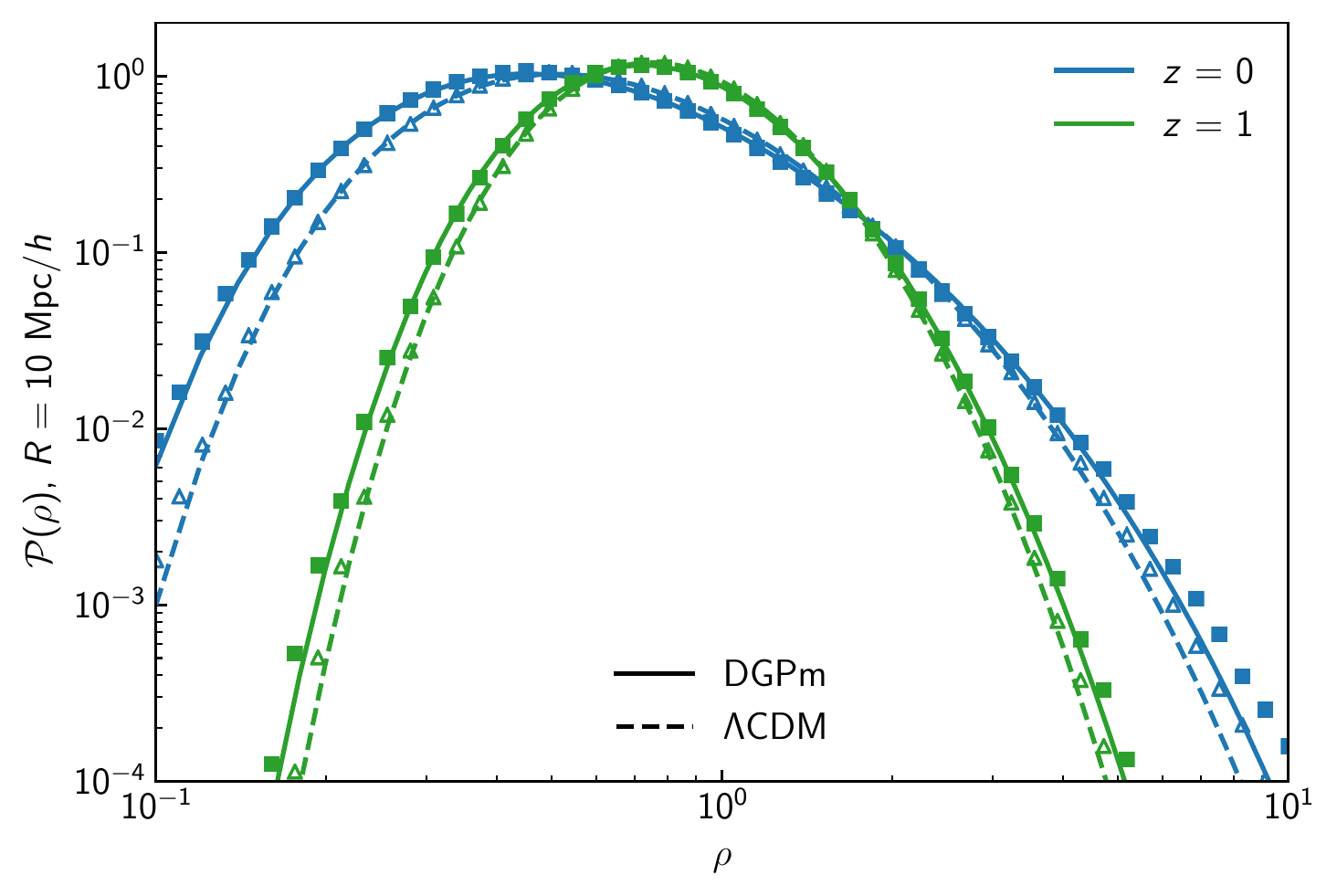}
    \quad
    \includegraphics[width=\columnwidth]{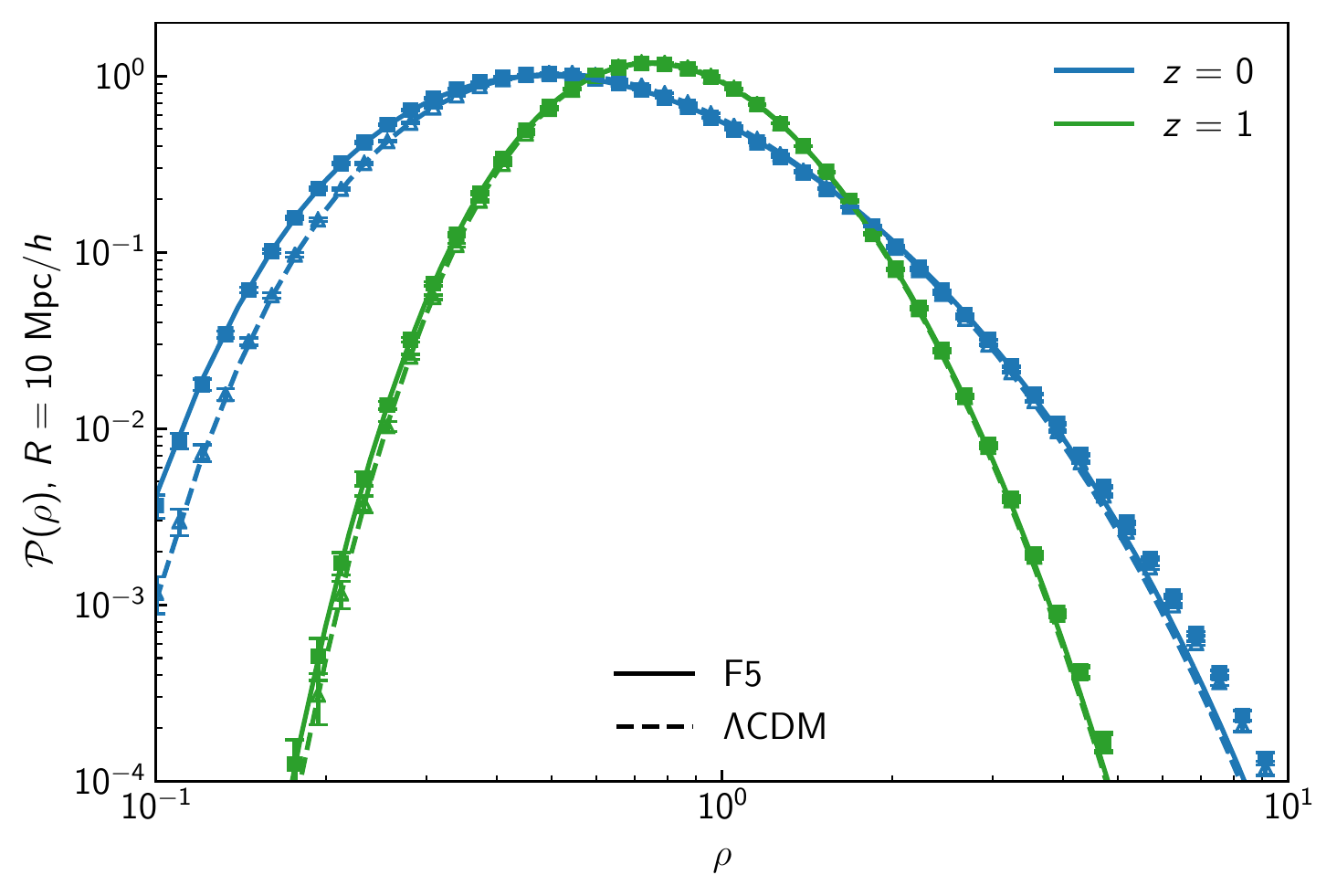}
    \caption{Matter PDF in spheres of radius $R = 10$ \Mpch{} at $z=0$ (blue) and $z=1$ (green) for \LCDM{} (dashed) and modified gravity (solid). \textit{Left:} Data points are the simulation measurements from a single realisation (triangles for \LCDM{} and squares for DGPm) and lines represent the theory predictions. For DGP the primary effect of the enhanced growth is that of increasing the variance of the distribution, which in turn results in heavier tails, i.e. more under/over-dense structures compared to the standard cosmology. \textit{Right:} data points and corresponding uncertainties are the mean and error on the mean measured from 8 realisations (triangles for \LCDM{} and squares for F5). Note that, contrary to the DGP cosmology, the modified growth in $f(R)$ gravity substantially affects the skewness of the distribution, thus leading to an asymmetric enhancement over \LCDM{}.}
    \label{fig:mg_pdf}
\end{figure*}

\begin{figure*}
    \centering
    \includegraphics[width=\textwidth]{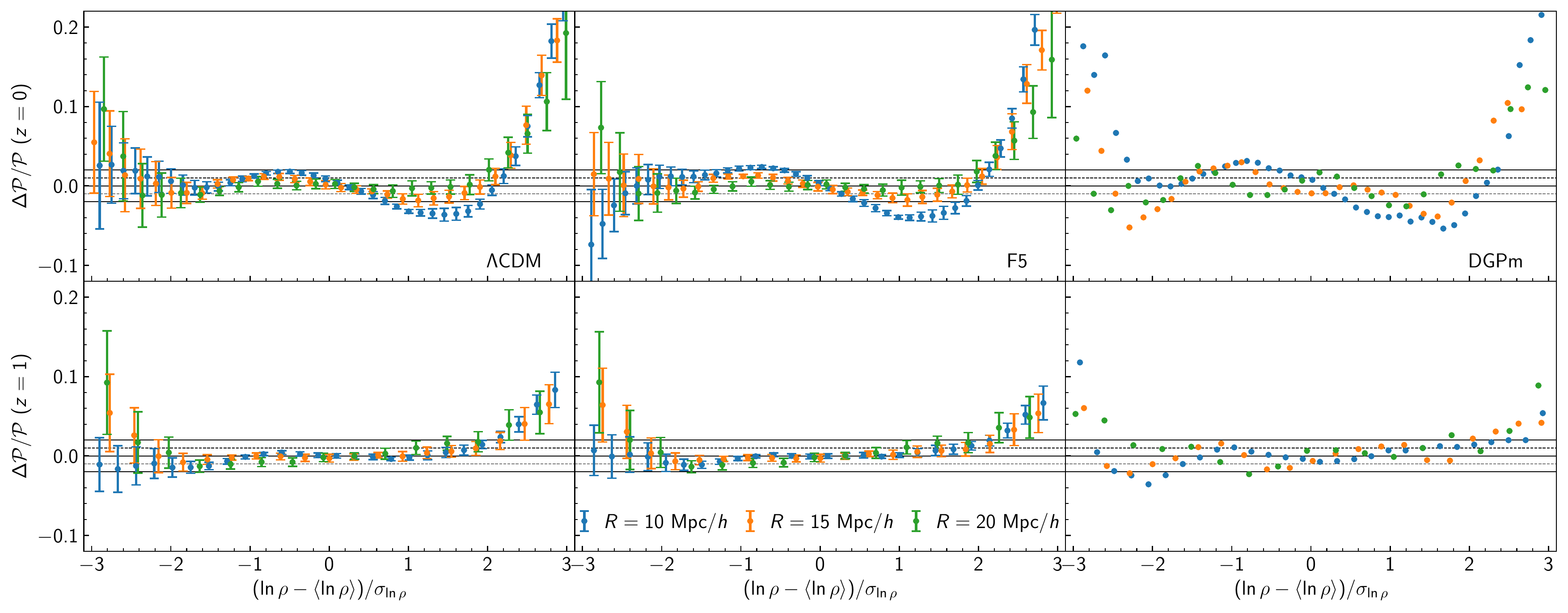}
    \caption{Residuals between the measured and predicted matter PDF normalised to the theory predictions for $z=0$ (top) and $z=1$ (bottom) in \LCDM{} (left), $f(R)$ gravity (centre) and DGP (right). When data points and error bars are both present they correspond to the mean and error on the mean across 8 realisations. Different colors indicate the radii of the spheres used for smoothing the density field, 10 \Mpch\ (blue), 15 \Mpch\ (orange) and 20 \Mpch\ (green). The solid and dashed lines mark 1\% and 2\% accuracy, respectively. Despite significant changes to the growth of structure, the accuracy of the modified gravity predictions based on the EdS spherical dynamics is comparable to that of the standard cosmology.}
    \label{fig:mg_residuals}
\end{figure*}

\subsection{Modified gravity}

% \begin{figure}
%     \centering
%     \includegraphics[width=\columnwidth]{./figures/dgp_residuals.pdf}
%     \caption{Same as Figure~\ref{fig:lcdm_residuals} for DGP gravity with $\rcH0 = 0.5$. Here data points are obtained from a single realisation.}
%     \label{fig:dgp_residuals}
% \end{figure}

\begin{figure*}
    \centering
    \includegraphics[width=\columnwidth]{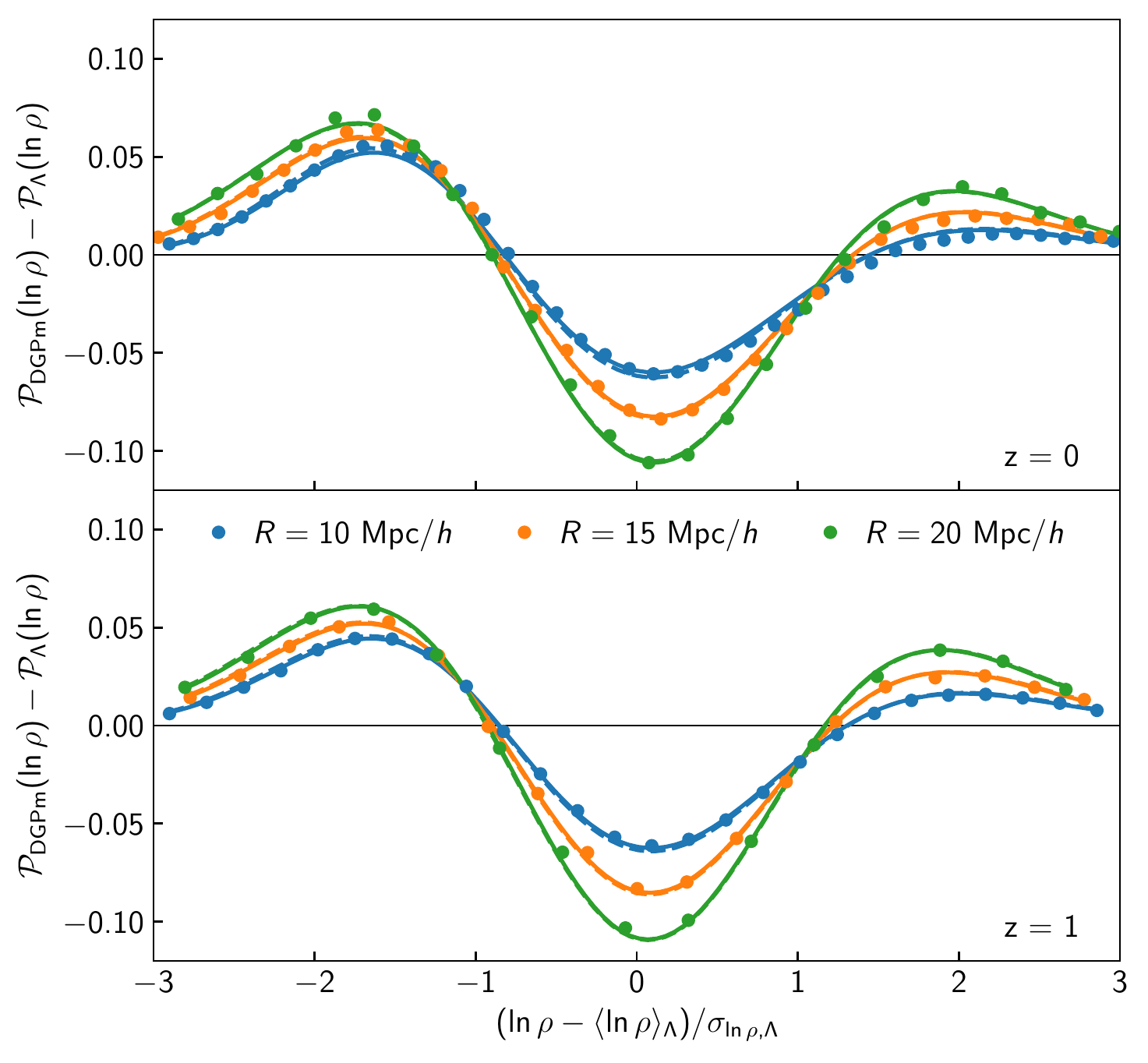}
    \quad
    \includegraphics[width=\columnwidth]{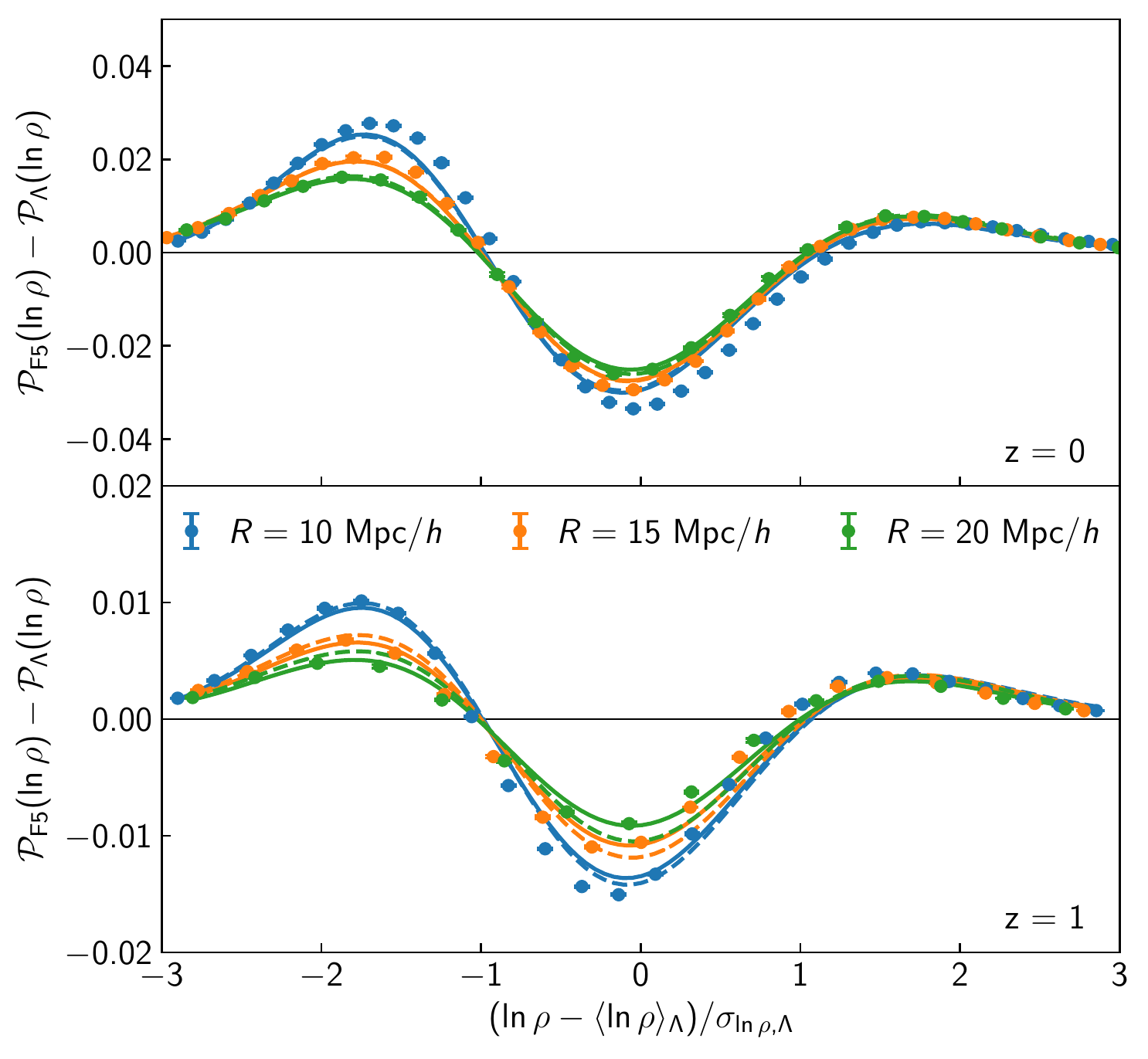}
    \caption{Simulated (data points) and predicted (lines) differences from \LCDM{} of the modified gravity matter PDF at $z=0$ (top) and $z=1$ (bottom). The density field is averaged in spheres of radius $R =$ 10 \Mpch\ (blue), 15 \Mpch\ (orange) and 20 \Mpch\ (green). Solid lines are obtained from the measured non-linear variance, $\sigma^2_\mu$, while dashed lines use Equation~\eqref{eq:siglogcos} in \pyLDT{}. The close agreement between the two type of predictions supports the use of the lognormal approximation for the non-linear variance. \textit{Left:} in DGP the shape of these changes is very similar to that induced by variations in $\sigma_8$ \citep[cf. Figure 8 in][]{Uhlemann:2020}. \textit{Right:} $f(R)$ gravity departures from the standard cosmology are more prominent for under-dense regions and become less significant with increasing smoothing radii owing to the finite range of the fifth force.}
    \label{fig:mg-lcdm_diff}
\end{figure*}

% \begin{figure}
%     \centering
%     \includegraphics[width=\columnwidth]{./figures/f5_v_lcdm_PDF_z.pdf}
%     \caption{Same as Figure~\ref{fig:dgp_pdf} for $f(R)$ gravity with $\fR0 = 10^{-5}$. The data points and corresponding uncertainties are the mean and error on the mean measured from 8 realisations. Note that here, in contrast to DGP gravity, the modified growth substantially affects the skewness of the distribution, thus leading to an asymmetric enhancement over \LCDM{}.}
%     \label{fig:fr_pdf}
% \end{figure}

% \begin{figure}
%     \centering
%     \includegraphics[width=\columnwidth]{./figures/fr_residuals.pdf}
%     \caption{Same as Figure~\ref{fig:lcdm_residuals} for $f(R)$ gravity with $\fR0 = 10^{-5}$.}
%     \label{fig:fr_residuals}
% \end{figure}

% \begin{figure}
%     \centering
%     \includegraphics[width=\columnwidth]{./figures/fr-lcdm_diff.pdf}
%     \caption{Caption}
%     \label{fig:fr-lcdm_diff}
% \end{figure}

As discussed in Section~\ref{sec:LDTwMG}, on mildly non-linear scales the Einstein-de Sitter dynamics approximates well the evolution of spherical top-hat density fluctuations even in cosmologies where the law of gravity deviates substantially from GR. Here, by using state-of-the-art simulations we assess how such an approximation impacts the accuracy of the LDT predictions for the matter PDF in two specific modified gravity scenarios, DGPm and F5 (see Section~\ref{sec:sims} for details). Equivalent results for DGPw and F6 can be found in Appendix~\ref{app:small_deviations}.

Figure~\ref{fig:mg_pdf} shows how the global shape of the PDF responds to scale-independent (left) or scale-dependent (right) modifications to the linear growth. As expected, when sharing the same initial conditions with \LCDM{} both PDFs approach the standard result at high redshifts and exhibit their largest deviations at low redshifts, and do so at a rate specific to the model under consideration. However, there are clear differences that reflect the infinite or finite range of the fifth force. In DGP, structures on all scales are subject to the same modification, and changes to the higher moments of the distribution are mainly driven by increases in the variance. This follows immediately from the expressions for the reduced cumulants in Equation~\eqref{eq:cumulants}--where the tree-level terms are identical for DGP and \LCDM{}--and will be explored in more detail below. In $f(R)$ gravity, instead, density fluctuations evolve in different gravity conditions depending on their size. For example, the present-day Compton wavelength in our F5 cosmology is approximately 8 \Mpch{} (and smaller at earlier times). Therefore, spherical over-densities reaching a final radius $R=10$ \Mpch{} experience very little fifth force for most of their collapse history. In contrast, spherical under-densities have sizes comparable to or smaller than the Compton wavelength (at the same epoch) and thus experience the fifth force in the later stages of their expansion (i.e. $z \lesssim 2$), with the emptiest regions experiencing a full 33\% enhancement of the gravitational force. It is this asymmetric behaviour that contributes to the increased skewness of the PDF in $f(R)$ gravity compared to \LCDM{} \citep[see also][]{Hellwing:2013}, our model Eq.~\eqref{eq:rate_new} can capture it thanks to the linear variance term probing different scales, $r = R\rho^{1/3}$, depending on the density of the sphere, $\rho$ (see also Eq.~\ref{eq:PsiNL}).

The central and right panels of Figure~\ref{fig:mg_residuals} present comparisons of the modified gravity predictions to the simulation measurements for different smoothing radii and redshifts. In all cases, the prescription described by Equation~\eqref{eq:rate_new} provides PDF predictions that are within a few percent from the simulations, which is consistent with the results for \LCDM{} (left panel). Note that the seemingly poorer performance for DGP is likely driven by sample variance, as we only have a single realisation for this cosmology. We also note that despite differences in $N$-body codes ({\sc Arepo} v {\sc Gadget-III}), mass-assignment schemes (DTFE v CiC), mass resolution ($m_{\rm p} \approx 10^{10} \, M_{\odot}/h$ v $m_{\rm p} \approx 8 \times 10^{10} \, M_{\odot}/h$), and number of realisations (8 v 100) the leftmost panels of Figure~\ref{fig:mg_residuals} illustrate that our measured PDFs are very much consistent with those of \citet{Uhlemann:2020} (see their figure 7), irrespective of smoothing radius and redshift.

Figure~\ref{fig:mg-lcdm_diff} shows in detail how the PDFs in the two modified gravity scenarios analysed in this work differ from their \LCDM{} counterparts. With a lowering of the peak compensated by heavier tails, DGP modifications (left panel) resemble very closely changes in the power spectrum normalisation \citep[cf. Figure 8 in][]{Uhlemann:2020}. This can be explained by the near equivalence between the boost of the linear matter power spectrum amplitude induced by the fifth force and an increase in $\sigma_8$. More complicated variations to the shape of the PDF in $f(R)$ gravity (right panel) follow from the combination of two effects: suppression of the non-linear variance compared to a DGP cosmology with a similar $\sigma_8$, and scale-mixing regulated by the redshift-dependent Compton wavelength. The former is a direct consequence of the chameleon screening mechanism acting on a broad range of scales, even in the mildly non-linear regime \citep[see, e.g.,][]{Cataneo:2019}; while the latter preferentially enhances the formation of density fluctuations with initial comoving size $R\rho^{1/3} \lesssim \lambda_{\rm C}(z)(1+z)$. At high redshifts and for large smoothing radii, this condition becomes increasingly difficult to satisfy for typical values of the density field (i.e. $| \ln\rho - \langle \ln\rho \rangle | < 3\sigma_{\ln\rho}$). As the PDF approaches the \LCDM{} result, the small residual deviations can be described by simple changes in the variance. The solid lines in both panels of Figure~\ref{fig:mg-lcdm_diff} represent the theory predictions with the log-density variances measured from the simulations, while the dashed lines use the log-normal approximation Equation~\eqref{eq:siglogcos} to compute the modified gravity $\sigma_\mu^2$ from that of the corresponding \LCDM{} cosmology. The LDT prescription, even when ignoring the impact of the fifth force on the evolution of spherical density fluctuations can capture deviations from GR remarkably well. As we shall see below, a detailed comparison to standard cosmologies sharing the same linear theory predictions (the so-called pseudo cosmologies) can help isolate very small effects that are characteristic of the non-standard interaction. 

\subsubsection{Pseudo cosmologies}\label{sec:pseudo_cosmo}
\label{subsec:pseudocos}
\begin{figure*}
    \centering
    \includegraphics[width=\columnwidth]{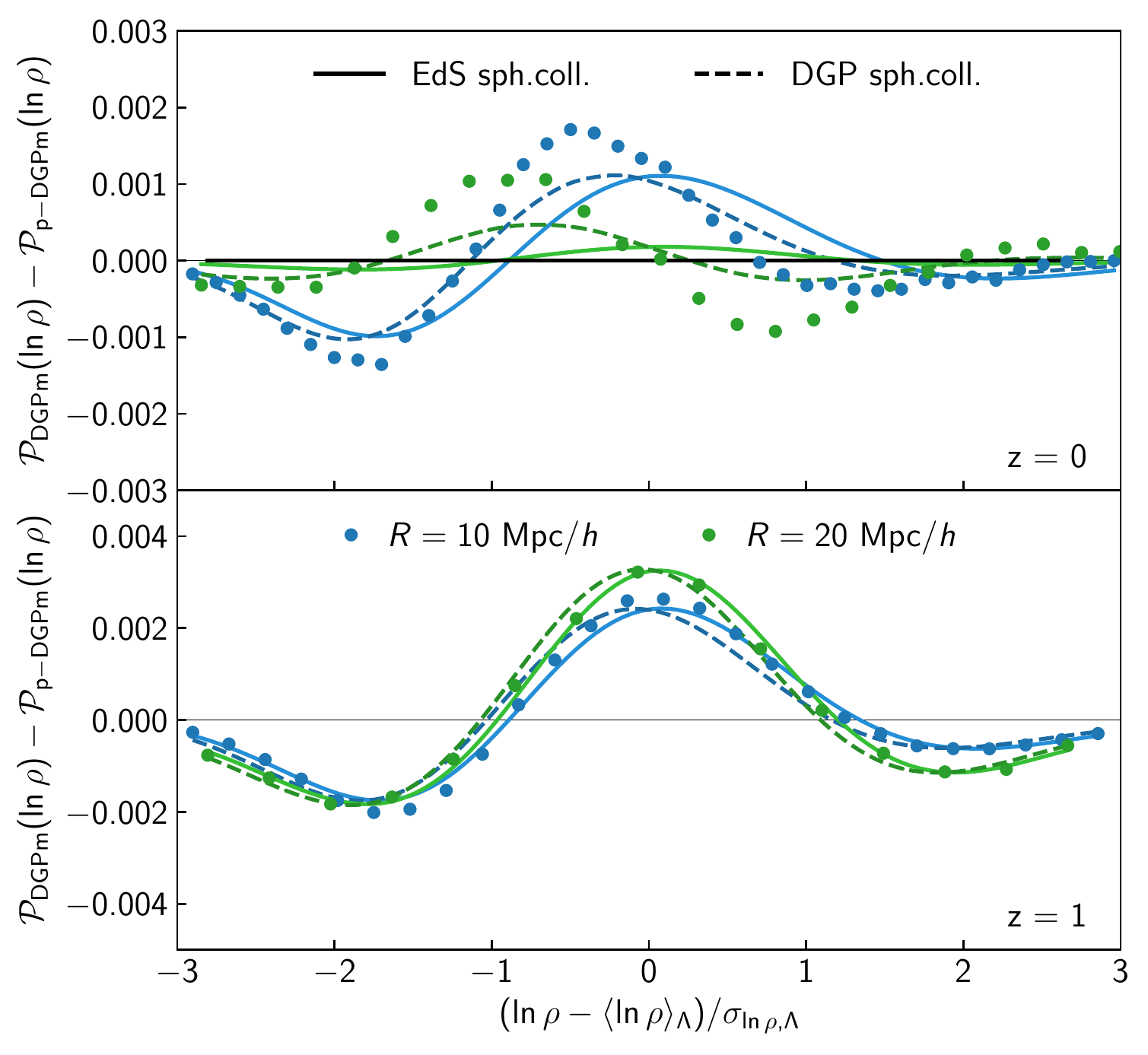}
    \quad
    \includegraphics[width=\columnwidth]{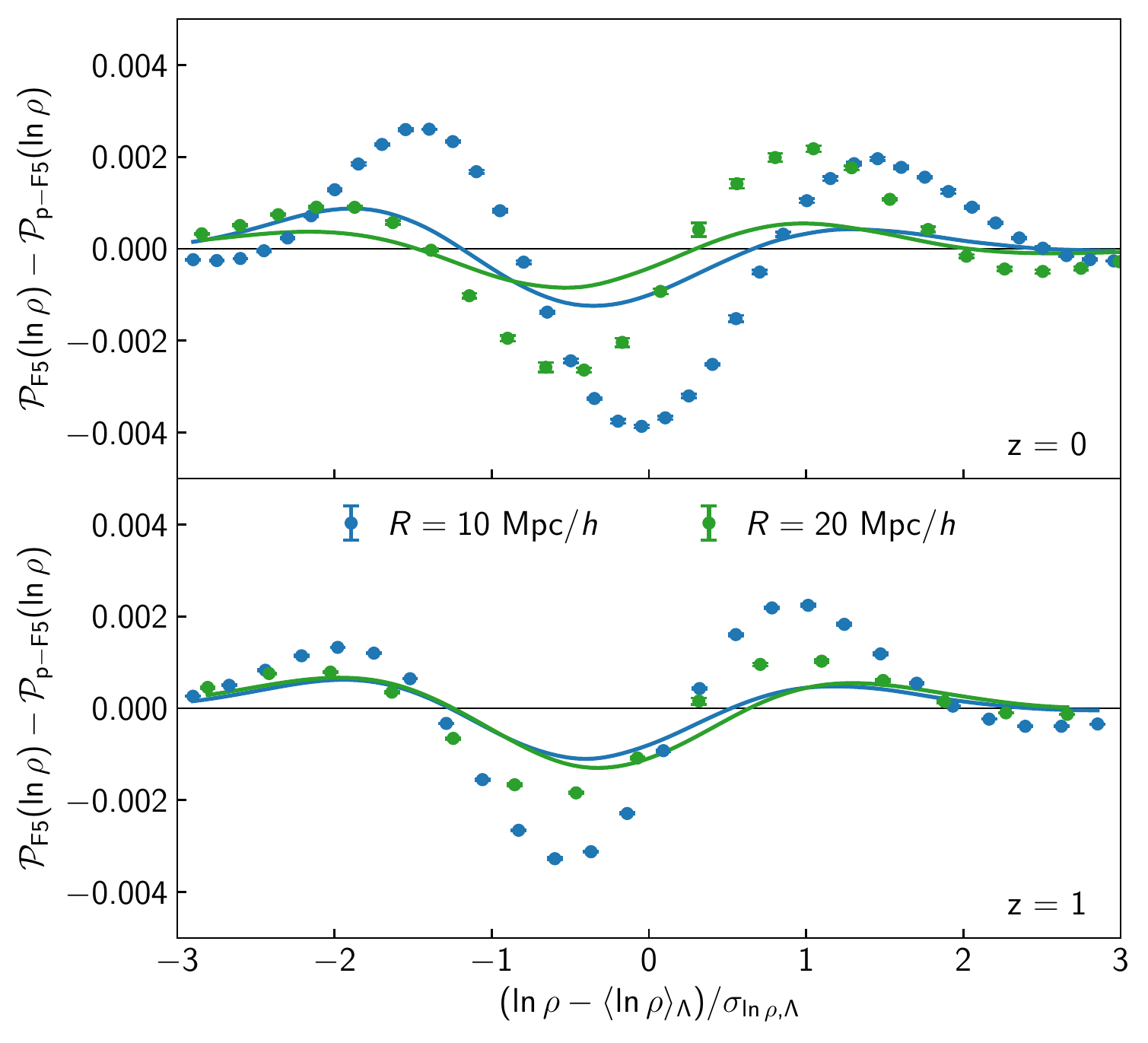}
    \caption{Simulated (data points) and predicted (lines) differences of the modified gravity matter PDFs from their pseudo-cosmology counterparts at $z=0$ (top) and $z=1$ (bottom). The density field is averaged in spheres of radius $R =$ 10 \Mpch\ (blue) and 20 \Mpch\ (green). Solid lines are obtained from the measured non-linear variance, $\sigma^2_\mu$, and the Einstein-de Sitter mapping shown in Figure~\ref{fig:tau}. Note that the amplitude of these differences is about ten times smaller than in Figure~\ref{fig:mg-lcdm_diff}, confirming the remarkable similarities between the pseudo and real cosmology on mildly non-linear scales. \textit{Left:} by replacing the EdS spherical evolution with that produced by the linearised DGP model (dashed lines) we can better predict some of the minute differences sourced by pure modifications to GR which are not captured by simple changes to $\sigma_8$ in a \LCDM{} cosmology. Remaining differences are likely the result of modelling inaccuracies in LDT and unaccounted for Vainshtein screening phenomenology. \textit{Right:} contrary to DGP, the violation of Birkhoff's theorem in $f(R)$ gravity precludes any attempt to find a simple solution to the evolution of spherical top-hat density perturbations even in the absence of screening mechanism \citep{Borisov:2012,Brax:2012,Kopp:2013}. Here we only show the predictions accounting for changes in the non-linear variance and note that both the finite range of the fifth force and chameleon screening slightly modify the spherical collapse dynamics, which in turn leads to small additional variations in the PDF.}
    \label{fig:mg-pmg_diff}
\end{figure*}

% \begin{figure}
%     \centering
%     \includegraphics[width=\columnwidth]{./figures/fr-pfr_diff.pdf}
%     \caption{Caption}
%     \label{fig:fr-pfr_diff}
% \end{figure}

%
% Following up on the work of \cite{Mead:2017}, \cite{Cataneo:2019} introduced this concept within the \emph{reaction} framework as a tool to mitigate the well-known inaccuracies of the halo model affecting the one- to two-halo transition in the non-linear matter power spectrum.
%
Although, by definition, for the pseudo cosmologies we have $\sigma_{\rm L}^{\rm pseudo}(R,z_{\rm f}) = \sigma_{\rm L}^{\rm real}(R,z_{\rm f})$ (see Eq.~\ref{eq:pseudo_cosmo}), new late-time physics affect the growth of structure beyond the linear regime. Therefore, the non-linear power spectrum of the pseudo cosmology differs from its real non-\LCDM{} counterpart and $\sigma_{\rm NL}^{\rm pseudo}(R,z_{\rm f}) \neq \sigma_{\rm NL}^{\rm real}(R,z_{\rm f})$. We can use this to compute the PDFs of the pseudo-MG cosmologies and compare them to the predictions for DGP and $f(R)$ gravity--given the identity in equation~\eqref{eq:pseudo_cosmo}, any significant difference not captured by a simple change to the non-linear variance will then signal modifications to the spherical dynamics due to the action of the fifth force. We recall that the linear power spectrum determines the scale-dependence of the linear variance and hence the density-dependence of the exponent of the PDF given by Equation~\eqref{eq:PsiNL}, while the nonlinear power spectrum determines the nonlinear variance and hence the width of the PDF. To a lesser extent, the nonlinear variance can also alter the scale-dependence of the PDF through its impact on the rescaling step in the PDF construction (see Equation~\eqref{eq:PDFnorm}).

Figure~\ref{fig:mg-pmg_diff} shows the difference between the real and pseudo cosmology PDFs for DGPm (left panel) and F5 (right panel). First, let us note that these differences are about an order of magnitude smaller than the departures of modified gravity from \LCDM{} shown in Figure~\ref{fig:mg-lcdm_diff}. In general, the two PDFs agree to percent level or better for $| \ln\rho - \langle \ln\rho \rangle | < 2\sigma_{\ln\rho}$. Thus, for densities not too far into the tails and in the mildly non-linear regime, the pseudo and real cosmology PDFs become indistinguishable for all intents and purposes. This result confirms the findings of \cite{Cataneo:2019} and extends them to statistics describing non-Gaussian properties of the density field. Our predictions using the Einstein-de Sitter spherical collapse for both the pseudo and the real MG cosmologies (solid lines) can partially explain the observed minute differences as changes in the variance of the distribution, especially at high redshifts. To gauge the contribution of the fifth force to the remaining unexplained difference, we also compute the PDFs for DGPm by including the linearised modification to the gravitational interaction (Equation~\ref{eq:eps_dgp}) into the dynamics of spherical top-hat density fluctuations (Equation \ref{eq:SCeqn}). These are shown as dashed lines in Figure~\ref{fig:mg-pmg_diff}. Although the modified non-linear evolution can better account for the differences between the real and pseudo cosmology, residuals associated with the neglected screening mechanism and intrinsic inaccuracies of the LDT formalism persist. To fully disentangle these two contributions one should run linearised modified gravity simulations \citep[akin to][]{Schmidt:2009,Koyama:2009}, which is, however, beyond the scope of this work. For the case of \fr shown in the right panel, a change in the variance (solid line) can only partially explain their observed differences. Modifications to the spherical collapse in \fr due to non-linear couplings even in the absence of screening \citep[such as modelled by][]{Brax:2012} are a potential source of the remaining discrepancy. The shape of the differences hints at an additional skewness with a slightly increased $S_3$ that cannot be captured by the EdS-based approximation in Equation~\eqref{eq:SCapprox}. The overall good agreement of the PDF in the real and pseudo cosmologies together with the successful prediction of their minute qualitative differences validate our PDF modelling assumptions for modified gravity.

\subsection{Evolving dark energy}

\begin{figure}
    \centering
    \includegraphics[width=\columnwidth]{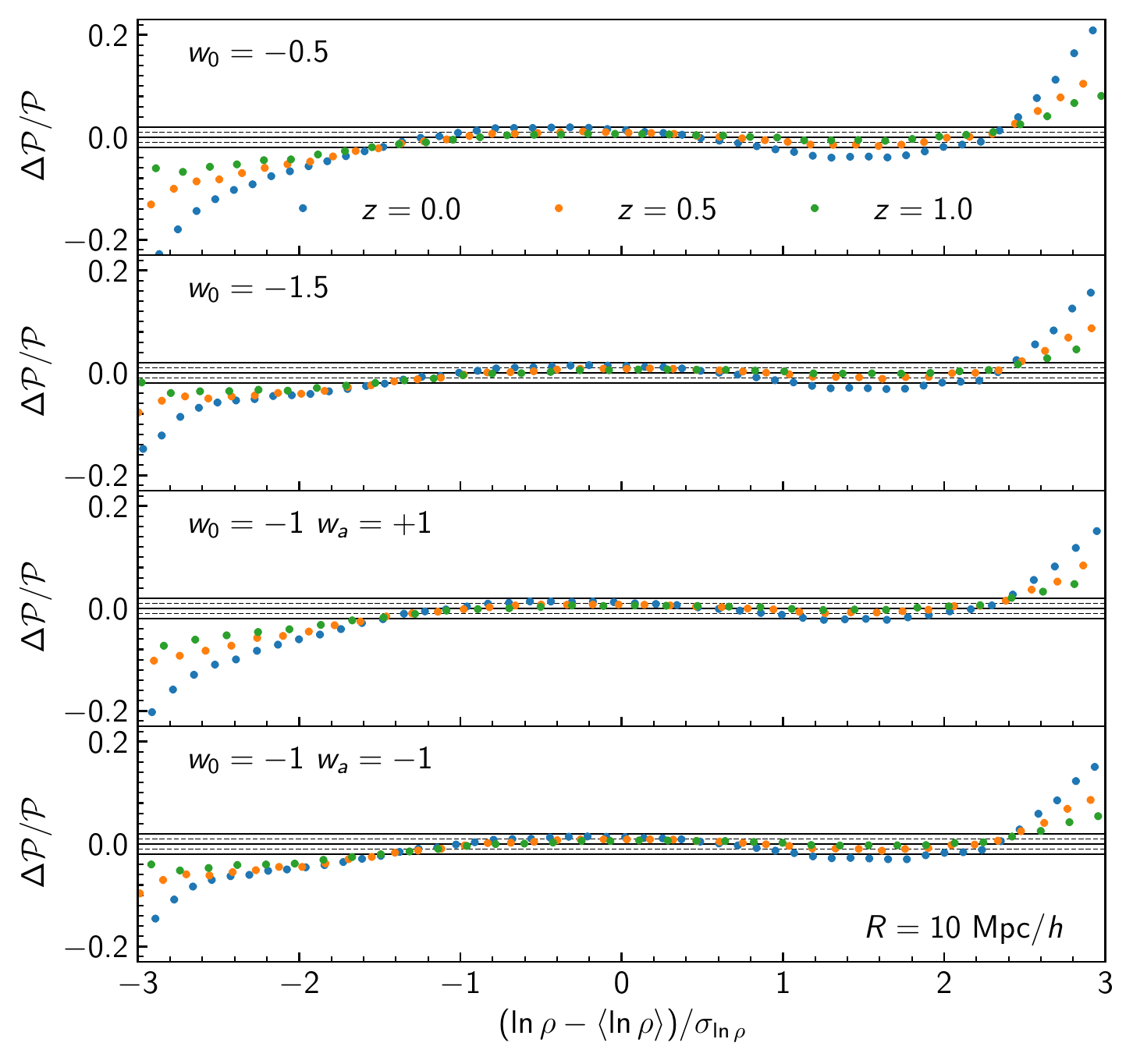}
    \caption{Residuals between the measured and predicted matter PDF in spheres of radius 10 \Mpch\ normalised to the theory predictions for various evolving dark energy cosmologies (from top to bottom). Different colors correspond to $z=0$ (blue), $z=0.5$ (orange) and $z=1$ (green). The solid and dashed lines mark the 1\% and 2\% accuracy, respectively. Note that for these simulations only one realisation is available, and the estimation of the smoothed density field differs from that used for the modified gravity simulations. Despite these differences the residuals are consistent with those for \LCDM{}, $f(R)$ gravity and DGP in Figure~\ref{fig:mg_residuals}.}
    \label{fig:wcdm_residuals}
\end{figure}

\begin{figure}
    \centering
    \includegraphics[width=\columnwidth]{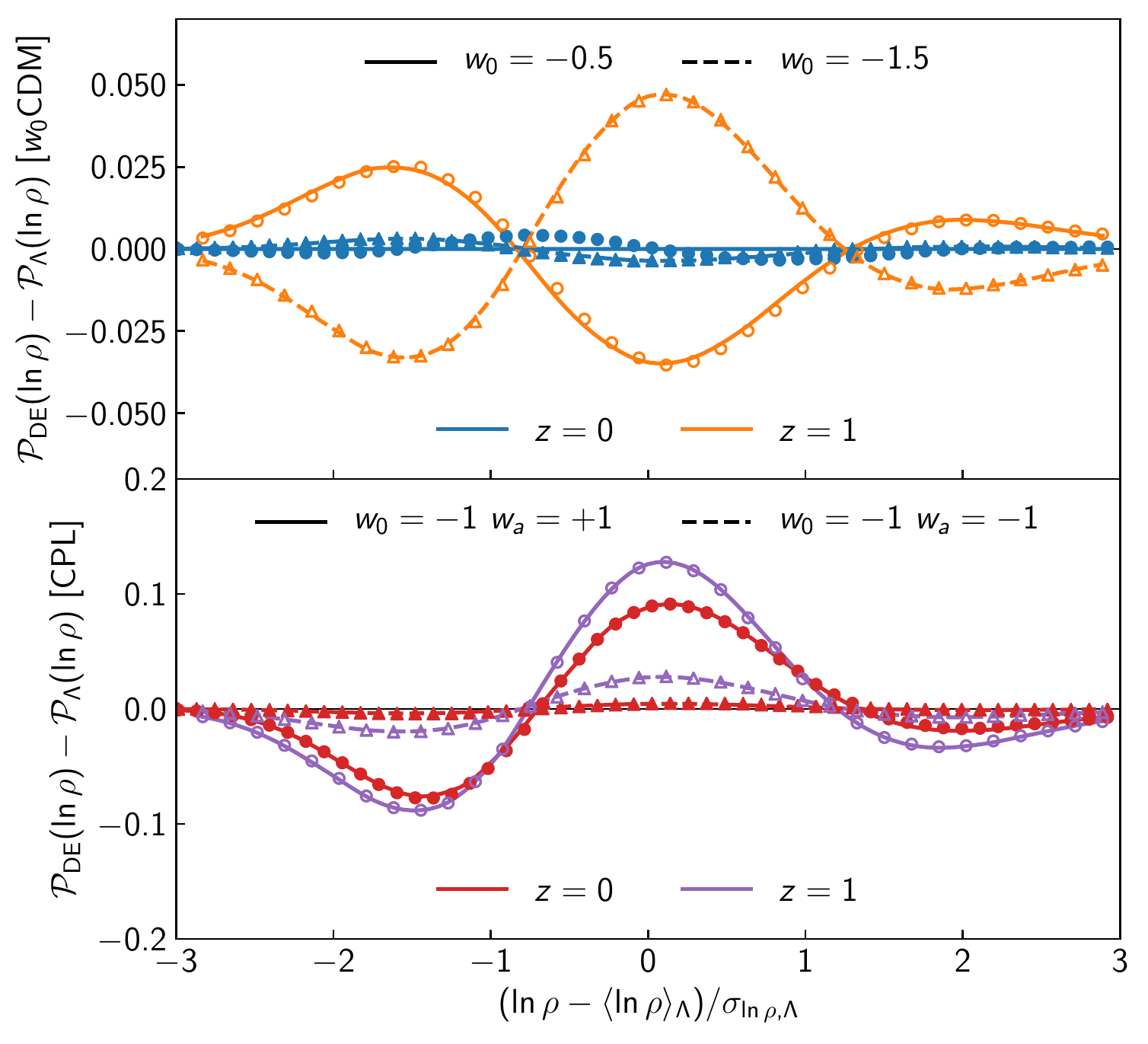}
    \caption{Measured (data points) and predicted (lines) differences from \LCDM{} of the $w_0$CDM (top) and the $w_0w_a$CDM (bottom) cosmologies at $z=0$ and $z=1$. The density field is averaged in spheres of radius $R =$ 10 \Mpch{} and the linear power spectra for all cosmologies (except $\{w_0, w_a\} = \{-1, +1\}$) are normalised such that $\sigma_8^{\rm DE}(z=0) = \sigma_8^{\Lambda}(z=0)$. Predictions are obtained from the measured non-linear variance, $\sigma^2_\mu$, together with Einstein-de Sitter spherical collapse. Except for $w_0 = -0.5$, knowledge of the non-linear variance is enough to accurately describe departures from the standard cosmology.}
    \label{fig:wcdm-lcdm_diffs}
\end{figure}

Analogously to modified gravity, the fractional deviations of the theory predictions from the simulation measurements shown in figure \ref{fig:wcdm_residuals} confirm that, despite neglecting the impact of dark energy on the spherical collapse, the LDT prescription in equation~\eqref{eq:rate_new} yields accuracies within a few percent for densities $| \ln\rho - \langle \ln\rho \rangle | < 2\sigma_{\ln\rho}$. Although the results presented here are only for density fields averaged in spheres of radius $R = 10$ \Mpch{}, we found similar or better performance for larger smoothing radii. Figure~\ref{fig:wcdm-lcdm_diffs} illustrates that in most cases deviations from the cosmological constant can be described very well by simple changes to the non-linear variance (lines). In fact, after fixing the standard cosmological parameters, the $w_0w_a$CDM and \LCDM{} cosmologies share the same shape of the linear matter power spectrum, and when using the Einstein-de Sitter approximation for the spherical dynamics the only degree of freedom left in equation~\eqref{eq:PsiNL} is the non-linear variance. However, the small yet visible discrepancies between theory and simulations for the $w_0 = -0.5$ cosmology suggest that in this extreme scenario the background expansion appreciably alters the spherical evolution and it should be taken into account to accurately predict the measured PDF deviations from \LCDM{} at both redshifts.

\subsection{Fisher forecasts} 
\label{sec:Fisher}

This section presents forecasts for DGP and $f(R)$ gravity and $w_0w_a$CDM combining the matter PDF and the matter power spectrum on mildly non-linear scales. For the MG models we determine the ability of future experiments to detect relatively small deviations from GR (i.e. F6 and DGPw) at a statistical significance $> 5\sigma$ (see Table~\ref{tab:MGdetection}), while for evolving DE we will be interested in the FoM using \LCDM\ as fiducial cosmology (see Table~\ref{tab:DEconstraints}). 

\subsubsection{Fisher formalism} 
To forecast the errors on a set of cosmological parameters, $\vec{\theta}$, we use the Fisher matrix formalism. The Fisher matrix given a (set of) summary statistics in the data vector $\vec{S}$ is defined as
\begin{equation}
F_{ij}= \sum_{\alpha,\beta}\frac{\partial S_\alpha}{\partial \theta_i}C^{-1}_{\alpha \beta}\frac{\partial S_\beta}{\partial \theta_j}~, \quad 
\label{eq:Fisher}
\end{equation}
where $S_\alpha$ is the $\alpha$-th element of the data vector $\vec{S}$ and $C^{-1}$ denotes the matrix-inverse of the covariance matrix $C$, whose components are 
\begin{equation}
\label{eq:covariance}
C_{\alpha \beta} = \langle (S_\alpha-\bar{S}_\alpha)(S_\beta - \bar{S}_\beta) \rangle\,,\quad \bar{S}_\alpha = \langle S_\alpha \rangle~.
\end{equation}
The parameter covariance matrix $\mathbf{C}(\vec{\theta})$ is then obtained as inverse of the Fisher matrix.
In the Fisher formalism, marginalisation over a subset of parameters is achieved by simply selecting the appropriate sub-elements of the parameter covariance. 

We consider three data vectors for our forecasts, corresponding to the three sets of constraints in Figures~\ref{fig:fisher_dgp}, \ref{fig:fisher_f6}, and \ref{fig:fisher_w}. These are the PDF alone, the matter power spectrum alone, and a stacked data vector which combines both the PDF and the matter power spectrum. For the PDF data vector, we only use the central region of the PDF around the peak (located in underdense regions), removing the lowest $3\%$ and highest 10\% of densities \citep[as advocated in][]{Uhlemann:2020}. We choose this approach in order to limit the impact of small-scale effects (like baryonic feedback, non-linear galaxy bias, shot noise and redshift-space distortions) that are more severe for rare events and would otherwise degrade the constraining power when moving from the 3D matter PDF to an actual observable like the spectroscopic tracer PDF.
For the matter power spectrum data vector, we limit ourselves to mildly non-linear scales up to $k_{\rm max}=0.2\, h$/Mpc to ensure the accuracy of theoretical derivatives from fitting functions, see Figure~\ref{fig:derivatives_Pk_validation}. We found the conservative scale cut for the power spectrum to be crucial to facilitate an agreement between parameter constraints and degeneracy directions from predicted and simulated derivatives, especially when considering the full set of cosmological parameters.
For all cosmological parameters, we compute partial derivatives from two-point finite differences
\begin{equation}
\frac{\partial \vec{S}}{\partial \theta}\simeq\frac{\vec{S}(\theta+d\theta)-\vec{S}(\theta-d\theta)}{2d\theta}\,.
\label{eq:derivatives}
\end{equation}
We rely on partial derivatives determined from theoretical predictions for the matter PDF from \pyLDT{} and the matter power spectrum from \texttt{ReACT} \citep{Bose:2020} combined with {\sc hmcode} \citep{Mead:2021}, which provides flexibility to compute constraints or the detection significance for extended models at the desired fiducial cosmology. The step sizes have been chosen to ensure convergence of the derivatives, and agree with the step sizes used in the {\sc Quijote} simulation suite for the set of $w_0$CDM parameters. The theory generated derivatives for $w_0$CDM parameters are validated with measurements from the {\sc Quijote} simulations in Appendix~\ref{app:small_deviations}. As discussed in Appendix~\ref{app:small_deviations}, we adopt Gaussian priors for  $\{\Ob,n_s\}$ to ensure compatibility of the matter power spectrum derivatives between simulations and theoretical predictions. The prior widths correspond to $\sigma[n_s] = 0.0041$ \citep{Planck:2018} and $\sigma[100\Omega_b h^2] = 0.052$ \citep{Cooke:2016, Abbott:2018}. 

In this work, we use the covariance matrix obtained from a set of 15000 simulations of the {\sc Quijote} $N$-body simulation suite \citep{Villaescusa-Navarro:2020} using the fiducial \LCDM{} cosmology ($\Om = 0.3175$,  $\Omega_{\rm b} = 0.049$, $H_0 = 68$ km/s/Mpc, $n_s = 0.96$, $\sigma_8 = 0.834$). The joint covariance matrix of the mildly non-linear matter PDF and the matter power spectrum is described in \cite{Uhlemann:2020}, see particularly their Figure~12. We make the approximation that the covariance matrix of the matter PDF and matter power spectrum in the mildly non-linear regime is independent of cosmology and theory of gravity and well-captured by the 15000 simulations of the {\sc Quijote} simulation suite. To mitigate potential effects of modified gravity on the covariance, we fix the standard cosmological parameters to the values of the fiducial {\sc Quijote} cosmology. In particular, we set $A_s=2.13\times 10^{-9}$ such that $\sigma_8$ increases only slightly for the modified gravity cosmologies, that is, by 1.6\% for F6 and 3.8\% for DGPw. As those are small perturbations from the fiducial \LCDM{} cosmology, they will only induce a small error on the true covariances and hence only marginally affect parameter constraints. As this error will affect both the PDF and power spectrum covariance in a similar way, comparisons of their respective constraining power are expected to be robust. For future high precision cosmology, covariance estimation for PDF-based observables from galaxy clustering and weak lensing can rely on tuned lognormal mocks \citep{Gruen:2018,Boyle_2020}, potentially complemented with predictions for effects induced by variations in the local mean density \citep{Jamieson_2020}. 
% from $\sigma_{8,\rm fid}^{\rm GR}=0.834$ to $\sigma_{8,\rm fid}^{\rm f6}=0.847$ (1.6\% higher) and $\sigma_{8,\rm fid}^{\rm nDGP}=0.866$ (3.8\% higher). 
To correct for a potential bias depending on  the size of the data vector $N_{S}$ compared to the number of simulations $N_{\rm sim}$, we multiply the inverse of the simulated covariance matrix by the Kaufman-Hartlap factor \citep{Kaufman67,Hartlap06}, $f_{\rm KH}=(N_{\rm sim} - 2 - N_{S})/(N_{\rm sim} - 1)$.  Since in our case the number of simulations for covariance estimation is very large (15000) compared to the maximal length of the data vector (218 for our three-redshift analysis of the PDF at three scales and the mildly non-linear power spectrum), this factor will be close to one throughout, $f_{\rm KH}\geq 0.985$. 
We mimic a Euclid-like effective comoving survey volume of $V \approx 20 \, ({\rm Gpc}/h)^3$ split across three redshift bins of equal width $\Delta z=0.2$ located at $z=0,0.5,1$ by
multiplying the covariance at each redshift with the ratio of the comoving shell volume to the simulation volume $V_{\rm sim}=1 \, ({\rm Gpc}/h)^3$. 
%\Cora{We visualise the resulting parameter covariances using the public Python package ChainConsumer \citep{ChainConsumer}.}

% \begin{itemize}
%     % \item \pyLDT{}
%     % \item Show DGPw forecast for different redshifts and radii, including the non-linear $P(k)$ as well.  Does the PDF help break some parameter degeneracies that improve the statistical significance of MG detection? 
%     % \item Show F6 forecast for different redshifts and radii, including the non-linear Pk as well. How much better can we do by including the PDF compared to Pk alone?
%     % \item Show $w_0w_a$CDM forecast for different redshifts and radii, including the non-linear $P(k)$ as well. FoM for various cases.  

% \end{itemize}

\subsubsection{Modified gravity}

\begin{figure}
    \centering
    \includegraphics[width=\columnwidth]{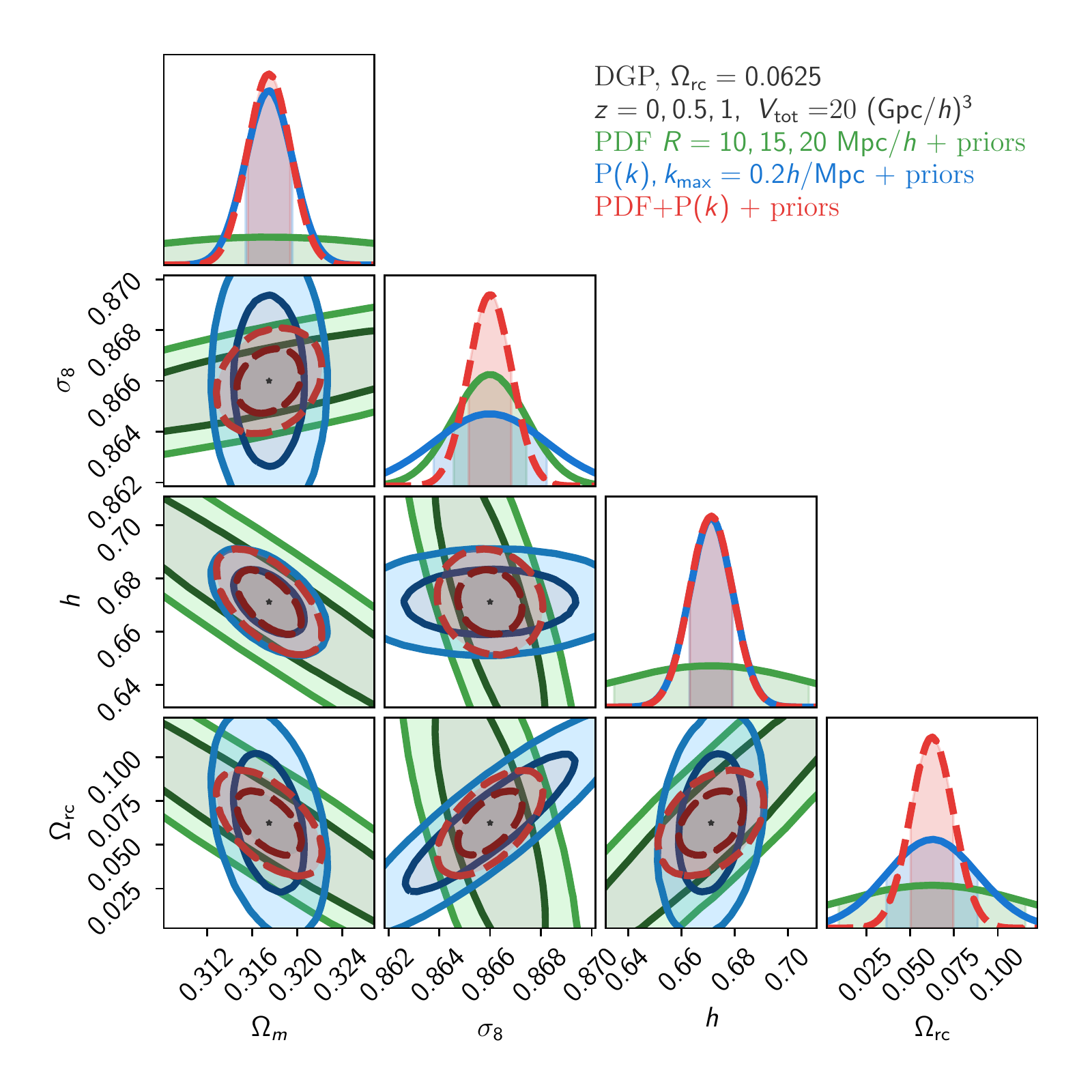}
    \caption{Marginalised Fisher forecast constraints on $\{\Om, \sigma_8, h, \Omega_{\rm rc}\}$ using external prior information on $n_s$ and $\Omega_b$ (as described in the text) for a DGPw fiducial cosmology. Contours correspond to the matter PDF at 3 scales and 3 redshifts (green), the matter power spectrum up to $k = 0.2 \ h \ \mathrm{Mpc}^{-1}$ (blue), and their combination, which includes the covariance between the PDF and power spectrum (red dashed).}
    \label{fig:fisher_dgp}
\end{figure}

\begin{figure}
    \centering
    \includegraphics[width=\columnwidth]{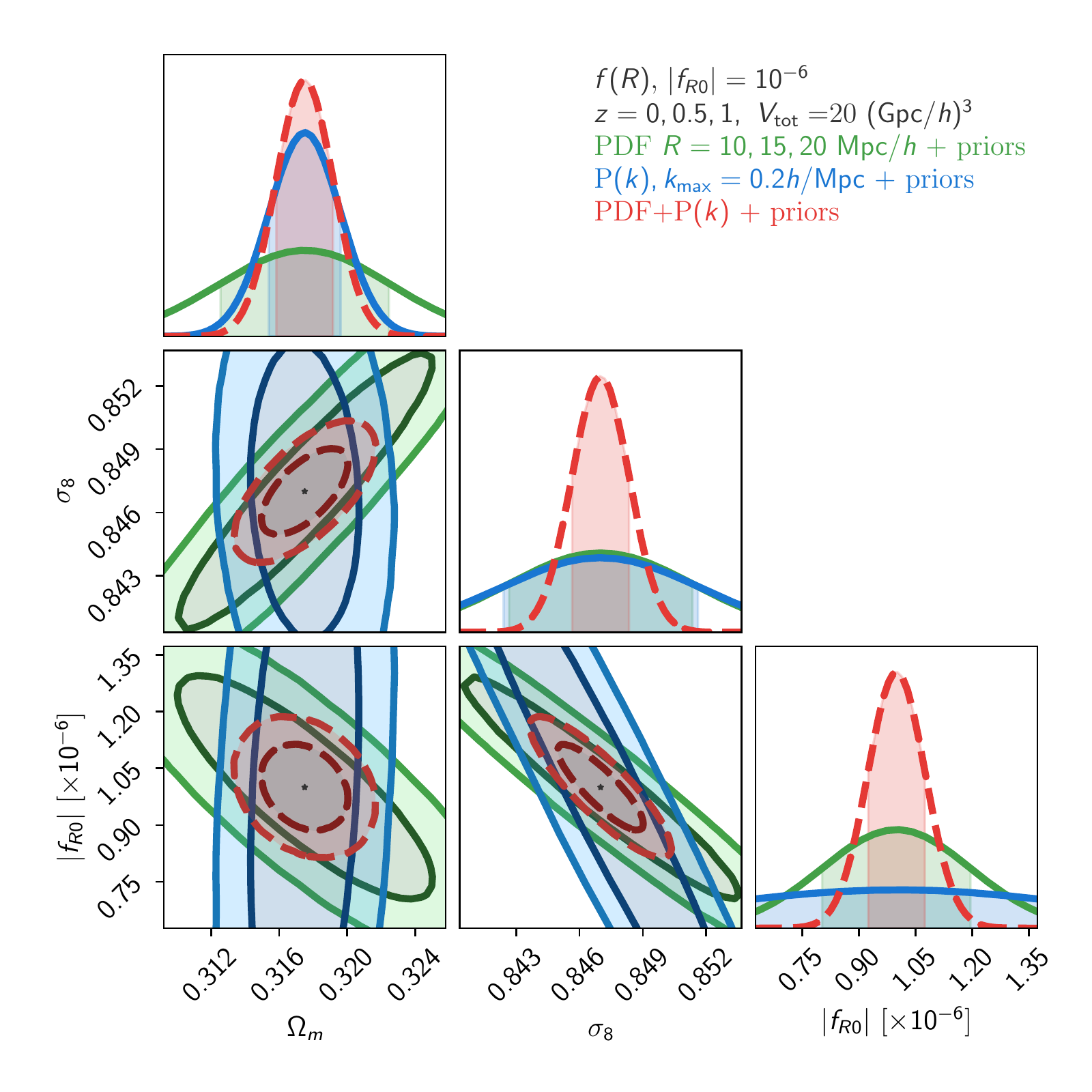}
    \caption{Marginalised Fisher forecast constraints on $\{\Om, \sigma_8, |f_{R0}| \}$ using external prior information on $n_s$ and $\Omega_b$ (as described in the text) for  an F6 fiducial cosmology. Contours correspond to the matter PDF at 3 scales and 3 redshifts (green), the matter power spectrum up to $k = 0.2 \ h \ \mathrm{Mpc}^{-1}$ (blue), and their combination, which includes the covariance between the PDF and the power spectrum (red dashed).}
    \label{fig:fisher_f6}
\end{figure}

We now compare the constraining power of the matter PDF to that of  the matter power spectrum  (with $k_{\rm max} = 0.2 \ h \ \rm Mpc^{-1}$) for DGP and $f(R)$ gravity. In all cases, the forecasts shown are marginalised over all remaining $\Lambda$CDM parameters.

Figures~\ref{fig:fisher_dgp}~and~\ref{fig:fisher_f6} show the Fisher forecasts for DGP and $f(R)$ cosmologies, respectively. Table~\ref{tab:MGdetection} summarises the detection significance for particular flavours of these modified gravity models expressed in units of  standard deviation from GR. In a universe where the growth of structure is governed by DGP gravity with $\Omega_{\rm rc} = 0.0625$, a $5\sigma$ detection of modified gravity can still be reached by combining the matter PDF with the matter power spectrum. Combining the PDF and power spectrum as complementary probes is beneficial in both MG scenarios. In particular, for DGP the matter PDF is important for constraining $\sigma_8$, while the power spectrum is important for obtaining the correct value of $\Om$. This is because while $\Om$ has a distinctive signature in the power spectrum (see Figure~\ref{fig:derivatives_Pk_validation}), the matter PDF is sensitive to the total matter density only through its impact on the skewness and the linear growth factor, $D(z)$. The anti-correlation between the Hubble parameter, $h$, and $\Om$ visible in Figure~\ref{fig:fisher_dgp} can be explained by their similar impact on the skewness of the PDF \citep[see Figure~9 in][]{Uhlemann:2020}. Evolving dark energy also presents this feature, although we do not show it in Figure~\ref{fig:fisher_w} as it does not create any unexpected degeneracy directions as in the DGP model.

\begin{figure}
    \centering
    \includegraphics[width=\columnwidth]{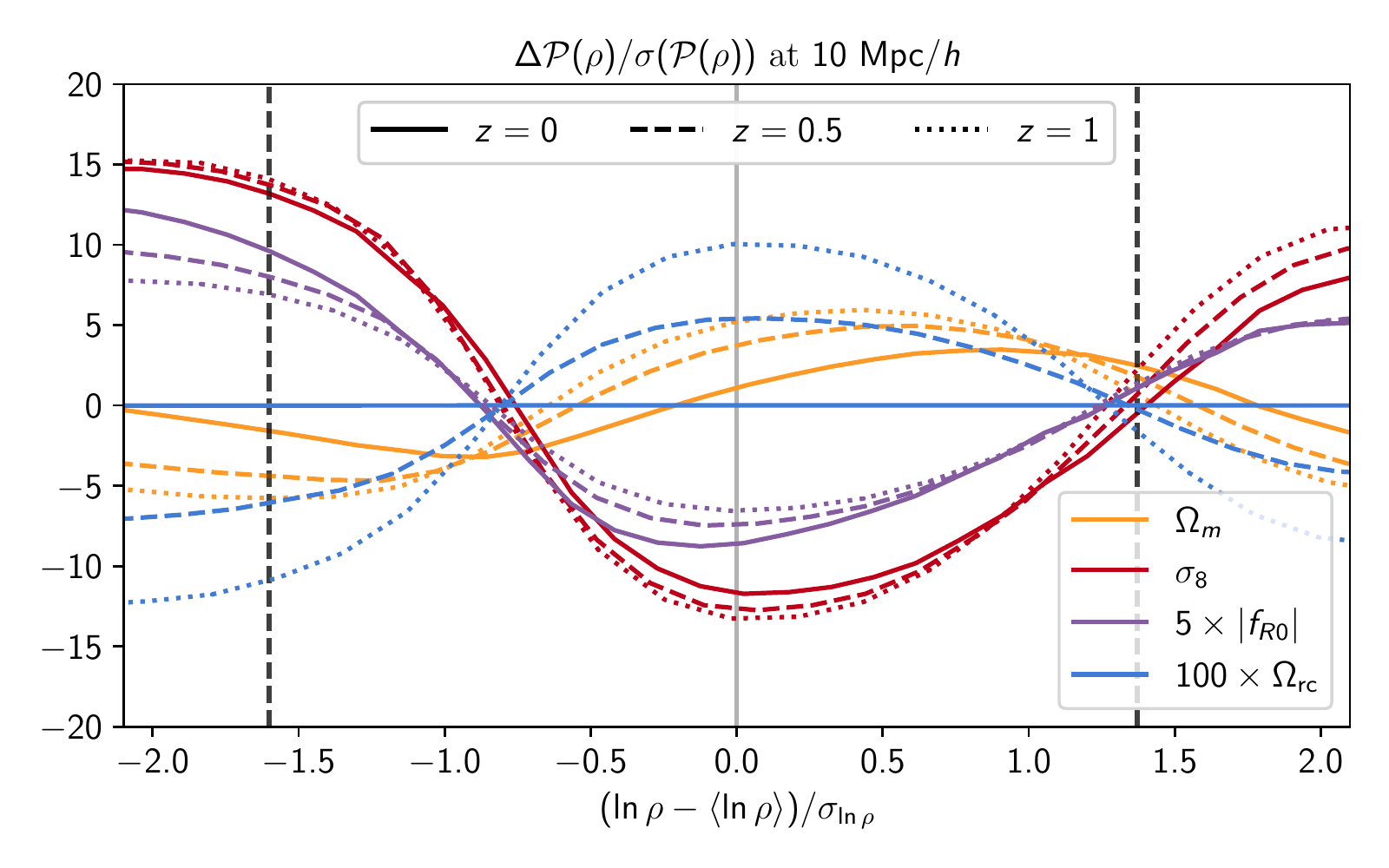}
    \caption{Comparison of PDF differences divided by the error on the PDF as estimated from the {\sc Quijote} simulations. Line style indicates the redshift, while colour indicates parameter being deviated. The vertical lines represent the region used at each redshift to construct the PDF data vector. While the shape of the $\sigma_8$ and $\Omega_{\rm rc}$ derivatives are similar, their different redshift dependence allows the degeneracy to be broken when combining redshifts. The $f_{R0}$ derivatives, in addition to having different redshift dependence, exhibit a skewness not present in the $\sigma_8$ derivatives, allowing significant information to be extracted even at a single redshift, including $z=0$. Note that the amplitudes between parameters should not be directly compared, as the $f_{R0}$ and $\Omega_{\rm rc}$ lines have been scaled up to be visible on the same scale as $\sigma_8$.}
    \label{fig:pdf_diff_comp_MG}
\end{figure}

The partial degeneracy in the PDF between $\sigma_8$ and the modified gravity parameters, $|f_{R0}|$ or $\Omega_{\rm rc}$, seen in Figures~\ref{fig:fisher_dgp}~and~\ref{fig:fisher_f6} is understood by noticing that the presence of modified gravity changes the width of the matter PDF, as can be seen in Figure~\ref{fig:pdf_diff_comp_MG}. However, the responses of the PDF to the presence of modified gravity or changes in $\sigma_8$ have different scale- and time-dependence, therefore by combining the information from different scales and redshifts we can break this degeneracy. 
Figure~\ref{fig:pdf_diff_comp_MG} shows that $\sigma_8$ and $\Omega_{\rm rc}$ have opposite effects on the PDF, which would lead to a positive correlation between these parameters. However, in Figure~\ref{fig:fisher_dgp} the $\sigma_8$--$\Omega_{\rm rc}$ plane shows an anti-correlation for the PDF, which is indirectly induced by the strong positive correlation between $\Omega_{\rm rc}$ and $h$. We checked that when $h$ is fixed to its fiducial value, rather than marginalised over, the PDF contours do indeed display a positive correlation between $\sigma_8$ and $\Omega_{\rm rc}$, as indicated by the derivatives in Figure~\ref{fig:pdf_diff_comp_MG}.

In the case of \fr, the matter PDF is particularly useful, reaching a $5\sigma$ detection before combining with the matter power spectrum. This is due to an additional skewness in the $|f_{R0}|$ derivatives sourced by the scale-dependent fifth force and the fact that the PDF holds information about deviations from $\Lambda$CDM even at redshift 0, unlike in DGP. However, Figure~\ref{fig:mg2-lcdm_diff} shows that using the non-linear variance predicted by Equation~\eqref{eq:siglogcos} is a better approximation in DGP than in $f(R)$ gravity, and we thus expect the forecasted constraints to be more reliable for the DGP model than for \fr .

\begin{table}
    \begin{tabular}{l|ccc}
    \hline
         ~ & F6 detection & DGPw detection\\\hline
        % PDF $\mathcal  P_R(\rho)$ 3 scales  & $3.38\sigma$ & $1.04\sigma$ & \\
         PDF, 3 scales + prior & $5.15\sigma$ & $1.17\sigma$ & \\
         %$P(k)$, $k_{\rm max} = 0.2h/\rm Mpc$ & $1.90\sigma$ & $1.12\sigma$ & \\
         $P(k)$, $k_{\rm max} = 0.2\ h /\rm Mpc$ + prior & $2.01\sigma$ & $2.42\sigma$ & \\
         %\hline % MNRAS only wants line breaks at top, bottom and after headings
         %PDF + $P(k)$  & $11.66\sigma$ & $5.16\sigma$ & \\
         PDF + $P(k)$ + prior & $13.40\sigma$ & $5.19\sigma$ & \\
    \hline
    \end{tabular}
    \caption{Detection significance for a fiducial $f(R)$ gravity model with $|f_{R0}|=10^{-6}$, and a fiducial DGP model with $\Omega_{\rm rc}=0.0625$. The constraints on $f(R)$ gravity from the PDF are stronger than in DGP owing to the additional skewness produced by the scale-dependent fifth force, which is visible in the $|f_{R0}|$ derivatives shown in Figure~\ref{fig:pdf_diff_comp_MG}. Moreover, unlike DGP, where the approximate theory PDF matches the $\Lambda$CDM prediction at $z=0$, in $f(R)$ gravity the PDF differs from that of the standard cosmology at low redshifts, which allows even more non-linear information to be extracted.}
    \label{tab:MGdetection}
\end{table}

\subsubsection{Evolving dark energy}

\begin{figure}
    \centering
     \includegraphics[width=\columnwidth]{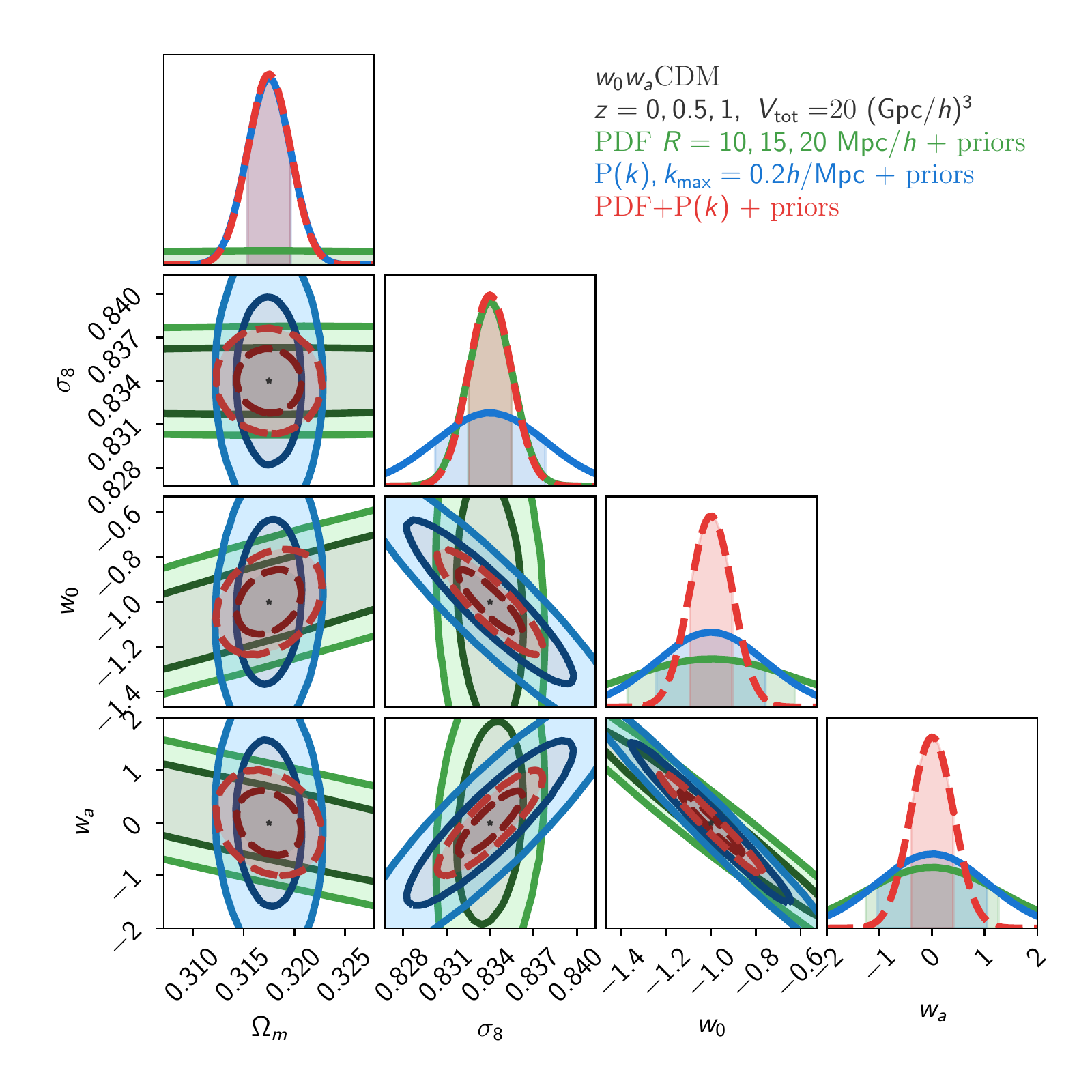}
    \caption{Fisher forecast constraints on $\{\Omega_m, \sigma_8, w_0, w_a\}$ (marginalised over $\{\Omega_b,n_s\}$ using the external prior described in the text) for the $w_0$CDM model around the fiducial {\sc Quijote} $\Lambda$CDM cosmology. Contours correspond to the matter PDF at 3 scales and 3 redshifts (green), the matter power spectrum up to $k = 0.2 \ h \ \mathrm{Mpc}^{-1}$ (blue), and their combination which includes the covariance between the PDF and power spectrum (red dashed).}
    \label{fig:fisher_w}
\end{figure}

In this section we consider a dark energy fluid with an equation of state described by Equation~\eqref{eq:DEeos}. Many of the features of the parameter constraints from the matter PDF and matter power spectrum are similar to the features seen for scale-independent modifications to GR. In particular, the matter PDF is much better at constraining $\sigma_8$ than the power spectrum, while the power spectrum more directly measures $\Om$, as can be seen in Figure~\ref{fig:fisher_w}. A summary of constraints on $\sigma_8$, $w_0$, and $w_a$, along with the dark energy Figure of Merit (FoM) is shown in Table~\ref{tab:DEconstraints}. The FoM is calculated from the inverse of the error ellipse area in the $w_0$-$w_a$ plane as
\begin{equation}
    \mathrm{FoM} = \frac{1}{\sqrt{\det(\mathbf{C}(w_0, w_a))}}\,,
\end{equation}
where $\mathbf{C}(w_0, w_a)$ is the parameter covariance matrix marginalised over all parameters except $w_0$ and $w_a$. The combined FoM for the matter PDF and matter power spectrum is a factor of 9 larger than the PDF alone, and  5 times better than the power spectrum when only information from the mildly non-linear regime is included. This combined FoM of 243 sits between the range of the pessimistic and optimistic predictions for combined galaxy clustering and weak lensing from \textit{Euclid} \citep[see Table~13 from][]{Euclid:2020}. The PDF is sufficient to measure $\sigma_8$ to sub-percent accuracy, with the inclusion of the power spectrum improving this constraint only marginally. 

Increasing either $w_0$ or $w_a$  increases the growth rate and hence the variances at $z>0$ (with marginal changes at $z=0$ due to fixed $\sigma_8$), which amounts to an anti-correlation between $w_0$ and $w_a$. Most of the other degeneracy directions in the $w_0w_a$CDM case can be understood by considerations of linear theory. Changing a single parameter at fixed $\sigma_8$ (or similarly when $\sigma_8$ is allowed to vary) induces a change in the growth rate. Suitable pairs of parameters can then produce growth rates close to the fiducial cosmology. For example, the positive correlation between $w_0$ and $\Om$ arises from the suppression of the growth rate by increasing $\Om$ while keeping $\sigma_8$ fixed. While one would expect $w_0$ and $w_a$ to vary in the same way with $\Om$ and $\sigma_8$, they in fact vary in opposing directions as shown in Figure~\ref{fig:fisher_w}. However, when $w_0$ is fixed to its fiducial value, rather than marginalised over, the contours do indeed flip in sign to the direction expected, suggesting that the tight anti-correlation in the $w_0$-$w_a$ plane dominates the other degeneracies.

\begin{table}
    \centering
    \begin{tabular}{l|cccc}
        \hline
         ~ & $\frac{\sigma[\sigma_8]}{\sigma_8^{\rm fid}}$ & $\sigma[w_0]$ &  $\sigma[w_a]$ & FoM\\\hline
         PDF, 3 scales + prior & 0.18\% & 0.37 & 1.25 & 27\\
         $P(k), k_{\rm max}=0.2 h/$Mpc + prior & 0.45\% & 0.24 & 1.03 & 50\\ %\hline % MNRAS wants only line at the top, bottom and after headings
         PDF + $P(k)$ + prior & 0.17\% & 0.09 & 0.40 & 243\\ 
         \hline
    \end{tabular}
    \caption{Constraints from mildly non-linear scales on $\sigma_8$, $w_0$ and $w_a$ derived including a prior on $\{\Ob,n_s\}$, as well as dark energy Figure of Merit (FoM) for the matter PDF, power spectrum and their combination.}
    \label{tab:DEconstraints}
\end{table}

\section{Summary and discussion}\label{sec:conclusions}

To harness the full statistical power of current and forthcoming galaxy surveys we must push past 2-point correlation functions. Gravitationally-driven non-Gaussianities are particularly sensitive to the late-time growth of structure. As a result, the full shape of the matter density PDF responds strongly to departures from GR and the cosmological constant making it a promising probe of new physics. In this work we built on the findings of \cite{Uhlemann16} with the aim of extending the large-deviation theory formalism for the 3D matter PDF to cosmologies with universally coupled fifth forces and non-standard expansion histories. As for \LCDM{}, our analytical predictions are derived from linear theory calculations and  spherical collapse dynamics, with the fiducial non-linear variance being a free parameter that can be measured from simulations. However, contrary to previous approaches \citep[cf.][]{Brax:2012}, we approximate the collapse or expansion of spherical top-hat fluctuations with an Einstein-de Sitter evolution, and showed that in the mildly non-linear regime this choice produces PDFs matching the simulations to better than a few percent around the peak of the distribution. Although in this work we analysed in great detail specific modified gravity and dark energy cosmologies, our method is readily applicable to more general models, as changes to the standard cosmology only enter the PDF through the linear matter power spectrum and the non-linear variance of the smoothed density field. We also implemented the LDT equations in \pyLDT{}, an easy-to-install and user-friendly Python package that enables fast calculations of the PDF of the spherically-averaged matter density field in \LCDM{}, modified gravity and evolving dark energy cosmologies. We employed \pyLDT{} in Fisher analyses of a \emph{Euclid}-like survey to estimate the additional information brought in by the matter PDF compared to 2-point statistics restricted to mildly non-linear scales. In all cases investigated the constraints on new physics (be it $G_{\rm eff} \neq G_{\rm Newton}$ or $w \neq -1$) from the combination of the matter PDF and power spectrum are substantially tighter than those obtained separately by the two statistics--a clear sign of complementarity \citep[see also][for massive neutrino cosmologies]{Uhlemann:2020}. For modified gravity, adding the matter PDF to the power spectrum can double the detection significance for the DGPw model to lift it above $5\sigma$ and increase the F6 detection significance sixfold as summarised in Table~\ref{tab:MGdetection}. For dark energy, combining the matter PDF with the power spectrum can also double our constraining power on the clustering amplitude, $\sigma_8$, and both of the dark energy equation of state parameters $w_0$ and $w_a$ as shown in Table~\ref{tab:DEconstraints}.

In spite of the idealised experimental set-up focusing on the statistics of the 3D matter field, our results are also encouraging for more realistic scenarios. The formalism described in this paper can be translated to galaxy survey observables accessible from weak lensing \citep{Barthelemy:2020,Boyle_2020,Thiele_2020}, galaxy clustering \citep{Repp_2020,Friedrich_2021} as well as their combination in density-split statistics \citep{Friedrich_2018,Gruen:2018}, which have been shown to be able to simultaneously extract galaxy bias, galaxy stochasticity and cosmological parameters. 
In particular, the LDT approach developed for the \LCDM{} lensing convergence PDF could be straightforwardly applied to the entire class of scalar-tensor theories with lensing potential $\Phi_{\rm lens}^{\rm MG} \approx \Phi_{\rm lens}^{\rm GR}$, which includes \fr and DGP. The PDF could also be useful for disentangling modified gravity and massive neutrinos \citep{Giocoli:2018}, but we leave the combination of those two scenarios for future work. Including the PDF of observable fields like cosmic shear or galaxy counts could help break degeneracies between astrophysical (e.g. baryonic feedback, intrinsic alignment, galaxy bias) and cosmological parameters present in the analyses of two-point statistics \citep{Patton_2017,Hadzhiyska:2021}.

\section*{Acknowledgements}

We are grateful to Marius Cautun for his support in setting up {\sc dtfe}, to Fabian Schmidt for sharing his 1D relaxation code at an early stage of this work, to Alexandre Barreira for giving access to his DGP simulations, to Jihye Shin for sharing the means and variances of the DE simulations, and to Alexander Mead for his support with {\sc hmcode}. We also thank Yanchuan Cai and Wojciech Hellwing for useful discussions. The figures in this work were created with {\sc matplotlib} \citep{Hunter:2007} and {\sc chaincosumer} \citep{Hinton:2016}, making use of the {\sc numpy} \citep{harris:2020} and {\sc scipy} \citep{Virtanen:2020} Python libraries. MC and CH acknowledge support from the European Research Council under grant number 647112. CH also acknowledges support from the Max Planck Society and the Alexander von Humboldt Foundation in the framework of the Max Planck-Humboldt Research Award endowed by the Federal Ministry of Education and Research. 
AG is supported by an EPSRC studentship under Project 2441314 from UK Research \& Innovation. 
CA and BL are supported by the European Research Council (ERC) through a starting Grant (ERC-StG-716532 PUNCA). BL is further supported by the UK Science and Technology Funding Council (STFC) Consolidated Grant No.~ST/I00162X/1 and ST/P000541/1.
This work used the DiRAC@Durham facility managed by the Institute for Computational Cosmology on behalf of the STFC DiRAC HPC Facility (\url{www.dirac.ac.uk}). The equipment was funded by BEIS capital funding via STFC capital grants ST/K00042X/1, ST/P002293/1, ST/R002371/1 and ST/S002502/1, Durham University and STFC operations grant ST/R000832/1. DiRAC is part of the National e-Infrastructure.

\section*{Data availability}
Our code to compute the matter PDF predictions is publicly available at  \url{https://github.com/mcataneo/pyLDT-cosmo}. The $f(R)$ gravity simulation data used in this paper may be available upon request to the corresponding author. The matter PDF measured from the Quijote simulations are publicly available at \url{https://quijote-simulations.readthedocs.io/en/latest/}. The matter PDF measurements for the dark energy cosmologies are publicly available at \url{https://astro.kias.re.kr/jhshin/}.



\bibliographystyle{mnras}
\bibliography{references}

\begin{thebibliography}{}
\makeatletter
\relax
\def\mn@urlcharsother{\let\do\@makeother \do\$\do\&\do\#\do\^\do\_\do\%\do\~}
\def\mn@doi{\begingroup\mn@urlcharsother \@ifnextchar [ {\mn@doi@}
  {\mn@doi@[]}}
\def\mn@doi@[#1]#2{\def\@tempa{#1}\ifx\@tempa\@empty \href
  {http://dx.doi.org/#2} {doi:#2}\else \href {http://dx.doi.org/#2} {#1}\fi
  \endgroup}
\def\mn@eprint#1#2{\mn@eprint@#1:#2::\@nil}
\def\mn@eprint@arXiv#1{\href {http://arxiv.org/abs/#1} {{\tt arXiv:#1}}}
\def\mn@eprint@dblp#1{\href {http://dblp.uni-trier.de/rec/bibtex/#1.xml}
  {dblp:#1}}
\def\mn@eprint@#1:#2:#3:#4\@nil{\def\@tempa {#1}\def\@tempb {#2}\def\@tempc
  {#3}\ifx \@tempc \@empty \let \@tempc \@tempb \let \@tempb \@tempa \fi \ifx
  \@tempb \@empty \def\@tempb {arXiv}\fi \@ifundefined
  {mn@eprint@\@tempb}{\@tempb:\@tempc}{\expandafter \expandafter \csname
  mn@eprint@\@tempb\endcsname \expandafter{\@tempc}}}

\bibitem[\protect\citeauthoryear{{Abbott} et~al.,}{{Abbott}
  et~al.}{2017}]{Abbott:2017}
{Abbott} B.~P.,  et~al., 2017, \mn@doi [\apjl] {10.3847/2041-8213/aa920c},
  \href {https://ui.adsabs.harvard.edu/abs/2017ApJ...848L..13A} {848, L13}

\bibitem[\protect\citeauthoryear{{Abbott} et~al.,}{{Abbott}
  et~al.}{2018}]{Abbott:2018}
{Abbott} T.~M.~C.,  et~al., 2018, \mn@doi [\mnras] {10.1093/mnras/sty1939},
  \href {https://ui.adsabs.harvard.edu/abs/2018MNRAS.480.3879A} {480, 3879}

\bibitem[\protect\citeauthoryear{{Abbott} et~al.,}{{Abbott}
  et~al.}{2019a}]{DES-y1MG:2019}
{Abbott} T.~M.~C.,  et~al., 2019a, \mn@doi [\prd] {10.1103/PhysRevD.99.123505},
  \href {https://ui.adsabs.harvard.edu/abs/2019PhRvD..99l3505A} {99, 123505}

\bibitem[\protect\citeauthoryear{{Abbott} et~al.,}{{Abbott}
  et~al.}{2019b}]{Abbott:2019}
{Abbott} B.~P.,  et~al., 2019b, \mn@doi [\prl]
  {10.1103/PhysRevLett.123.011102}, \href
  {https://ui.adsabs.harvard.edu/abs/2019PhRvL.123a1102A} {123, 011102}

\bibitem[\protect\citeauthoryear{{Aiola} et~al.,}{{Aiola}
  et~al.}{2020}]{Aiola:2020}
{Aiola} S.,  et~al., 2020, \mn@doi [\jcap] {10.1088/1475-7516/2020/12/047},
  \href {https://ui.adsabs.harvard.edu/abs/2020JCAP...12..047A} {2020, 047}

\bibitem[\protect\citeauthoryear{{Alam} et~al.,}{{Alam}
  et~al.}{2021}]{Alam:2021}
{Alam} S.,  et~al., 2021, \mn@doi [\prd] {10.1103/PhysRevD.103.083533}, \href
  {https://ui.adsabs.harvard.edu/abs/2021PhRvD.103h3533A} {103, 083533}

\bibitem[\protect\citeauthoryear{{Amon} et~al.,}{{Amon}
  et~al.}{2018}]{Amon_2018_Eg}
{Amon} A.,  et~al., 2018, \mn@doi [\mnras] {10.1093/mnras/sty1624}, \href
  {https://ui.adsabs.harvard.edu/abs/2018MNRAS.479.3422A} {479, 3422}

\bibitem[\protect\citeauthoryear{{Arnold}, {Leo}  \& {Li}}{{Arnold}
  et~al.}{2019}]{arnold2019}
{Arnold} C.,  {Leo} M.,   {Li} B.,  2019, \mn@doi [Nature Astronomy]
  {10.1038/s41550-019-0823-y}, \href
  {https://ui.adsabs.harvard.edu/abs/2019NatAs.tmp..386A} {p.~386}

\bibitem[\protect\citeauthoryear{Barreira, Bose  \& Li}{Barreira
  et~al.}{2015}]{Barreira:2015xvp}
Barreira A.,  Bose S.,   Li B.,  2015, \mn@doi [JCAP]
  {10.1088/1475-7516/2015/12/059}, 12, 059

\bibitem[\protect\citeauthoryear{{Barthelemy}, {Codis}, {Uhlemann},
  {Bernardeau}  \& {Gavazzi}}{{Barthelemy} et~al.}{2020}]{Barthelemy:2020}
{Barthelemy} A.,  {Codis} S.,  {Uhlemann} C.,  {Bernardeau} F.,   {Gavazzi} R.,
   2020, \mn@doi [\mnras] {10.1093/mnras/staa053}, \href
  {https://ui.adsabs.harvard.edu/abs/2020MNRAS.492.3420B} {492, 3420}

\bibitem[\protect\citeauthoryear{{Bellini} \& {Sawicki}}{{Bellini} \&
  {Sawicki}}{2014}]{Bellini:2014}
{Bellini} E.,  {Sawicki} I.,  2014, \mn@doi [\jcap]
  {10.1088/1475-7516/2014/07/050}, \href
  {https://ui.adsabs.harvard.edu/abs/2014JCAP...07..050B} {2014, 050}

\bibitem[\protect\citeauthoryear{{Bellini}, {Sawicki}  \&
  {Zumalac{\'a}rregui}}{{Bellini} et~al.}{2020}]{Bellini:2020}
{Bellini} E.,  {Sawicki} I.,   {Zumalac{\'a}rregui} M.,  2020, \mn@doi [\jcap]
  {10.1088/1475-7516/2020/02/008}, \href
  {https://ui.adsabs.harvard.edu/abs/2020JCAP...02..008B} {2020, 008}

\bibitem[\protect\citeauthoryear{{Bernardeau}}{{Bernardeau}}{1994}]{Bernardeau94}
{Bernardeau} F.,  1994, \mn@doi [\apj] {10.1086/174121}, \href
  {http://adsabs.harvard.edu/abs/1994ApJ...427...51B} {427, 51}

\bibitem[\protect\citeauthoryear{Bernardeau \& Brax}{Bernardeau \&
  Brax}{2011}]{Bernardeau_2011}
Bernardeau F.,  Brax P.,  2011, \mn@doi [Journal of Cosmology and Astroparticle
  Physics] {10.1088/1475-7516/2011/06/019}, 2011, 019–019

\bibitem[\protect\citeauthoryear{{Bernardeau} \& {Reimberg}}{{Bernardeau} \&
  {Reimberg}}{2016}]{LDPinLSS}
{Bernardeau} F.,  {Reimberg} P.,  2016, \mn@doi [\prd]
  {10.1103/PhysRevD.94.063520}, \href
  {https://ui.adsabs.harvard.edu/abs/2016PhRvD..94f3520B} {94, 063520}

\bibitem[\protect\citeauthoryear{{Bernardeau}, {Colombi}, {Gazta{\~n}aga}  \&
  {Scoccimarro}}{{Bernardeau} et~al.}{2002}]{Bernardeau:2002}
{Bernardeau} F.,  {Colombi} S.,  {Gazta{\~n}aga} E.,   {Scoccimarro} R.,  2002,
  \mn@doi [\physrep] {10.1016/S0370-1573(02)00135-7}, \href
  {https://ui.adsabs.harvard.edu/abs/2002PhR...367....1B} {367, 1}

\bibitem[\protect\citeauthoryear{{Bernardeau}, {Pichon}  \&
  {Codis}}{{Bernardeau} et~al.}{2014}]{Bernardeau14}
{Bernardeau} F.,  {Pichon} C.,   {Codis} S.,  2014, \mn@doi [\prd]
  {10.1103/PhysRevD.90.103519}, \href
  {http://adsabs.harvard.edu/abs/2014PhRvD..90j3519B} {90, 103519}

\bibitem[\protect\citeauthoryear{Blas, Lesgourgues  \& Tram}{Blas
  et~al.}{2011}]{CLASS}
Blas D.,  Lesgourgues J.,   Tram T.,  2011, \mn@doi [Journal of Cosmology and
  Astroparticle Physics] {10.1088/1475-7516/2011/07/034}, 2011, 034–034

\bibitem[\protect\citeauthoryear{{Bocquet}, {Heitmann}, {Habib}, {Lawrence},
  {Uram}, {Frontiere}, {Pope}  \& {Finkel}}{{Bocquet}
  et~al.}{2020}]{Bocquet:2020}
{Bocquet} S.,  {Heitmann} K.,  {Habib} S.,  {Lawrence} E.,  {Uram} T.,
  {Frontiere} N.,  {Pope} A.,   {Finkel} H.,  2020, \mn@doi [\apj]
  {10.3847/1538-4357/abac5c}, \href
  {https://ui.adsabs.harvard.edu/abs/2020ApJ...901....5B} {901, 5}

\bibitem[\protect\citeauthoryear{{Borisov}, {Jain}  \& {Zhang}}{{Borisov}
  et~al.}{2012}]{Borisov:2012}
{Borisov} A.,  {Jain} B.,   {Zhang} P.,  2012, \mn@doi [\prd]
  {10.1103/PhysRevD.85.063518}, \href
  {https://ui.adsabs.harvard.edu/abs/2012PhRvD..85f3518B} {85, 063518}

\bibitem[\protect\citeauthoryear{{Bose}, {Cataneo}, {Tr{\"o}ster}, {Xia},
  {Heymans}  \& {Lombriser}}{{Bose} et~al.}{2020a}]{Bose:2020}
{Bose} B.,  {Cataneo} M.,  {Tr{\"o}ster} T.,  {Xia} Q.,  {Heymans} C.,
  {Lombriser} L.,  2020a, \mn@doi [\mnras] {10.1093/mnras/staa2696}, \href
  {https://ui.adsabs.harvard.edu/abs/2020MNRAS.498.4650B} {498, 4650}

\bibitem[\protect\citeauthoryear{{Bose}, {Byun}, {Lacasa}, {Moradinezhad
  Dizgah}  \& {Lombriser}}{{Bose} et~al.}{2020b}]{Bose:2020b}
{Bose} B.,  {Byun} J.,  {Lacasa} F.,  {Moradinezhad Dizgah} A.,   {Lombriser}
  L.,  2020b, \mn@doi [\jcap] {10.1088/1475-7516/2020/02/025}, \href
  {https://ui.adsabs.harvard.edu/abs/2020JCAP...02..025B} {2020, 025}

\bibitem[\protect\citeauthoryear{{Boyle}, {Uhlemann}, {Friedrich},
  {Barthelemy}, {Codis}, {Bernardeau}, {Giocoli}  \& {Baldi}}{{Boyle}
  et~al.}{2020}]{Boyle_2020}
{Boyle} A.,  {Uhlemann} C.,  {Friedrich} O.,  {Barthelemy} A.,  {Codis} S.,
  {Bernardeau} F.,  {Giocoli} C.,   {Baldi} M.,  2020, arXiv e-prints, \href
  {https://ui.adsabs.harvard.edu/abs/2020arXiv201207771B} {p. arXiv:2012.07771}

\bibitem[\protect\citeauthoryear{{Brax} \& {Valageas}}{{Brax} \&
  {Valageas}}{2012}]{Brax:2012}
{Brax} P.,  {Valageas} P.,  2012, \mn@doi [\prd] {10.1103/PhysRevD.86.063512},
  \href {https://ui.adsabs.harvard.edu/abs/2012PhRvD..86f3512B} {86, 063512}

\bibitem[\protect\citeauthoryear{{Casarini}, {Bonometto}, {Tessarotto}  \&
  {Corasaniti}}{{Casarini} et~al.}{2016}]{Casarini:2016}
{Casarini} L.,  {Bonometto} S.~A.,  {Tessarotto} E.,   {Corasaniti} P.~S.,
  2016, \mn@doi [\jcap] {10.1088/1475-7516/2016/08/008}, \href
  {https://ui.adsabs.harvard.edu/abs/2016JCAP...08..008C} {2016, 008}

\bibitem[\protect\citeauthoryear{{Cataneo}, {Rapetti}, {Lombriser}  \&
  {Li}}{{Cataneo} et~al.}{2016}]{Cataneo:2016}
{Cataneo} M.,  {Rapetti} D.,  {Lombriser} L.,   {Li} B.,  2016, \mn@doi [\jcap]
  {10.1088/1475-7516/2016/12/024}, \href
  {https://ui.adsabs.harvard.edu/abs/2016JCAP...12..024C} {2016, 024}

\bibitem[\protect\citeauthoryear{{Cataneo}, {Lombriser}, {Heymans}, {Mead},
  {Barreira}, {Bose}  \& {Li}}{{Cataneo} et~al.}{2019}]{Cataneo:2019}
{Cataneo} M.,  {Lombriser} L.,  {Heymans} C.,  {Mead} A.~J.,  {Barreira} A.,
  {Bose} S.,   {Li} B.,  2019, \mn@doi [\mnras] {10.1093/mnras/stz1836}, \href
  {https://ui.adsabs.harvard.edu/abs/2019MNRAS.488.2121C} {488, 2121}

\bibitem[\protect\citeauthoryear{{Cautun} \& {van de Weygaert}}{{Cautun} \&
  {van de Weygaert}}{2011}]{Cautun:2011}
{Cautun} M.~C.,  {van de Weygaert} R.,  2011, arXiv e-prints, \href
  {https://ui.adsabs.harvard.edu/abs/2011arXiv1105.0370C} {p. arXiv:1105.0370}

\bibitem[\protect\citeauthoryear{{Chevallier} \& {Polarski}}{{Chevallier} \&
  {Polarski}}{2001}]{Chevallier:2001}
{Chevallier} M.,  {Polarski} D.,  2001, \mn@doi [International Journal of
  Modern Physics D] {10.1142/S0218271801000822}, \href
  {https://ui.adsabs.harvard.edu/abs/2001IJMPD..10..213C} {10, 213}

\bibitem[\protect\citeauthoryear{{Chudaykin}, {Dolgikh}  \&
  {Ivanov}}{{Chudaykin} et~al.}{2021}]{Chudaykin:2021}
{Chudaykin} A.,  {Dolgikh} K.,   {Ivanov} M.~M.,  2021, \mn@doi [\prd]
  {10.1103/PhysRevD.103.023507}, \href
  {https://ui.adsabs.harvard.edu/abs/2021PhRvD.103b3507C} {103, 023507}

\bibitem[\protect\citeauthoryear{{Codis}, {Pichon}, {Bernardeau}, {Uhlemann}
  \& {Prunet}}{{Codis} et~al.}{2016}]{Codis:2016}
{Codis} S.,  {Pichon} C.,  {Bernardeau} F.,  {Uhlemann} C.,   {Prunet} S.,
  2016, \mn@doi [\mnras] {10.1093/mnras/stw1084}, \href
  {https://ui.adsabs.harvard.edu/abs/2016MNRAS.460.1549C} {460, 1549}

\bibitem[\protect\citeauthoryear{{Contarini}, {Marulli}, {Moscardini},
  {Veropalumbo}, {Giocoli}  \& {Baldi}}{{Contarini}
  et~al.}{2021}]{Contarini:2021}
{Contarini} S.,  {Marulli} F.,  {Moscardini} L.,  {Veropalumbo} A.,  {Giocoli}
  C.,   {Baldi} M.,  2021, \mn@doi [\mnras] {10.1093/mnras/stab1112}, \href
  {https://ui.adsabs.harvard.edu/abs/2021MNRAS.504.5021C} {504, 5021}

\bibitem[\protect\citeauthoryear{{Cooke}, {Pettini}, {Nollett}  \&
  {Jorgenson}}{{Cooke} et~al.}{2016}]{Cooke:2016}
{Cooke} R.~J.,  {Pettini} M.,  {Nollett} K.~M.,   {Jorgenson} R.,  2016,
  \mn@doi [\apj] {10.3847/0004-637X/830/2/148}, \href
  {https://ui.adsabs.harvard.edu/abs/2016ApJ...830..148C} {830, 148}

\bibitem[\protect\citeauthoryear{{Crisostomi}, {Lewandowski}  \&
  {Vernizzi}}{{Crisostomi} et~al.}{2020}]{Crisostomi:2020}
{Crisostomi} M.,  {Lewandowski} M.,   {Vernizzi} F.,  2020, \mn@doi [\prd]
  {10.1103/PhysRevD.101.123501}, \href
  {https://ui.adsabs.harvard.edu/abs/2020PhRvD.101l3501C} {101, 123501}

\bibitem[\protect\citeauthoryear{{Crocce}, {Pueblas}  \&
  {Scoccimarro}}{{Crocce} et~al.}{2006}]{lpt}
{Crocce} M.,  {Pueblas} S.,   {Scoccimarro} R.,  2006, \mn@doi [\mnras]
  {10.1111/j.1365-2966.2006.11040.x}, \href
  {https://ui.adsabs.harvard.edu/abs/2006MNRAS.373..369C} {373, 369}

\bibitem[\protect\citeauthoryear{{Cusin}, {Lewandowski}  \& {Vernizzi}}{{Cusin}
  et~al.}{2018}]{Cusin:2018}
{Cusin} G.,  {Lewandowski} M.,   {Vernizzi} F.,  2018, \mn@doi [\jcap]
  {10.1088/1475-7516/2018/04/005}, \href
  {https://ui.adsabs.harvard.edu/abs/2018JCAP...04..005C} {2018, 005}

\bibitem[\protect\citeauthoryear{{DES Collaboration} et~al.,}{{DES
  Collaboration} et~al.}{2021}]{DES-y3kp:2021}
{DES Collaboration} et~al., 2021, arXiv e-prints, \href
  {https://ui.adsabs.harvard.edu/abs/2021arXiv210513549D} {p. arXiv:2105.13549}

\bibitem[\protect\citeauthoryear{{De Felice} \& {Tsujikawa}}{{De Felice} \&
  {Tsujikawa}}{2010}]{DeFelice:2010}
{De Felice} A.,  {Tsujikawa} S.,  2010, \mn@doi [Living Reviews in Relativity]
  {10.12942/lrr-2010-3}, \href
  {https://ui.adsabs.harvard.edu/abs/2010LRR....13....3D} {13, 3}

\bibitem[\protect\citeauthoryear{{Di Valentino} et~al.,}{{Di Valentino}
  et~al.}{2021a}]{DiValentino:2021b}
{Di Valentino} E.,  et~al., 2021a, \mn@doi [Classical and Quantum Gravity]
  {10.1088/1361-6382/ac086d}, \href
  {https://ui.adsabs.harvard.edu/abs/2021CQGra..38o3001D} {38, 153001}

\bibitem[\protect\citeauthoryear{{Di Valentino} et~al.,}{{Di Valentino}
  et~al.}{2021b}]{DiValentino:2021}
{Di Valentino} E.,  et~al., 2021b, \mn@doi [Astroparticle Physics]
  {10.1016/j.astropartphys.2021.102604}, \href
  {https://ui.adsabs.harvard.edu/abs/2021APh...13102604D} {131, 102604}

\bibitem[\protect\citeauthoryear{{Douspis}, {Salvati}  \& {Aghanim}}{{Douspis}
  et~al.}{2019}]{Douspis:2019}
{Douspis} M.,  {Salvati} L.,   {Aghanim} N.,  2019, arXiv e-prints, \href
  {https://ui.adsabs.harvard.edu/abs/2019arXiv190105289D} {p. arXiv:1901.05289}

\bibitem[\protect\citeauthoryear{{Dutcher} et~al.,}{{Dutcher}
  et~al.}{2021}]{Dutcher:2021}
{Dutcher} D.,  et~al., 2021, \mn@doi [\prd] {10.1103/PhysRevD.104.022003},
  \href {https://ui.adsabs.harvard.edu/abs/2021PhRvD.104b2003D} {104, 022003}

\bibitem[\protect\citeauthoryear{{Dvali}, {Gabadadze}  \& {Porrati}}{{Dvali}
  et~al.}{2000}]{Dvali:2000}
{Dvali} G.,  {Gabadadze} G.,   {Porrati} M.,  2000, \mn@doi [Physics Letters B]
  {10.1016/S0370-2693(00)00669-9}, \href
  {https://ui.adsabs.harvard.edu/abs/2000PhLB..485..208D} {485, 208}

\bibitem[\protect\citeauthoryear{{Euclid Collaboration} et~al.,}{{Euclid
  Collaboration} et~al.}{2020}]{Euclid:2020}
{Euclid Collaboration} et~al., 2020, \mn@doi [\aap]
  {10.1051/0004-6361/202038071}, \href
  {https://ui.adsabs.harvard.edu/abs/2020A&A...642A.191E} {642, A191}

\bibitem[\protect\citeauthoryear{{Euclid Collaboration} et~al.,}{{Euclid
  Collaboration} et~al.}{2021}]{EE2:2021}
{Euclid Collaboration} et~al., 2021, \mn@doi [\mnras] {10.1093/mnras/stab1366},
  \href {https://ui.adsabs.harvard.edu/abs/2021MNRAS.505.2840E} {505, 2840}

\bibitem[\protect\citeauthoryear{{Falck}, {Koyama}  \& {Zhao}}{{Falck}
  et~al.}{2015}]{Falck:2015}
{Falck} B.,  {Koyama} K.,   {Zhao} G.-B.,  2015, \mn@doi [\jcap]
  {10.1088/1475-7516/2015/07/049}, \href
  {https://ui.adsabs.harvard.edu/abs/2015JCAP...07..049F} {2015, 049}

\bibitem[\protect\citeauthoryear{{Fang}, {Li}  \& {Zhao}}{{Fang}
  et~al.}{2017}]{Fang:2017}
{Fang} W.,  {Li} B.,   {Zhao} G.-B.,  2017, \mn@doi [\prl]
  {10.1103/PhysRevLett.118.181301}, \href
  {https://ui.adsabs.harvard.edu/abs/2017PhRvL.118r1301F} {118, 181301}

\bibitem[\protect\citeauthoryear{{Ferreira}}{{Ferreira}}{2019}]{Ferreira:2019}
{Ferreira} P.~G.,  2019, \mn@doi [\araa] {10.1146/annurev-astro-091918-104423},
  \href {https://ui.adsabs.harvard.edu/abs/2019ARA&A..57..335F} {57, 335}

\bibitem[\protect\citeauthoryear{Friedrich et~al.,}{Friedrich
  et~al.}{2018}]{Friedrich_2018}
Friedrich O.,  et~al., 2018, \mn@doi [Physical Review D]
  {10.1103/physrevd.98.023508}, 98

\bibitem[\protect\citeauthoryear{{Friedrich}, {Uhlemann},
  {Villaescusa-Navarro}, {Baldauf}, {Manera}  \& {Nishimichi}}{{Friedrich}
  et~al.}{2020}]{Friedrich20pNG}
{Friedrich} O.,  {Uhlemann} C.,  {Villaescusa-Navarro} F.,  {Baldauf} T.,
  {Manera} M.,   {Nishimichi} T.,  2020, \mn@doi [\mnras]
  {10.1093/mnras/staa2160}, \href
  {https://ui.adsabs.harvard.edu/abs/2020MNRAS.498..464F} {498, 464}

\bibitem[\protect\citeauthoryear{Friedrich, Halder, Boyle, Uhlemann, Britt,
  Codis, Gruen  \& Hahn}{Friedrich et~al.}{2021}]{Friedrich_2021}
Friedrich O.,  Halder A.,  Boyle A.,  Uhlemann C.,  Britt D.,  Codis S.,  Gruen
  D.,   Hahn C.,  2021, The PDF perspective on the tracer-matter connection:
  Lagrangian bias and non-Poissonian shot noise (\mn@eprint {arXiv}
  {2107.02300})

\bibitem[\protect\citeauthoryear{{Frusciante} \& {Perenon}}{{Frusciante} \&
  {Perenon}}{2020}]{Frusciante:2020}
{Frusciante} N.,  {Perenon} L.,  2020, \mn@doi [\physrep]
  {10.1016/j.physrep.2020.02.004}, \href
  {https://ui.adsabs.harvard.edu/abs/2020PhR...857....1F} {857, 1}

\bibitem[\protect\citeauthoryear{{Giocoli}, {Baldi}  \& {Moscardini}}{{Giocoli}
  et~al.}{2018}]{Giocoli:2018}
{Giocoli} C.,  {Baldi} M.,   {Moscardini} L.,  2018, \mn@doi [\mnras]
  {10.1093/mnras/sty2465}, \href
  {https://ui.adsabs.harvard.edu/abs/2018MNRAS.481.2813G} {481, 2813}

\bibitem[\protect\citeauthoryear{{Gleyzes}, {Langlois}, {Piazza}  \&
  {Vernizzi}}{{Gleyzes} et~al.}{2013}]{Gleyzes:2013}
{Gleyzes} J.,  {Langlois} D.,  {Piazza} F.,   {Vernizzi} F.,  2013, \mn@doi
  [\jcap] {10.1088/1475-7516/2013/08/025}, \href
  {https://ui.adsabs.harvard.edu/abs/2013JCAP...08..025G} {2013, 025}

\bibitem[\protect\citeauthoryear{{Gruen} et~al.,}{{Gruen}
  et~al.}{2018}]{Gruen:2018}
{Gruen} D.,  et~al., 2018, \mn@doi [\prd] {10.1103/PhysRevD.98.023507}, \href
  {https://ui.adsabs.harvard.edu/abs/2018PhRvD..98b3507G} {98, 023507}

\bibitem[\protect\citeauthoryear{{Hadzhiyska}, {Liu}, {Somerville},
  {Gabrielpillai}, {Bose}, {Eisenstein}  \& {Hernquist}}{{Hadzhiyska}
  et~al.}{2021}]{Hadzhiyska:2021}
{Hadzhiyska} B.,  {Liu} S.,  {Somerville} R.~S.,  {Gabrielpillai} A.,  {Bose}
  S.,  {Eisenstein} D.,   {Hernquist} L.,  2021, arXiv e-prints, \href
  {https://ui.adsabs.harvard.edu/abs/2021arXiv210800006H} {p. arXiv:2108.00006}

\bibitem[\protect\citeauthoryear{{Hagstotz}, {Costanzi}, {Baldi}  \&
  {Weller}}{{Hagstotz} et~al.}{2019}]{Hagstotz:2019}
{Hagstotz} S.,  {Costanzi} M.,  {Baldi} M.,   {Weller} J.,  2019, \mn@doi
  [\mnras] {10.1093/mnras/stz1051}, \href
  {https://ui.adsabs.harvard.edu/abs/2019MNRAS.486.3927H} {486, 3927}

\bibitem[\protect\citeauthoryear{{Hamana} et~al.,}{{Hamana}
  et~al.}{2020}]{Hamana:2020}
{Hamana} T.,  et~al., 2020, \mn@doi [\pasj] {10.1093/pasj/psz138}, \href
  {https://ui.adsabs.harvard.edu/abs/2020PASJ...72...16H} {72, 16}

\bibitem[\protect\citeauthoryear{Harnois-Deraps, Giblin  \&
  Joachimi}{Harnois-Deraps et~al.}{2019}]{slics}
Harnois-Deraps J.,  Giblin B.,   Joachimi B.,  2019, \mn@doi [Astron.
  Astrophys.] {10.1051/0004-6361/201935912}, 631, A160

\bibitem[\protect\citeauthoryear{Harris et~al.,}{Harris
  et~al.}{2020}]{harris:2020}
Harris C.~R.,  et~al., 2020, \mn@doi [Nature] {10.1038/s41586-020-2649-2}, 585,
  357

\bibitem[\protect\citeauthoryear{Hartlap, Simon  \& Schneider}{Hartlap
  et~al.}{2006}]{Hartlap06}
Hartlap J.,  Simon P.,   Schneider P.,  2006, \mn@doi [Astronomy \&
  Astrophysics] {10.1051/0004-6361:20066170}, 464, 399–404

\bibitem[\protect\citeauthoryear{{Heitmann}, {Lawrence}, {Kwan}, {Habib}  \&
  {Higdon}}{{Heitmann} et~al.}{2014}]{Heitmann:2014}
{Heitmann} K.,  {Lawrence} E.,  {Kwan} J.,  {Habib} S.,   {Higdon} D.,  2014,
  \mn@doi [\apj] {10.1088/0004-637X/780/1/111}, \href
  {https://ui.adsabs.harvard.edu/abs/2014ApJ...780..111H} {780, 111}

\bibitem[\protect\citeauthoryear{{Hellwing}, {Li}, {Frenk}  \&
  {Cole}}{{Hellwing} et~al.}{2013}]{Hellwing:2013}
{Hellwing} W.~A.,  {Li} B.,  {Frenk} C.~S.,   {Cole} S.,  2013, \mn@doi
  [\mnras] {10.1093/mnras/stt1430}, \href
  {https://ui.adsabs.harvard.edu/abs/2013MNRAS.435.2806H} {435, 2806}

\bibitem[\protect\citeauthoryear{{Hellwing}, {Koyama}, {Bose}  \&
  {Zhao}}{{Hellwing} et~al.}{2017}]{Hellwing:2017}
{Hellwing} W.~A.,  {Koyama} K.,  {Bose} B.,   {Zhao} G.-B.,  2017, \mn@doi
  [\prd] {10.1103/PhysRevD.96.023515}, \href
  {https://ui.adsabs.harvard.edu/abs/2017PhRvD..96b3515H} {96, 023515}

\bibitem[\protect\citeauthoryear{{Heymans} et~al.,}{{Heymans}
  et~al.}{2021}]{Heymans:2021}
{Heymans} C.,  et~al., 2021, \mn@doi [\aap] {10.1051/0004-6361/202039063},
  \href {https://ui.adsabs.harvard.edu/abs/2021A&A...646A.140H} {646, A140}

\bibitem[\protect\citeauthoryear{{Hinton}}{{Hinton}}{2016}]{Hinton:2016}
{Hinton} S.~R.,  2016, \mn@doi [The Journal of Open Source Software]
  {10.21105/joss.00045}, \href
  {http://adsabs.harvard.edu/abs/2016JOSS....1...45H} {1, 00045}

\bibitem[\protect\citeauthoryear{{Horndeski}}{{Horndeski}}{1974}]{Horndeski:1974}
{Horndeski} G.~W.,  1974, \mn@doi [International Journal of Theoretical
  Physics] {10.1007/BF01807638}, \href
  {https://ui.adsabs.harvard.edu/abs/1974IJTP...10..363H} {10, 363}

\bibitem[\protect\citeauthoryear{{Hu} \& {Sawicki}}{{Hu} \&
  {Sawicki}}{2007}]{Hu:2007}
{Hu} W.,  {Sawicki} I.,  2007, \mn@doi [\prd] {10.1103/PhysRevD.76.064004},
  \href {https://ui.adsabs.harvard.edu/abs/2007PhRvD..76f4004H} {76, 064004}

\bibitem[\protect\citeauthoryear{{Hu}, {Raveri}, {Frusciante}  \&
  {Silvestri}}{{Hu} et~al.}{2014}]{Hu:2014}
{Hu} B.,  {Raveri} M.,  {Frusciante} N.,   {Silvestri} A.,  2014, \mn@doi
  [\prd] {10.1103/PhysRevD.89.103530}, \href
  {https://ui.adsabs.harvard.edu/abs/2014PhRvD..89j3530H} {89, 103530}

\bibitem[\protect\citeauthoryear{Hunter}{Hunter}{2007}]{Hunter:2007}
Hunter J.~D.,  2007, \mn@doi [Computing in Science \& Engineering]
  {10.1109/MCSE.2007.55}, 9, 90

\bibitem[\protect\citeauthoryear{{Ishak}}{{Ishak}}{2019}]{Ishak:2019}
{Ishak} M.,  2019, \mn@doi [Living Reviews in Relativity]
  {10.1007/s41114-018-0017-4}, \href
  {https://ui.adsabs.harvard.edu/abs/2019LRR....22....1I} {22, 1}

\bibitem[\protect\citeauthoryear{Ivanov, Kaurov  \& Sibiryakov}{Ivanov
  et~al.}{2019}]{Ivanov_2019}
Ivanov M.~M.,  Kaurov A.~A.,   Sibiryakov S.,  2019, \mn@doi [Journal of
  Cosmology and Astroparticle Physics] {10.1088/1475-7516/2019/03/009}, 2019,
  009–009

\bibitem[\protect\citeauthoryear{Jamieson \& Loverde}{Jamieson \&
  Loverde}{2020}]{Jamieson_2020}
Jamieson D.,  Loverde M.,  2020, \mn@doi [Physical Review D]
  {10.1103/physrevd.102.123546}, 102

\bibitem[\protect\citeauthoryear{{Kaufman}}{{Kaufman}}{1967}]{Kaufman67}
{Kaufman} G.~M.,  1967, Report No. 6710, Center for Operations Research and
  Econometrics, Catholic University of Louvain, Heverlee, Belgium

\bibitem[\protect\citeauthoryear{{Kopp}, {Appleby}, {Achitouv}  \&
  {Weller}}{{Kopp} et~al.}{2013}]{Kopp:2013}
{Kopp} M.,  {Appleby} S.~A.,  {Achitouv} I.,   {Weller} J.,  2013, \mn@doi
  [\prd] {10.1103/PhysRevD.88.084015}, \href
  {https://ui.adsabs.harvard.edu/abs/2013PhRvD..88h4015K} {88, 084015}

\bibitem[\protect\citeauthoryear{Koyama}{Koyama}{2018}]{Koyama:2018}
Koyama K.,  2018, \mn@doi [Int. J. Mod. Phys. D] {10.1142/S0218271818480012},
  27, 1848001

\bibitem[\protect\citeauthoryear{{Koyama}, {Taruya}  \& {Hiramatsu}}{{Koyama}
  et~al.}{2009}]{Koyama:2009}
{Koyama} K.,  {Taruya} A.,   {Hiramatsu} T.,  2009, \mn@doi [\prd]
  {10.1103/PhysRevD.79.123512}, \href
  {https://ui.adsabs.harvard.edu/abs/2009PhRvD..79l3512K} {79, 123512}

\bibitem[\protect\citeauthoryear{{Kratochvil}, {Lim}, {Wang}, {Haiman}, {May}
  \& {Huffenberger}}{{Kratochvil} et~al.}{2012}]{Kratochvil:2012}
{Kratochvil} J.~M.,  {Lim} E.~A.,  {Wang} S.,  {Haiman} Z.,  {May} M.,
  {Huffenberger} K.,  2012, \mn@doi [\prd] {10.1103/PhysRevD.85.103513}, \href
  {https://ui.adsabs.harvard.edu/abs/2012PhRvD..85j3513K} {85, 103513}

\bibitem[\protect\citeauthoryear{{Lam} \& {Li}}{{Lam} \& {Li}}{2012}]{Lam_2012}
{Lam} T.~Y.,  {Li} B.,  2012, \mn@doi [\mnras]
  {10.1111/j.1365-2966.2012.21746.x}, \href
  {https://ui.adsabs.harvard.edu/abs/2012MNRAS.426.3260L} {426, 3260}

\bibitem[\protect\citeauthoryear{{Lee} et~al.,}{{Lee} et~al.}{2021}]{Lee:2021}
{Lee} S.,  et~al., 2021, arXiv e-prints, \href
  {https://ui.adsabs.harvard.edu/abs/2021arXiv210414515L} {p. arXiv:2104.14515}

\bibitem[\protect\citeauthoryear{Lewis, Challinor  \& Lasenby}{Lewis
  et~al.}{2000}]{CAMB}
Lewis A.,  Challinor A.,   Lasenby A.,  2000, \mn@doi [\apj] {10.1086/309179},
  538, 473

\bibitem[\protect\citeauthoryear{Li \& Efstathiou}{Li \&
  Efstathiou}{2012}]{LiEfstathiou_2012}
Li B.,  Efstathiou G.,  2012, \mn@doi [Monthly Notices of the Royal
  Astronomical Society] {10.1111/j.1365-2966.2011.20404.x}, 421, 1431–1442

\bibitem[\protect\citeauthoryear{Li, Zhao, Teyssier  \& Koyama}{Li
  et~al.}{2012a}]{Li:2011vk}
Li B.,  Zhao G.-B.,  Teyssier R.,   Koyama K.,  2012a, \mn@doi [JCAP]
  {10.1088/1475-7516/2012/01/051}, 01, 051

\bibitem[\protect\citeauthoryear{Li, Zhao  \& Koyama}{Li
  et~al.}{2012b}]{Li_2012halosvoidsfR}
Li B.,  Zhao G.-B.,   Koyama K.,  2012b, \mn@doi [Monthly Notices of the Royal
  Astronomical Society] {10.1111/j.1365-2966.2012.20573.x}, 421, 3481–3487

\bibitem[\protect\citeauthoryear{Li, Zhao  \& Koyama}{Li
  et~al.}{2013}]{Li:2013nua}
Li B.,  Zhao G.-B.,   Koyama K.,  2013, \mn@doi [JCAP]
  {10.1088/1475-7516/2013/05/023}, 05, 023

\bibitem[\protect\citeauthoryear{{Linder}}{{Linder}}{2003}]{Linder:2003}
{Linder} E.~V.,  2003, \mn@doi [\prl] {10.1103/PhysRevLett.90.091301}, \href
  {https://ui.adsabs.harvard.edu/abs/2003PhRvL..90i1301L} {90, 091301}

\bibitem[\protect\citeauthoryear{{Linder}}{{Linder}}{2005}]{Linder:2005}
{Linder} E.~V.,  2005, \mn@doi [\prd] {10.1103/PhysRevD.72.043529}, \href
  {https://ui.adsabs.harvard.edu/abs/2005PhRvD..72d3529L} {72, 043529}

\bibitem[\protect\citeauthoryear{{Liu}, {Valogiannis}, {Battaglia}  \&
  {Bean}}{{Liu} et~al.}{2021}]{Liu:2021}
{Liu} R.,  {Valogiannis} G.,  {Battaglia} N.,   {Bean} R.,  2021, arXiv
  e-prints, \href {https://ui.adsabs.harvard.edu/abs/2021arXiv210108728L} {p.
  arXiv:2101.08728}

\bibitem[\protect\citeauthoryear{{Lombriser}}{{Lombriser}}{2018}]{Lombriser:2018}
{Lombriser} L.,  2018, \mn@doi [International Journal of Modern Physics D]
  {10.1142/S0218271818480024}, \href
  {https://ui.adsabs.harvard.edu/abs/2018IJMPD..2748002L} {27, 1848002}

\bibitem[\protect\citeauthoryear{{Lombriser}, {Hu}, {Fang}  \&
  {Seljak}}{{Lombriser} et~al.}{2009}]{Lombriser:2009}
{Lombriser} L.,  {Hu} W.,  {Fang} W.,   {Seljak} U.,  2009, \mn@doi [\prd]
  {10.1103/PhysRevD.80.063536}, \href
  {https://ui.adsabs.harvard.edu/abs/2009PhRvD..80f3536L} {80, 063536}

\bibitem[\protect\citeauthoryear{{Lombriser}, {Li}, {Koyama}  \&
  {Zhao}}{{Lombriser} et~al.}{2013}]{Lombriser:2013}
{Lombriser} L.,  {Li} B.,  {Koyama} K.,   {Zhao} G.-B.,  2013, \mn@doi [\prd]
  {10.1103/PhysRevD.87.123511}, \href
  {https://ui.adsabs.harvard.edu/abs/2013PhRvD..87l3511L} {87, 123511}

\bibitem[\protect\citeauthoryear{Mandal \& Nadkarni-Ghosh}{Mandal \&
  Nadkarni-Ghosh}{2020}]{Mandal_2020}
Mandal A.,  Nadkarni-Ghosh S.,  2020, \mn@doi [Monthly Notices of the Royal
  Astronomical Society] {10.1093/mnras/staa2073}, 498, 355–372

\bibitem[\protect\citeauthoryear{{McClintock} et~al.,}{{McClintock}
  et~al.}{2019}]{McClintock:2019}
{McClintock} T.,  et~al., 2019, \mn@doi [\apj] {10.3847/1538-4357/aaf568},
  \href {https://ui.adsabs.harvard.edu/abs/2019ApJ...872...53M} {872, 53}

\bibitem[\protect\citeauthoryear{{Mead}}{{Mead}}{2017}]{Mead:2017}
{Mead} A.~J.,  2017, \mn@doi [\mnras] {10.1093/mnras/stw2312}, \href
  {https://ui.adsabs.harvard.edu/abs/2017MNRAS.464.1282M} {464, 1282}

\bibitem[\protect\citeauthoryear{{Mead}, {Heymans}, {Lombriser}, {Peacock},
  {Steele}  \& {Winther}}{{Mead} et~al.}{2016}]{Mead:2016}
{Mead} A.~J.,  {Heymans} C.,  {Lombriser} L.,  {Peacock} J.~A.,  {Steele}
  O.~I.,   {Winther} H.~A.,  2016, \mn@doi [\mnras] {10.1093/mnras/stw681},
  \href {https://ui.adsabs.harvard.edu/abs/2016MNRAS.459.1468M} {459, 1468}

\bibitem[\protect\citeauthoryear{{Mead}, {Brieden}, {Tr{\"o}ster}  \&
  {Heymans}}{{Mead} et~al.}{2021}]{Mead:2021}
{Mead} A.~J.,  {Brieden} S.,  {Tr{\"o}ster} T.,   {Heymans} C.,  2021, \mn@doi
  [\mnras] {10.1093/mnras/stab082}, \href
  {https://ui.adsabs.harvard.edu/abs/2021MNRAS.502.1401M} {502, 1401}

\bibitem[\protect\citeauthoryear{{Muir} et~al.,}{{Muir}
  et~al.}{2021}]{Muir:2021}
{Muir} J.,  et~al., 2021, \mn@doi [\prd] {10.1103/PhysRevD.103.023528}, \href
  {https://ui.adsabs.harvard.edu/abs/2021PhRvD.103b3528M} {103, 023528}

\bibitem[\protect\citeauthoryear{{Munshi}}{{Munshi}}{2017}]{Munshi:2017}
{Munshi} D.,  2017, \mn@doi [\jcap] {10.1088/1475-7516/2017/01/049}, \href
  {https://ui.adsabs.harvard.edu/abs/2017JCAP...01..049M} {2017, 049}

\bibitem[\protect\citeauthoryear{Munshi \& McEwen}{Munshi \&
  McEwen}{2020}]{Munshi:2020}
Munshi D.,  McEwen J.~D.,  2020, \mn@doi [Mon. Not. Roy. Astron. Soc.]
  {10.1093/mnras/staa2706}, 498, 5299

\bibitem[\protect\citeauthoryear{{Nicolis} \& {Rattazzi}}{{Nicolis} \&
  {Rattazzi}}{2004}]{Nicolis:2004}
{Nicolis} A.,  {Rattazzi} R.,  2004, \mn@doi [Journal of High Energy Physics]
  {10.1088/1126-6708/2004/06/059}, \href
  {https://ui.adsabs.harvard.edu/abs/2004JHEP...06..059N} {2004, 059}

\bibitem[\protect\citeauthoryear{{Nishimichi}, {Bernardeau}  \&
  {Taruya}}{{Nishimichi} et~al.}{2017}]{Nishimichi17}
{Nishimichi} T.,  {Bernardeau} F.,   {Taruya} A.,  2017, \mn@doi [\prd]
  {10.1103/PhysRevD.96.123515}, \href
  {https://ui.adsabs.harvard.edu/abs/2017PhRvD..96l3515N} {96, 123515}

\bibitem[\protect\citeauthoryear{{Park}, {Zurek}  \& {Watson}}{{Park}
  et~al.}{2010}]{Park:2010}
{Park} M.,  {Zurek} K.~M.,   {Watson} S.,  2010, \mn@doi [\prd]
  {10.1103/PhysRevD.81.124008}, \href
  {https://ui.adsabs.harvard.edu/abs/2010PhRvD..81l4008P} {81, 124008}

\bibitem[\protect\citeauthoryear{Patton, Blazek, Honscheid, Huff, Melchior,
  Ross  \& Suchyta}{Patton et~al.}{2017}]{Patton_2017}
Patton K.,  Blazek J.,  Honscheid K.,  Huff E.,  Melchior P.,  Ross A.~J.,
  Suchyta E.,  2017, \mn@doi [Monthly Notices of the Royal Astronomical
  Society] {10.1093/mnras/stx1626}, 472, 439

\bibitem[\protect\citeauthoryear{{Peacock} \& {Smith}}{{Peacock} \&
  {Smith}}{2014}]{halofit}
{Peacock} J.~A.,  {Smith} R.~E.,  2014, {HALOFIT: Nonlinear distribution of
  cosmological mass and galaxies} (\mn@eprint {ascl} {1402.032})

\bibitem[\protect\citeauthoryear{{Peel}, {Pettorino}, {Giocoli}, {Starck}  \&
  {Baldi}}{{Peel} et~al.}{2018}]{Peel:2018}
{Peel} A.,  {Pettorino} V.,  {Giocoli} C.,  {Starck} J.-L.,   {Baldi} M.,
  2018, \mn@doi [\aap] {10.1051/0004-6361/201833481}, \href
  {https://ui.adsabs.harvard.edu/abs/2018A&A...619A..38P} {619, A38}

\bibitem[\protect\citeauthoryear{{Perico}, {Voivodic}, {Lima}  \&
  {Mota}}{{Perico} et~al.}{2019}]{Perico:2019}
{Perico} E. L.~D.,  {Voivodic} R.,  {Lima} M.,   {Mota} D.~F.,  2019, arXiv
  e-prints, \href {https://ui.adsabs.harvard.edu/abs/2019arXiv190512450P} {p.
  arXiv:1905.12450}

\bibitem[\protect\citeauthoryear{{Perivolaropoulos} \&
  {Skara}}{{Perivolaropoulos} \& {Skara}}{2021}]{Perivolaropoulos:2021}
{Perivolaropoulos} L.,  {Skara} F.,  2021, arXiv e-prints, \href
  {https://ui.adsabs.harvard.edu/abs/2021arXiv210505208P} {p. arXiv:2105.05208}

\bibitem[\protect\citeauthoryear{{Perlmutter} et~al.,}{{Perlmutter}
  et~al.}{1999}]{Perlmutter:1999}
{Perlmutter} S.,  et~al., 1999, \mn@doi [\apj] {10.1086/307221}, \href
  {https://ui.adsabs.harvard.edu/abs/1999ApJ...517..565P} {517, 565}

\bibitem[\protect\citeauthoryear{{Planck Collaboration} et~al.,}{{Planck
  Collaboration} et~al.}{2020}]{Planck:2018}
{Planck Collaboration} et~al., 2020, \mn@doi [\aap]
  {10.1051/0004-6361/201833910}, \href
  {https://ui.adsabs.harvard.edu/abs/2020A&A...641A...6P} {641, A6}

\bibitem[\protect\citeauthoryear{{Pogosian}, {Raveri}, {Koyama}, {Martinelli},
  {Silvestri}  \& {Zhao}}{{Pogosian} et~al.}{2021}]{Pogosian:2021}
{Pogosian} L.,  {Raveri} M.,  {Koyama} K.,  {Martinelli} M.,  {Silvestri} A.,
  {Zhao} G.-B.,  2021, arXiv e-prints, \href
  {https://ui.adsabs.harvard.edu/abs/2021arXiv210712992P} {p. arXiv:2107.12992}

\bibitem[\protect\citeauthoryear{{Ramachandra}, {Valogiannis}, {Ishak},
  {Heitmann}  \& {LSST Dark Energy Science Collaboration}}{{Ramachandra}
  et~al.}{2021}]{Ramachandra:2021}
{Ramachandra} N.,  {Valogiannis} G.,  {Ishak} M.,  {Heitmann} K.,   {LSST Dark
  Energy Science Collaboration} 2021, \mn@doi [\prd]
  {10.1103/PhysRevD.103.123525}, \href
  {https://ui.adsabs.harvard.edu/abs/2021PhRvD.103l3525R} {103, 123525}

\bibitem[\protect\citeauthoryear{{Raveri} et~al.,}{{Raveri}
  et~al.}{2021}]{Raveri:2021}
{Raveri} M.,  et~al., 2021, arXiv e-prints, \href
  {https://ui.adsabs.harvard.edu/abs/2021arXiv210712990R} {p. arXiv:2107.12990}

\bibitem[\protect\citeauthoryear{Repp \& Szapudi}{Repp \&
  Szapudi}{2020}]{Repp_2020}
Repp A.,  Szapudi I.,  2020, \mn@doi [Monthly Notices of the Royal Astronomical
  Society: Letters] {10.1093/mnrasl/slaa139}, 498, L125–L129

\bibitem[\protect\citeauthoryear{{Riess} et~al.,}{{Riess}
  et~al.}{1998}]{Riess:1998}
{Riess} A.~G.,  et~al., 1998, \mn@doi [\aj] {10.1086/300499}, \href
  {https://ui.adsabs.harvard.edu/abs/1998AJ....116.1009R} {116, 1009}

\bibitem[\protect\citeauthoryear{{Sahl{\'e}n}}{{Sahl{\'e}n}}{2019}]{Sahlen:2019}
{Sahl{\'e}n} M.,  2019, \mn@doi [\prd] {10.1103/PhysRevD.99.063525}, \href
  {https://ui.adsabs.harvard.edu/abs/2019PhRvD..99f3525S} {99, 063525}

\bibitem[\protect\citeauthoryear{{Schaap} \& {van de Weygaert}}{{Schaap} \&
  {van de Weygaert}}{2000}]{Schaap:2000}
{Schaap} W.~E.,  {van de Weygaert} R.,  2000, \aap, \href
  {https://ui.adsabs.harvard.edu/abs/2000A&A...363L..29S} {363, L29}

\bibitem[\protect\citeauthoryear{{Schmidt}}{{Schmidt}}{2009}]{Schmidt:2009}
{Schmidt} F.,  2009, \mn@doi [\prd] {10.1103/PhysRevD.80.123003}, \href
  {https://ui.adsabs.harvard.edu/abs/2009PhRvD..80l3003S} {80, 123003}

\bibitem[\protect\citeauthoryear{{Schmidt}, {Lima}, {Oyaizu}  \&
  {Hu}}{{Schmidt} et~al.}{2009}]{Schmidt:2009b}
{Schmidt} F.,  {Lima} M.,  {Oyaizu} H.,   {Hu} W.,  2009, \mn@doi [\prd]
  {10.1103/PhysRevD.79.083518}, \href
  {https://ui.adsabs.harvard.edu/abs/2009PhRvD..79h3518S} {79, 083518}

\bibitem[\protect\citeauthoryear{{Schmidt}, {Hu}  \& {Lima}}{{Schmidt}
  et~al.}{2010}]{Schmidt:2010}
{Schmidt} F.,  {Hu} W.,   {Lima} M.,  2010, \mn@doi [\prd]
  {10.1103/PhysRevD.81.063005}, \href
  {https://ui.adsabs.harvard.edu/abs/2010PhRvD..81f3005S} {81, 063005}

\bibitem[\protect\citeauthoryear{{Shin}, {Kim}, {Pichon}, {Jeong}  \&
  {Park}}{{Shin} et~al.}{2017}]{Shin:2017}
{Shin} J.,  {Kim} J.,  {Pichon} C.,  {Jeong} D.,   {Park} C.,  2017, \mn@doi
  [\apj] {10.3847/1538-4357/aa74b9}, \href
  {https://ui.adsabs.harvard.edu/abs/2017ApJ...843...73S} {843, 73}

\bibitem[\protect\citeauthoryear{{Shirasaki}, {Nishimichi}, {Li}  \&
  {Higuchi}}{{Shirasaki} et~al.}{2017}]{Shirasaki:2017}
{Shirasaki} M.,  {Nishimichi} T.,  {Li} B.,   {Higuchi} Y.,  2017, \mn@doi
  [\mnras] {10.1093/mnras/stw3254}, \href
  {https://ui.adsabs.harvard.edu/abs/2017MNRAS.466.2402S} {466, 2402}

\bibitem[\protect\citeauthoryear{{Simpson} et~al.,}{{Simpson}
  et~al.}{2013}]{Simpson:2013}
{Simpson} F.,  et~al., 2013, \mn@doi [\mnras] {10.1093/mnras/sts493}, \href
  {https://ui.adsabs.harvard.edu/abs/2013MNRAS.429.2249S} {429, 2249}

\bibitem[\protect\citeauthoryear{{Song} et~al.,}{{Song}
  et~al.}{2015}]{Song:2015}
{Song} Y.-S.,  et~al., 2015, \mn@doi [\prd] {10.1103/PhysRevD.92.043522}, \href
  {https://ui.adsabs.harvard.edu/abs/2015PhRvD..92d3522S} {92, 043522}

\bibitem[\protect\citeauthoryear{{Springel}}{{Springel}}{2010}]{springel2010}
{Springel} V.,  2010, \mn@doi [\mnras] {10.1111/j.1365-2966.2009.15715.x},
  \href {https://ui.adsabs.harvard.edu/#abs/2010MNRAS.401..791S} {401, 791}

\bibitem[\protect\citeauthoryear{{Takahashi}, {Sato}, {Nishimichi}, {Taruya}
  \& {Oguri}}{{Takahashi} et~al.}{2012}]{Takahashi:2012}
{Takahashi} R.,  {Sato} M.,  {Nishimichi} T.,  {Taruya} A.,   {Oguri} M.,
  2012, \mn@doi [\apj] {10.1088/0004-637X/761/2/152}, \href
  {https://ui.adsabs.harvard.edu/abs/2012ApJ...761..152T} {761, 152}

\bibitem[\protect\citeauthoryear{Teyssier}{Teyssier}{2002}]{Teyssier:2001cp}
Teyssier R.,  2002, \mn@doi [Astron. Astrophys.] {10.1051/0004-6361:20011817},
  385, 337

\bibitem[\protect\citeauthoryear{Thiele, Hill  \& Smith}{Thiele
  et~al.}{2020}]{Thiele_2020}
Thiele L.,  Hill J.~C.,   Smith K.~M.,  2020, \mn@doi [Physical Review D]
  {10.1103/physrevd.102.123545}, 102

\bibitem[\protect\citeauthoryear{Touchette}{Touchette}{2012}]{Touchette_2012}
Touchette H.,  2012, A basic introduction to large deviations: Theory,
  applications, simulations (\mn@eprint {arXiv} {1106.4146})

\bibitem[\protect\citeauthoryear{{Tr{\"o}ster} et~al.,}{{Tr{\"o}ster}
  et~al.}{2021}]{Troester:2021}
{Tr{\"o}ster} T.,  et~al., 2021, \mn@doi [\aap] {10.1051/0004-6361/202039805},
  \href {https://ui.adsabs.harvard.edu/abs/2021A&A...649A..88T} {649, A88}

\bibitem[\protect\citeauthoryear{{Uhlemann}, {Codis}, {Pichon}, {Bernardeau}
  \& {Reimberg}}{{Uhlemann} et~al.}{2016}]{Uhlemann16}
{Uhlemann} C.,  {Codis} S.,  {Pichon} C.,  {Bernardeau} F.,   {Reimberg} P.,
  2016, \mn@doi [\mnras] {10.1093/mnras/stw1074}, \href
  {http://adsabs.harvard.edu/abs/2016MNRAS.460.1529U} {460, 1529}

\bibitem[\protect\citeauthoryear{{Uhlemann} et~al.,}{{Uhlemann}
  et~al.}{2018a}]{Uhlemann:2018b}
{Uhlemann} C.,  et~al., 2018a, \mn@doi [\mnras] {10.1093/mnras/stx2616}, \href
  {https://ui.adsabs.harvard.edu/abs/2018MNRAS.473.5098U} {473, 5098}

\bibitem[\protect\citeauthoryear{{Uhlemann}, {Pajer}, {Pichon}, {Nishimichi},
  {Codis}  \& {Bernardeau}}{{Uhlemann} et~al.}{2018b}]{Uhlemann18pNG}
{Uhlemann} C.,  {Pajer} E.,  {Pichon} C.,  {Nishimichi} T.,  {Codis} S.,
  {Bernardeau} F.,  2018b, \mn@doi [\mnras] {10.1093/mnras/stx2623}, \href
  {https://ui.adsabs.harvard.edu/abs/2018MNRAS.474.2853U} {474, 2853}

\bibitem[\protect\citeauthoryear{{Uhlemann}, {Friedrich},
  {Villaescusa-Navarro}, {Banerjee}  \& {Codis}}{{Uhlemann}
  et~al.}{2020}]{Uhlemann:2020}
{Uhlemann} C.,  {Friedrich} O.,  {Villaescusa-Navarro} F.,  {Banerjee} A.,
  {Codis} S.,  2020, \mn@doi [\mnras] {10.1093/mnras/staa1155}, \href
  {https://ui.adsabs.harvard.edu/abs/2020MNRAS.495.4006U} {495, 4006}

\bibitem[\protect\citeauthoryear{{Valageas}}{{Valageas}}{2002}]{Valageas02}
{Valageas} P.,  2002, \mn@doi [\aap] {10.1051/0004-6361:20011663}, \href
  {http://adsabs.harvard.edu/cgi-bin/nph-bib_query?bibcode=2002A%26A...382..412V&db_key=AST}
  {382, 412}

\bibitem[\protect\citeauthoryear{{Vazsonyi}, {Taylor}, {Valogiannis},
  {Ramachandra}, {Fert{\'e}}  \& {Rhodes}}{{Vazsonyi}
  et~al.}{2021}]{Vazsonyi:2021}
{Vazsonyi} L.,  {Taylor} P.~L.,  {Valogiannis} G.,  {Ramachandra} N.~S.,
  {Fert{\'e}} A.,   {Rhodes} J.,  2021, arXiv e-prints, \href
  {https://ui.adsabs.harvard.edu/abs/2021arXiv210710277V} {p. arXiv:2107.10277}

\bibitem[\protect\citeauthoryear{{Verza}, {Pisani}, {Carbone}, {Hamaus}  \&
  {Guzzo}}{{Verza} et~al.}{2019}]{Verza:2019}
{Verza} G.,  {Pisani} A.,  {Carbone} C.,  {Hamaus} N.,   {Guzzo} L.,  2019,
  \mn@doi [\jcap] {10.1088/1475-7516/2019/12/040}, \href
  {https://ui.adsabs.harvard.edu/abs/2019JCAP...12..040V} {2019, 040}

\bibitem[\protect\citeauthoryear{{Villaescusa-Navarro}
  et~al.,}{{Villaescusa-Navarro} et~al.}{2020}]{Villaescusa-Navarro:2020}
{Villaescusa-Navarro} F.,  et~al., 2020, \mn@doi [\apjs]
  {10.3847/1538-4365/ab9d82}, \href
  {https://ui.adsabs.harvard.edu/abs/2020ApJS..250....2V} {250, 2}

\bibitem[\protect\citeauthoryear{Virtanen et~al.,}{Virtanen
  et~al.}{2020}]{Virtanen:2020}
Virtanen P.,  et~al., 2020, \mn@doi [Nature Methods]
  {10.1038/s41592-019-0686-2}, \href {https://rdcu.be/b08Wh} {17, 261}

\bibitem[\protect\citeauthoryear{{Weinberger}, {Springel}  \&
  {Pakmor}}{{Weinberger} et~al.}{2020}]{weinberger2020}
{Weinberger} R.,  {Springel} V.,   {Pakmor} R.,  2020, \mn@doi [\apjs]
  {10.3847/1538-4365/ab908c}, \href
  {https://ui.adsabs.harvard.edu/abs/2020ApJS..248...32W} {248, 32}

\bibitem[\protect\citeauthoryear{Wen, Kemball  \& Saslaw}{Wen
  et~al.}{2020}]{Wen_2020}
Wen D.,  Kemball A.~J.,   Saslaw W.~C.,  2020, \mn@doi [The Astrophysical
  Journal] {10.3847/1538-4357/ab6d6f}, 890, 160

\bibitem[\protect\citeauthoryear{{Will}}{{Will}}{2014}]{Will:2014}
{Will} C.~M.,  2014, \mn@doi [Living Reviews in Relativity]
  {10.12942/lrr-2014-4}, \href
  {https://ui.adsabs.harvard.edu/abs/2014LRR....17....4W} {17, 4}

\bibitem[\protect\citeauthoryear{{Winther}, {Casas}, {Baldi}, {Koyama}, {Li},
  {Lombriser}  \& {Zhao}}{{Winther} et~al.}{2019}]{Winther:2019}
{Winther} H.~A.,  {Casas} S.,  {Baldi} M.,  {Koyama} K.,  {Li} B.,  {Lombriser}
  L.,   {Zhao} G.-B.,  2019, \mn@doi [\prd] {10.1103/PhysRevD.100.123540},
  \href {https://ui.adsabs.harvard.edu/abs/2019PhRvD.100l3540W} {100, 123540}

\bibitem[\protect\citeauthoryear{{Yamauchi}, {Yokoyama}  \&
  {Tashiro}}{{Yamauchi} et~al.}{2017}]{Yamauchi:2017}
{Yamauchi} D.,  {Yokoyama} S.,   {Tashiro} H.,  2017, \mn@doi [\prd]
  {10.1103/PhysRevD.96.123516}, \href
  {https://ui.adsabs.harvard.edu/abs/2017PhRvD..96l3516Y} {96, 123516}

\bibitem[\protect\citeauthoryear{{Zhao}}{{Zhao}}{2014}]{Zhao:2014}
{Zhao} G.-B.,  2014, \mn@doi [\apjs] {10.1088/0067-0049/211/2/23}, \href
  {https://ui.adsabs.harvard.edu/abs/2014ApJS..211...23Z} {211, 23}

\bibitem[\protect\citeauthoryear{{Zumalac{\'a}rregui}, {Bellini}, {Sawicki},
  {Lesgourgues}  \& {Ferreira}}{{Zumalac{\'a}rregui}
  et~al.}{2017}]{Zumalacarregui:2017}
{Zumalac{\'a}rregui} M.,  {Bellini} E.,  {Sawicki} I.,  {Lesgourgues} J.,
  {Ferreira} P.~G.,  2017, \mn@doi [\jcap] {10.1088/1475-7516/2017/08/019},
  \href {https://ui.adsabs.harvard.edu/abs/2017JCAP...08..019Z} {2017, 019}

\makeatother
\end{thebibliography}

%\begin{thebibliography}{99}
%\end{thebibliography}

%%%%%%%%%%%%%%%%%%%%%%%%%%%%%%%%%%%%%%%%%%%%%%%%%%

%%%%%%%%%%%%%%%%% APPENDICES %%%%%%%%%%%%%%%%%%%%%

\appendix

\section{The impact of mass-assignment schemes on the PDF}
\label{sec:CiCvDTFE}

\begin{figure}
    \centering
    \includegraphics[width=\columnwidth]{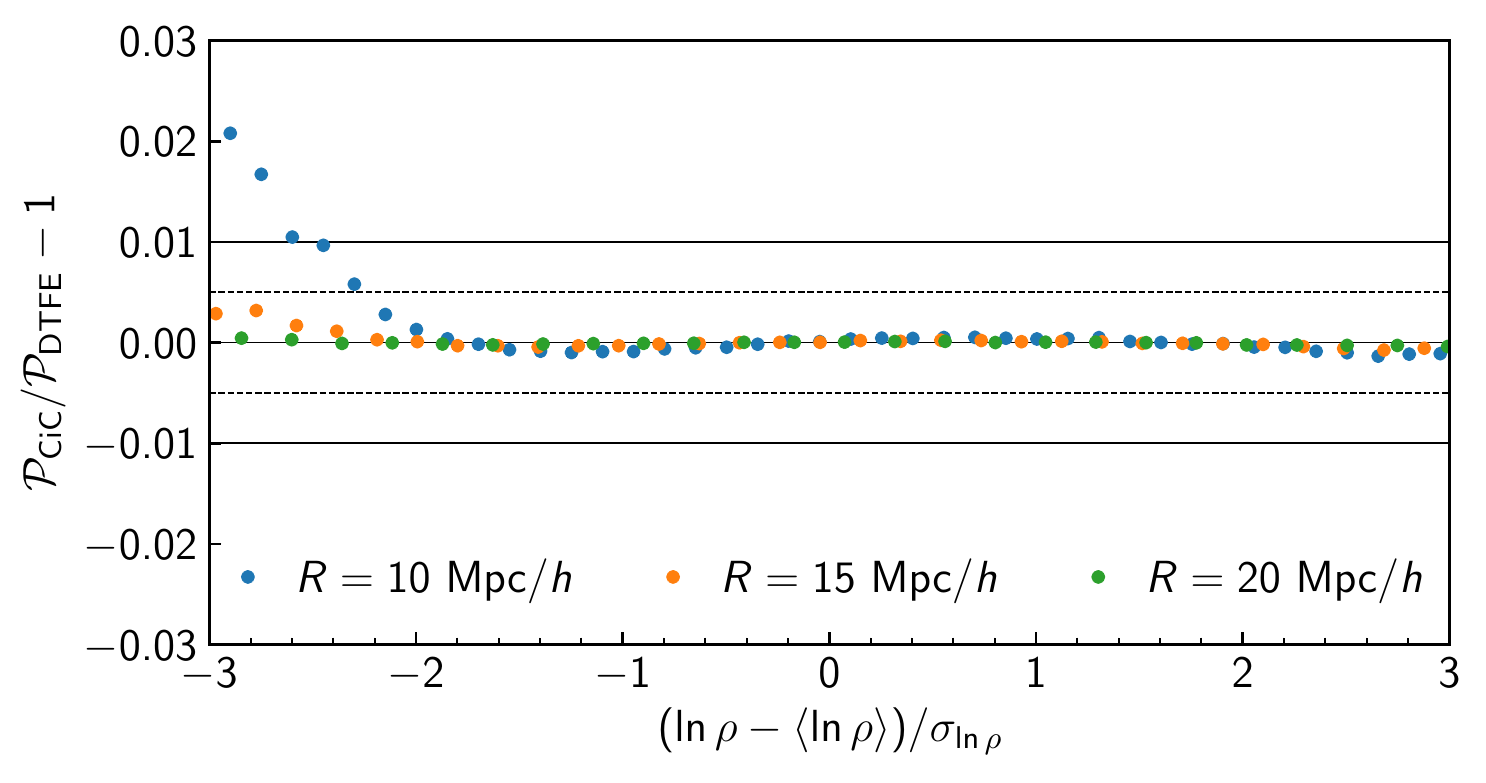}
    \caption{Relative deviation between the matter PDF constructed from the Cloud-in-Cell (CiC) mass-assignment scheme and that based on the Delaunay tassellation (DTFE). Data points represent measurements from a single $z=0$ snapshot of a \LCDM{} simulation. The agreement between the two mass-assignment schemes is excellent over the entire range of densities relevant for this work.}
    \label{fig:cic_v_dtfe}
\end{figure}

The various methods designed to interpolate the simulated density field on a grid may lead to differences between the measured PDFs large enough to potentially bias the predictive accuracy of a particular theoretical framework. In Figure~\ref{fig:cic_v_dtfe} we compare two such popular methods--the Cloud-in-Cell algorithm and the Delaunay tassellation field estimator--using as summary statistic the PDF extracted from a single snapshot after applying top-hat filters with three different smoothing radii. Reassuringly, the distributions agree to better than 1\% for all densities but the rarest under-densities, thus validating the performance of LDT discussed in Section~\ref{sec:results} and previously presented in \cite{Uhlemann:2020}.

\section{Means and variances of the simulated non-linear density field}
\label{app:stats}

\begin{table*}
\begin{tabular}{@{}cccc|ccc|ccc@{}}
\toprule
               & \multicolumn{3}{c|}{$\Lambda$CDM}                        & \multicolumn{3}{c|}{F5}                                  & \multicolumn{3}{c}{DGPm}                                 \\
               & $\sigma_\rho^2$ & $\sigma_\mu^2$ & $\langle \mu \rangle$ & $\sigma_\rho^2$ & $\sigma_\mu^2$ & $\langle \mu \rangle$ & $\sigma_\rho^2$ & $\sigma_\mu^2$ & $\langle \mu \rangle$ \\ \midrule
$R=10$ Mpc/$h$ &                 &                &                       &                 &                &                       &                 &                &                       \\ \midrule
$z=0$          & 0.567           & 0.392          & -0.205                & 0.612           & 0.428          & -0.223                & 0.716           & 0.465          & -0.246                \\
$z=1$          & 0.195           & 0.167          & -0.0836               & 0.199           & 0.171          & -0.0857               & 0.223           & 0.188          & -0.0953               \\ \midrule
$R=15$ Mpc/$h$ &                 &                &                       &                 &                &                       &                 &                &                       \\ \midrule
$z=0$          & 0.276           & 0.233          & -0.118                & 0.291           & 0.248          & -0.126                & 0.345           & 0.282          & -0.144                \\
$z=1$          & 0.0993          & 0.093          & -0.0468               & 0.101           & 0.0945         & -0.0475               & 0.114           & 0.106          & -0.0532               \\ \midrule
$R=20$ Mpc/$h$ &                 &                &                       &                 &                &                       &                 &                &                       \\ \midrule
$z=0$          & 0.163           & 0.149          & -0.0745               & 0.17            & 0.157          & -0.078                & 0.204           & 0.184          & -0.0922               \\
$z=1$          & 0.0598          & 0.0579         & -0.0285               & 0.0603          & 0.0586         & -0.0288               & 0.0684          & 0.0661         & -0.033                \\ \bottomrule
\end{tabular}
\caption{Measured variances and means of the smoothed density and log-density field for the standard cosmology, $f(R)$ gravity with $|f_{R0}| = 10^{-5}$ and DGP gravity with $\Orc = 0.25$. All values for \LCDM{} and F5 are the average over eight realisations.}
\label{tab:summary_stats}
\end{table*}

Table~\ref{tab:summary_stats} lists variances and means extracted from the simulations for various smoothing radii and redshifts, which we used to produce the \LCDM{} and modified gravity results presented in Section~\ref{sec:results}. The corresponding quantities for the dark energy cosmologies can be requested to the authors of \cite{Shin:2017}.

\section{Validation of LDT predictions for small deviations from \texorpdfstring{\LCDM{}}{LCDM} fiducial}
\label{app:small_deviations}

\subsection{Modified gravity}

\begin{figure*}
    \centering
    \includegraphics[width=0.7\textwidth]{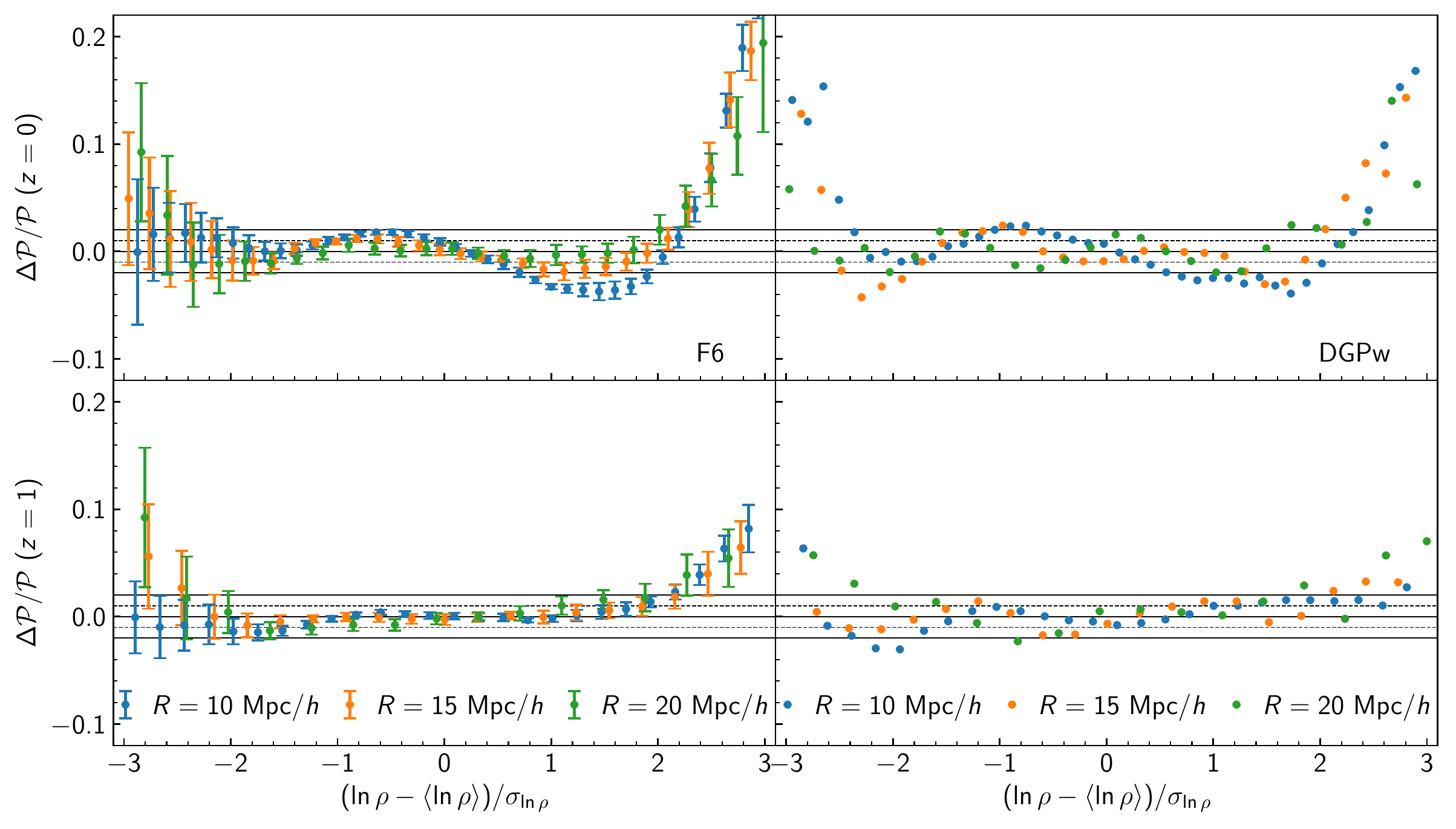}
    \caption{Residuals between the measured and predicted matter PDF normalised to the theory predictions for $z=0$ (top) and $z=1$ (bottom) in $f(R)$ gravity with $|f_{R0}| = 10^{-6}$ (left) and DGP with $\Orc = 0.0625$ (right). Different colors indicate the radii of the spheres used for smoothing the density field, 10 \Mpch\ (blue), 15 \Mpch\ (orange) and 20 \Mpch\ (green). The solid and dashed lines mark the 1\% and 2\% accuracy, respectively.}
    \label{fig:mg2_residuals}
\end{figure*}

\begin{figure*}
    \centering
    \includegraphics[width=0.9\columnwidth]{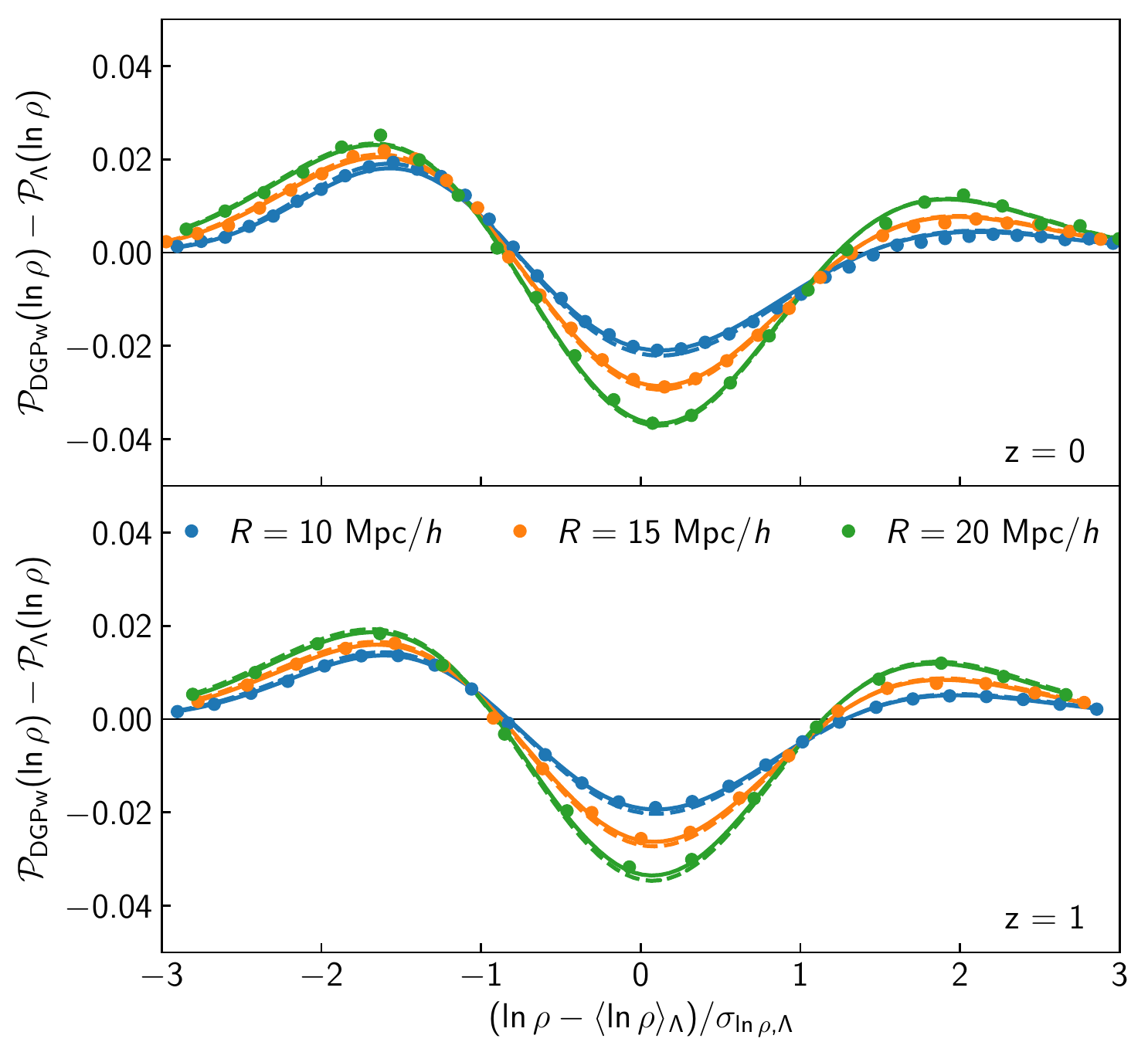}
    \quad
    \includegraphics[width=0.9\columnwidth]{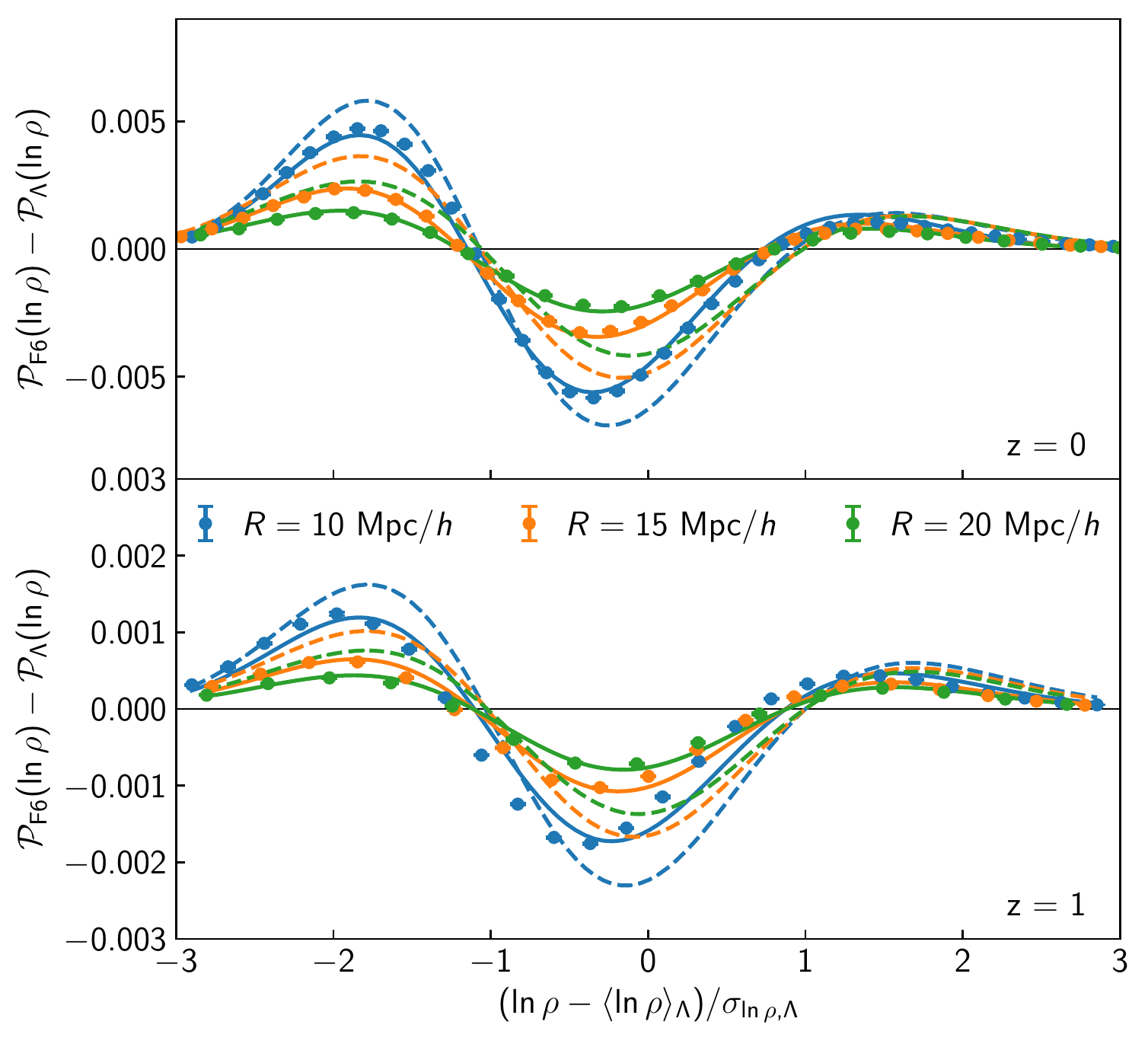}
    \caption{Measured (data points) and predicted (lines) responses of the matter PDF to modified gravity at $z=0$ (top) and $z=1$ (bottom). The density field is averaged in spheres of radius $R =$ 10 \Mpch\ (blue), 15 \Mpch\ (orange) and 20 \Mpch\ (green). Solid lines are obtained from the measured non-linear variance, $\sigma^2_\mu$, while dashed lines from its approximation Equation~\eqref{eq:siglogcos} used in \pyLDT{}. \textit{Left:} differences from \LCDM{} for DGP gravity with $\Orc = 0.0625$. \textit{Right:} same as left panel for $f(R)$ gravity with $|f_{R0}| = 10^{-6}$.}
    \label{fig:mg2-lcdm_diff}
\end{figure*}

Our theoretical prediction for the matter PDF have been validated with numerical simulations in the main text for the two modified gravity models F5 and DGPm (see Figure~\ref{fig:mg-lcdm_diff}).
Figures~\ref{fig:mg2_residuals} and \ref{fig:mg2-lcdm_diff} illustrate the accuracy of the theoretical predictions for F6 and DGPw in the form of residuals from the simulations and departures from \LCDM{}, as well as the impact of using the log-normal approximation (Equation~\ref{eq:siglogcos}) in  \pyLDT{}.

% \section{Validation of analytical and simulated derivatives on Fisher forecasts}
% \label{sec:validation_of_derivatives}

\subsection{Dark energy}

Our theoretical prediction for the matter PDF have been validated with numerical simulations in the main text for large changes in the parametrised dark energy equation of state (see Figure~\ref{fig:wcdm-lcdm_diffs}). A similar comparison for all \LCDM{} parameters has been provided in \cite{Uhlemann:2020}, see Figures~8~and~9 therein. Here we provide complementary results for the full set of $w_0$CDM parameters available from the {\sc Quijote} simulations \citep{Villaescusa-Navarro:2020}.

\begin{figure}
    \centering
    \includegraphics[width=\columnwidth]{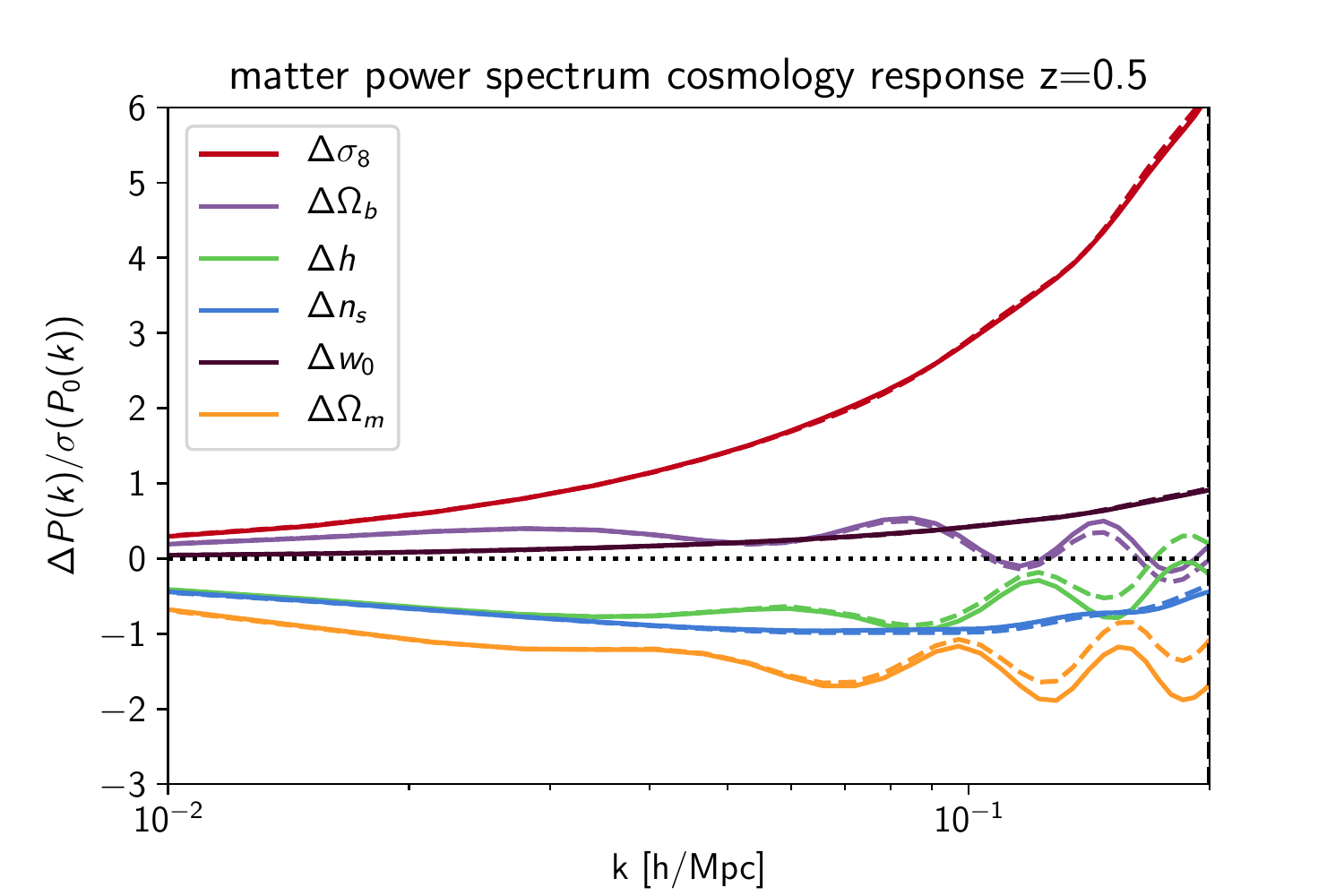}
    \caption{Validation of matter power spectrum derivatives up to $k_{\rm max}=0.2$ \hMpc{} at $z=0.5$ as obtained from {\sc hmcode} (dashed lines) compared to measurements in the {\sc Quijote} simulation suite (solid lines) for the full set of 6 cosmological parameters for $w_0$CDM. We show the results as a signal-to-noise like ratio of the differences in the matter power spectrum, $\Delta P(k)$, and the expected error on the fiducial power spectrum from the diagonal of the covariance matrix, $\sigma(P_0(k))$. }
    \label{fig:derivatives_Pk_validation}
\end{figure}

To further validate our joint matter PDF and matter power spectrum constraints, we compare theoretical power spectrum derivatives from {\sc hmcode} to measured derivatives from the {\sc Quijote} simulations in Figure~\ref{fig:derivatives_Pk_validation}. When limiting ourselves to mildly non-linear scales $k<k_{\rm max}=0.2 \ h/$Mpc we find good agreement between the two. We notice slight discrepancies for some parameters that turn out to be unimportant when constraining just a few parameters, but hampering agreement between theory and simulation matter power spectrum in a Fisher forecast simultaneously varying all $w_0$CDM parameters. To mitigate this minor issue for the power spectrum, we decided to include an external prior on $\{\Ob, n_s\}$, as described in the main text.

% \begin{figure}
% \includegraphics[width=\columnwidth]{figures/diffPDFlnwcorrHRplotz0predsig.pdf}

% \caption{The fractional difference of the PDF as measured (data points) and predicted (solid lines) for a change in the dark energy equation of state $w_0=-1\pm 0.05$ for $R=10,15,20$ Mpc/h (blue,green,red) and the fiducial model (with equal $\sigma_8$) as a function of density at redshifts $z=0,0.5,1$ (solid, dashed, dotted). The gray vertical lines indicate the region used for the Fisher analysis in Section~\ref{sec:Fisher}.}
%   \label{fig:diffPDFsimvstheo_w0}
% \end{figure}

\begin{figure}
    \centering
    \includegraphics[width=\columnwidth]{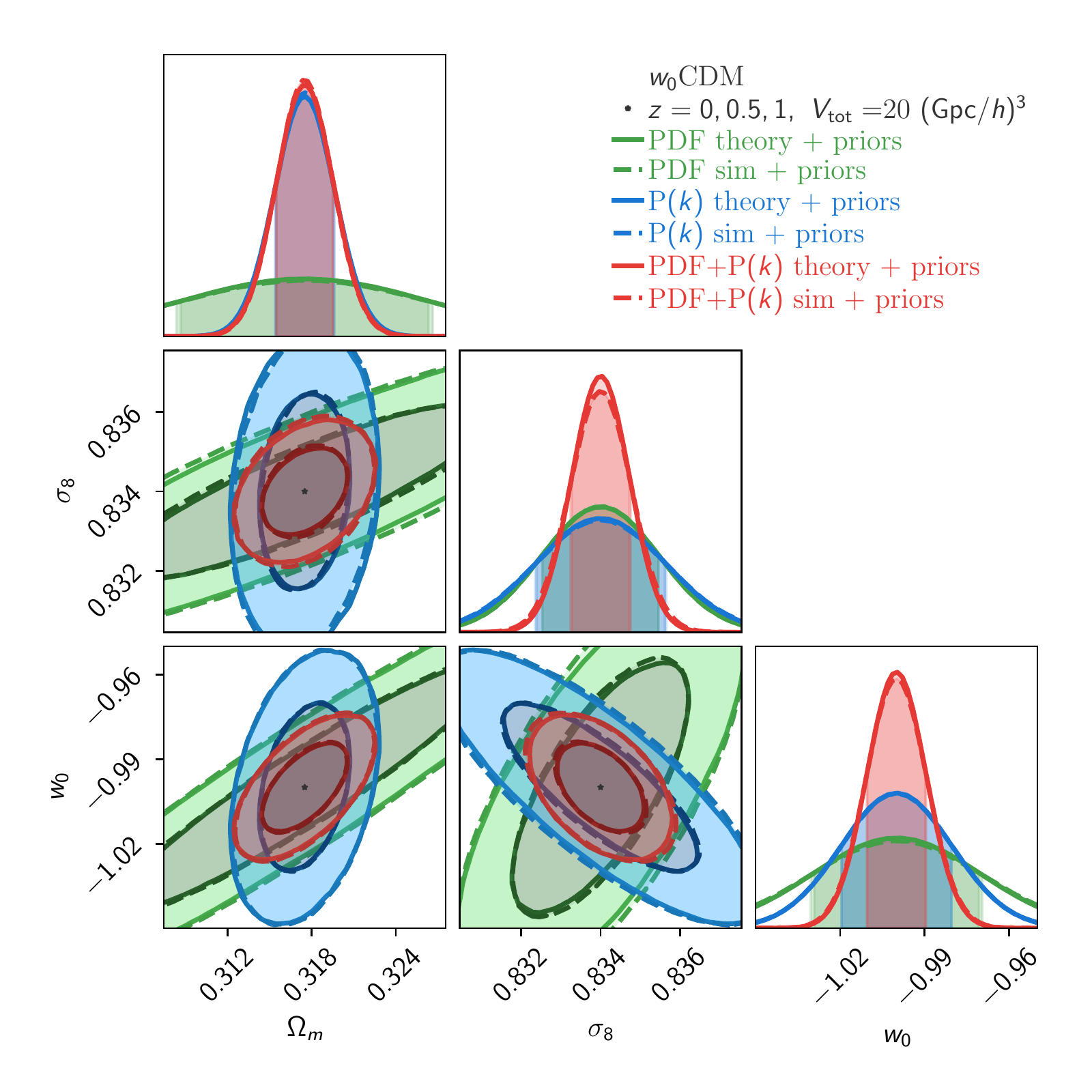}
    \caption{Marginalised constraints on $\Om$, $\sigma_8$ and $w_0$ from a $w_0$CDM Fisher forecast obtained from using the theory derivatives for the matter PDF and power spectrum (solid lines) or the simulated derivatives from the {\sc Quijote} simulation suite (dashed lines), both when including a prior on $\{\Ob,n_s\}$. This validates the robustness of our theoretical predictions used for the forecasts in the main text.}
    \label{fig:fisher_sim_vs_theory}
\end{figure}

Using this prior, we successfully validated the constraints obtained using our theoretical derivatives against simulations by performing a Fisher forecast with all six $w_0$CDM parameters for which derivatives are available from the {\sc Quijote} simulation suite. In Figure~\ref{fig:fisher_sim_vs_theory} we demonstrate that we obtain virtually identical results for both the degeneracy directions and the individual parameter constraints when marginalised over all other parameters.

%%%%%%%%%%%%%%%%%%%%%%%%%%%%%%%%%%%%%%%%%%%%%%%%%%

% Don't change these lines
\bsp	% typesetting comment
\label{lastpage}
\end{document}